\documentclass[reprint,aps,prapplied]{revtex4-2}
\usepackage{dcolumn}%
\usepackage{bm}
\usepackage{graphicx} 
\usepackage{amsmath} 
\usepackage{amssymb} 
\usepackage{enumitem}
\usepackage{tabularx}
\usepackage{soul}
\usepackage{lipsum}
\usepackage[mathscr]{eucal}
\usepackage{dsfont}
\usepackage{comment}
\usepackage{physics}
\usepackage[cal=boondoxo]{mathalfa}
\usepackage{hyperref}
\usepackage{xcolor}

\DeclareMathOperator\arctanh{arctanh}
\DeclareMathOperator\Var{Var}

\renewcommand{\vec}[1]{{\mathbf #1}}

\newcommand{\be}{\begin{equation}}
\newcommand{\ee}{\end{equation}}

\begin{document}

\title{Harnessing coherent-wave control for sensing applications}

\author{Pablo Jara$^1$}
\email{pxjdbk@mst.edu}
\author{Arthur Goetschy$^2$}
\author{Hui Cao$^3$}
\author{Alexey Yamilov$^1$}
\email{yamilov@mst.edu}

\affiliation{$^1$Department of Physics, Missouri University of Science and Technology, Rolla, Missouri 65409, USA\\
$^2$ESPCI ParisTech, PSL Research University, CNRS, Institut Langevin, F-75005 Paris, France\\
$^3$Department of Applied Physics, Yale University, New Haven, Connecticut 06520, USA}

\date{\today}

\begin{abstract}
Imaging techniques such as functional near-infrared spectroscopy (fNIRS) and diffuse optical tomography (DOT) achieve deep, non-invasive sensing in turbid media, but they are constrained by the photon budget, as most of the injected light is lost to scattering before reaching the detector.
Wavefront shaping (WFS) can enhance signal strength via interference at specific locations within scattering media, enhancing light-matter interactions and potentially extending the penetration depth of these techniques.
Interpreting the resulting measurements rests on the knowledge of optical sensitivity -- a relationship between detected signal changes and perturbations at a specific location inside the medium. 
However, conventional diffusion-based sensitivity models rely on assumptions that become invalid under coherent illumination. 
In this work, we develop a microscopic theory for optical sensitivity that captures the inherent interference effects that diffusion theory necessarily neglects. 
We show analytically that, under disorder averaging with random illumination, the microscopic and diffusive descriptions coincide.
Beyond this limit, our framework identifies WFS strategies that enhance sensitivity.
We demonstrate that the input state obtained through phase conjugation at a given point inside the system leads to the largest enhancement of optical sensitivity but requires an input wavefront that depends on the target position.
In sharp contrast, the maximum remission eigenchannel, corresponding to the largest eigenvalue of the monochromatic remission matrix, leads to a global enhancement of the sensitivity map with a fixed input wavefront. 
This global enhancement equals to remission enhancement and preserves the spatial distribution of the sensitivity, making it compatible with existing DOT reconstruction algorithms.  
Our results, validated through extensive numerical simulations, establish the theoretical foundation for integrating wavefront control with diffuse optical imaging, enabling deeper tissue penetration through improved signal strength in biomedical applications.
\end{abstract}

\maketitle

\section{Introduction\label{sec:intro}}

\noindent In optics, acoustics, seismology, microwave physics, and other fields, scattered waves serve as versatile non-invasive probes for a stand-off characterization of complex scattering systems~\cite{1978_Ishimaru,1980_Aki,2002_Pike_Sabatier,1995_Chew_Scattering_book}. 
While both electromagnetic waves and sound waves are extensively used in medical imaging~\cite{2022_Bertolotti_Imaging_review}, optical methods operating in the near-infrared regime offer unique advantages for probing biological tissue, combining molecular specificity with deep penetration~\cite{2007_Wang}. 
Ballistic or nearly-ballistic photon imaging methods such as e.g. optical coherence tomography (OCT) offer high spatial resolution but they are limited to essentially superficial tissue layers~\cite{2001_Bouma_OCT}. 
Diffuse optical tomography (DOT) and functional near-infrared spectroscopy (fNIRS) extract information from multiply scattered light enabling interrogation of much larger depths~\cite{1995_Yodh_DOT,2001_Boas_DOT,2010_Durduran_DOT,2014_Scholkmann_fNIRS,2023_Vidal_fNIRS_DOT_perspectives} aided by reduced attenuation in the near-infrared spectral transparency window in biological materials~\cite{1977_Jobsis}. 
These techniques operate in {\it remission geometry} -- it involves injecting the near-infrared light into tissue at source position, while collecting the multiply scattered photons that emerge at a detector placed at different location on the tissue surface. 
The detected light signals from multiple source-detector pairs are then used to reconstruct the spatial distribution of tissue optical properties through inverse problem techniques\cite{1999_Arridge_DOT_inverse_problem,2009_Arridge_Schotland_DOT_inverse_problem}.
Therefore, the success of the DOT-based approaches hinges on two key principles~\cite{2010_Durduran_DOT}: 
(1) the ability to predict how local changes in tissue properties modulate the detected signal (sensitivity)~\cite{2024_Blaney_sensitivity_review}, and 
(2) the ability to reconstruct the photon propagation paths connecting sources to detectors (inverse problem). 
The omni-directional nature of diffuse light propagation leads to a decrease of signal strength with both depth and source-detector separation~\cite{2001_Boas_DOT}, whereas the maximum input intensity is limited by tissue damage thresholds~\cite{2017_Tsai_Damage_threshold}. 
Fixed photon budget ultimately constrains the maximum depth accessible by DOT-based approaches~\cite{2015_Mora_Depth_limit,2016_Pifferi_Depth_limit}. 

Wavefront shaping (WFS) techniques~\cite{2012_Mosk_Wavefront_shaping_review,2015_Yu_Wavefront_Shaping_Review,2017_Rotter_Gigan_review,2022_Gigan_Roadmap,2022_Cao_Mosk_Rotter_review} have emerged as promising avenue for enhancing signal strength at depth~\cite{2020_Yoon_Review,2022_Bender_Depth_Targeted_Energy_Deposition}. 
This is expected to enhance light-matter interaction and hence increase modulation of the remitted signal. However, the field/intensity patterns created inside the scattering medium can no longer be adequately described by diffusion theory~\cite{2011_Mosk,2015_Genack_Eigenchannels_Inside,2016_Sarma_Open_Channels,2016_Ojambati_Fundamental_Mode_Experiment,2021_Carminati_Schotland}, nor can the relationship between signal modulation and the underlying perturbation be justified~\cite{2022_Bender_Coherent_Enhancement} using traditional diffusion-based approaches~\cite{2010_Durduran_DOT}. 
These observations put into question the applicability of diffusion-based interpretation of the sensitivity map when used in conjunction with WFS input. 

\begin{figure}
\includegraphics[width=\linewidth]{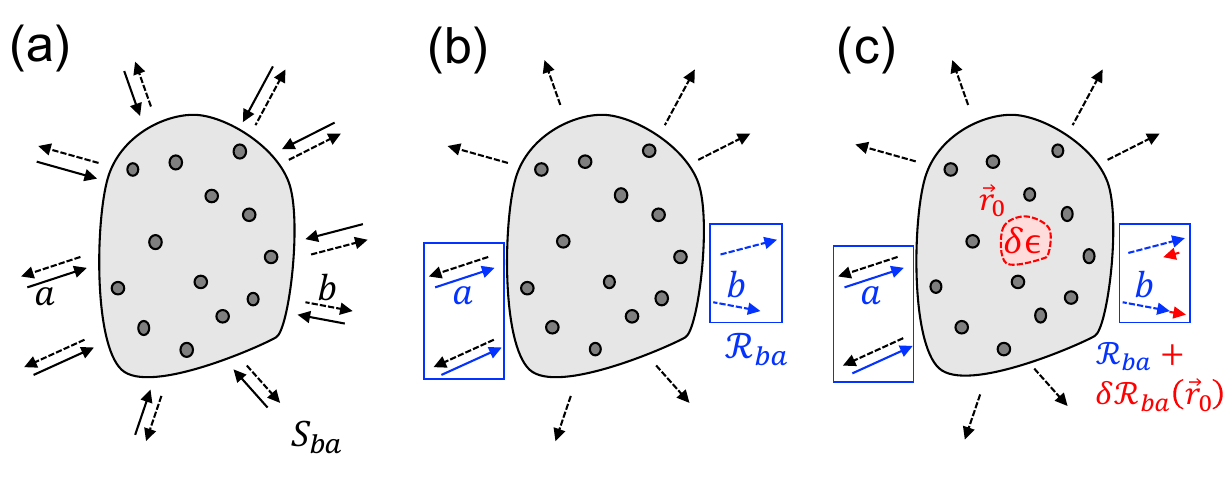}
\vskip -0.4cm
\caption{Schematic of scattering approach. 
(a) Scattering matrix element $S_{ba}$ relates field amplitudes of the incident mode $a$ (solid line arrows) and outgoing mode $b$ (dashed arrows). 
(b) Experimentally, one has access to a finite number of spatial channels (blue arrows), described by matrix ${\cal R}_{ba}$, which is a subset of $S$. 
(c) Addition of a small perturbation in the dielectric properties inside the system results in a change $\delta{\cal R}_{ba}(\vec{r}_{0})$, which depends on its location $\vec{r}_{0}$.
\label{fig:geometry_sketch_generic}}
\end{figure}

To formally describe wave transport through complex media, it is useful to adopt a scattering framework. Element $S_{ba}$ of a scattering matrix describes amplitude of  the wave scattered by system into spatial channel $b$ (e.g. wavevector direction), upon a coherent excitation via channel $a$~\cite{2006_Sheng,1997_Beenakker}, see Fig.~\ref{fig:geometry_sketch_generic}a. Commonly, experimental setup for interrogation of the scattering system involves detection of a subset of all incoming and outgoing waves~\cite{2017_Rotter_Gigan_review}, ${\cal R}_{ba}$, Fig.~\ref{fig:geometry_sketch_generic}b. Sensitivity analysis~\cite{1997_Schotland,1999_Arridge_DOT_inverse_problem,2009_Arridge_Schotland_DOT_inverse_problem,2010_Durduran_DOT} is a powerful tool that aims to characterize the variations in local properties of complex medium, e.g. in real or imaginary part of the dielectric function $\delta\epsilon$ at a spatial position ${\bf r}_0$, through detection of changes $\delta{\cal R}_{ba}({\bf r}_0)$ in scattered amplitudes, Fig.~\ref{fig:geometry_sketch_generic}c. 

An important insight into problem has been obtained recently in Ref.~\cite{2022_Bender_Coherent_Enhancement}, where wavefront control technique was applied to the continuous wave (CW) matrix ${\cal R}$ defined specifically in the remission geometry -- a configuration involving a semi-infinite medium where both source and detector lie on the same side of the sample. 
In equation ${\bf E}_\text{out}={\cal R}\,{\bf E}_\text{in}$, the matrix operator relates the incident field ${\bf E}_\text{in}$ and remitted field ${\bf E}_\text{out}$, which can be decomposed into $N_1$ and $N_2$ spatial channels, e.g. flux-carrying modes of the optical fibers in DOT. 
As could be expected, the numerical and experimental results indeed demonstrated a significant (an order of magnitude) enhancement of the remitted signal by exciting the highest remission eigenchannel -- input wavefront given by the eigenvector of ${\cal R}^\dagger{\cal R}$, corresponding to its largest eigenvalue. 
Importantly, a new {\it microscopic} formulation based on scalar wave equation was proposed to obtain the sensitivity map~\cite{2022_Bender_Coherent_Enhancement}. 
Remarkably, it was found that: {\it spatial distribution} of the sensitivity map for the maximum remission eigenchannel (MRE) remained the same as that for random input, while the {\it magnitude} of the sensitivity enhancement was nearly equal to the magnitude of the remission enhancement. 
These findings suggest a possibility of extending the diffusion-based DOT treatments to include WFS excitation. 
Indeed, if sensitivity magnitude could be computed theoretically, while the inverse-problem solution could use the same sensitivity map found in diffusion model, coherent enhancement of remitted signal would immediately translate into deeper non-invasive sensing and imaging in biological media. 

While results in our previous work~\cite{2022_Bender_Coherent_Enhancement} provided an important clue into how optical sensitivity is affected by coherent control, a comprehensive study is required to formulate new approach and validate its applicability. 
First, the microscopic theory must be rigorously tested against the widely-used diffusion-based approach, e.g. for random input excitation the field-based approach should reduce to diffusion theory. 
Second, dependence of the sensitivity on key system parameters like transport mean free path and source-detector geometry needs systematic verification. 
It also remains unclear what the optimal strategy for enhancing optical sensitivity is within the context of wavefront shaping. 
Furthermore, since coherent WFS can manipulate both amplitude, phase and polarization of incident waves, extension of the scalar wave theory to incorporate full vectorial nature of the electromagnetic waves is essential for a complete theoretical description of optical sensitivity. 

In this work, we present a comprehensive study of CW optical sensitivity in complex media, focusing on the role of wavefront shaping in enhancing detection capabilities. 
In Sec.~\ref{sec:sensitivity_analytical_scalar}, we derive analytical expressions for sensitivity using a microscopic approach that remains valid even when coherent wavefront shaping is employed. 
We establish a connection between optical sensitivity and phase conjugation~\cite{2008_Yaqoob_Phase_conjugation,2000_Fink}, demonstrating how tailored input wavefronts can maximize the system’s response to perturbations. 
In Sec.~\ref{sec:numerical_simulation_scalar}, we perform numerical simulations to validate our theoretical predictions, comparing sensitivity enhancements achieved through different excitation schemes. 
We further investigate how sensitivity enhancement scales with system parameters such as the transport mean free path and the source-detector geometry. 
In Sec.~\ref{sec:sensitivity_analytical_vector}, we extend our analysis to vector electromagnetic waves and demonstrate that the key results remain consistent with those derived for scalar waves. 
Our findings provide a theoretical foundation for understanding sensitivity in coherent wave transport, offer insights into different wavefront shaping strategies, and pave way for next-generation diffuse optical imaging techniques.
    
\section{Optical sensitivity: analytical results for scalar waves\label{sec:sensitivity_analytical_scalar}}

\noindent In this section, we consider scalar waves and derive analytical expressions for the continuous-wave (CW) optical sensitivity using a unified approach requiring minimal assumptions. In Sec.~\ref{sec:field_sensitivity} we present a derivation of sensitivity that applies to arbitrary system geometries and does not rely on diffusion approximation. We then establish in Sec.~\ref{sec:phase_conjugation} the connection between optical sensitivity and phase conjugation, describing construction of such a wavefront and demonstrating enhancement of sensitivity. In Sec.~\ref{sec:sensitivity_diffusion_approximation} we develop the disorder-averaged intensity sensitivity model, which requires the additional assumptions that the input wavefront is uncorrelated with the disordered system and that wave transport is diffusive. We prove analytically in Sec.~\ref{sec:equivalence} that under random excitation, these two models lead to identical results in diffusive media. We conclude with Sec.~\ref{sec:normalized_sensitivity} by introducing normalized sensitivity and discuss its advantages in sensitivity analyses.

\subsection{Microscopic approach to optical sensitivity\label{sec:field_sensitivity}}
\noindent We consider CW monochromatic scalar field $E_0(\vec{r})$ in a medium with dielectric function $\epsilon(\vec{r})$, representing one realization of disorder. The unperturbed field satisfies the wave equation 
\begin{equation}
    \left[\nabla^2 + k^2\epsilon(\vec{r})\right]E_0(\vec{r}) = 0,
    \label{EqWave1}
\end{equation}
where $k=2\pi/\lambda_0$ is wavenumber in vacuum, and $\epsilon(\vec{r})$ is a spatially varying dielectric function of the medium. A small perturbation $\delta\epsilon(\vec{r})$ near $\vec{r}_0$ modifies the wave equation, resulting in a new solution $E(\vec{r})$. The change in the field $\delta E(\vec{r}) \equiv E(\vec{r}) - E_0(\vec{r})$ can be written using the Lippmann-Schwinger equation: 
\begin{equation}
    \delta E(\vec{r}) = \int d\vec{r}'\,G_0(\vec{r},\vec{r}')\,{\cal V}(\vec{r}')\,E(\vec{r}').
    \label{EqLS}
\end{equation}
Here we adopted the language of the scattering theory with $G_0=(k^2-H_0)^{-1}$ being the (retarded) Green's function of the original wave equation, with $H_0 = -\nabla^2 - k^2[\epsilon(\vec{r}) - 1]$ and ${\cal V}(\vec{r}) = -k^2\delta\epsilon(\vec{r})$. 
In the Born approximation and assuming a point-like perturbation at $\vec{r}_0$, we set $\delta\epsilon(\vec{r}) = \delta\epsilon\,\delta V\,\delta(\vec{r} - \vec{r}_0)$, yielding 
\begin{equation}
    \delta E(\vec{r}) \simeq\tilde{\cal V}\, G_0(\vec{r}, \vec{r}_0) E_0(\vec{r}_0),
    \label{EqBornField}
\end{equation}
where $\tilde{\cal V}=-k^2\delta\epsilon \,\delta V$ is the strength of the perturbation and $\delta V$ represents an effective volume associated with the perturbation.

The above expression, given in the position representation $E(\vec{r})=\braket{\vec{r}}{E}$, also holds in any representation $E(\psi)=\braket{\psi}{E}$. 
This allows us to cast problem in language of optical tomography. 
To that end, we evaluate the variation of flux $F$ received by a detector, see Fig.~\ref{fig:geometry_sketch_generic}c, in the basis of flux-normalized output modes $\{\ket{\psi_b}\}$,
\begin{equation}
\delta F = 
\sum_{b=1}^{N_2} \vert \braket{\psi_b}{E}\vert^2 - \vert\braket{\psi_b}{E_0}\vert^2\simeq 
\sum_{b=1}^{N_2} 2\text{Re}\left[\delta E_b\,u_b^* \right] ,
\label{EqSolFlux}
\end{equation}
where $u_b= \braket{\psi_b}{E_0}$ in the output field measured in the channel $b$ and  
\begin{equation}
\delta E_b = \frac{\tilde{{\cal V}}}{2i}\,\phi_b({\vec{r}_0})\,E_0(\vec{r}_0).
\label{EqDeltaEb}
\end{equation}
Here, $\phi_b({\vec{r}_0})=2i\mel{\psi_b}{G_0}{\vec{r}_0}$ represents the field in channel $b$ when excited by a point-like source at $\vec{r}_0$~\cite{1981_Fisher_Lee_Relation_between_conductivity_and_transmission_matrix}. By reciprocity, it is also the field at $\vec{r}_0$ when channel $b$ is excited from the output port. As flux-normalized states $\psi_b(\vec{r})$ are related to intensity normalized states $\chi_b(\vec{r})$ according to  $\psi_b(\vec{r})= \chi_b(\vec{r})/\sqrt{k_b^z}$, where $k_b^z$ is the longitudinal component of the momentum of $\chi_b(\vec{r})$, we can also write    $\phi_b({\vec{r}_0})=2i\sqrt{k_b^z}\mel{\vec{r}_0}{G_0}{ \chi_b }=2ik_b^z\mel{\vec{r}_0}{G_0}{\psi_b}$. 
Substituting Eq.~\eqref{EqDeltaEb} into Eq.~\eqref{EqSolFlux} gives
\be
\delta F\simeq\text{Re}\left[-i\,\tilde{{\cal V}}\,\phi(\vec{r}_0)\,E_0(\vec{r}_0)\right],
\label{EqSolExact}
\ee
where we introduced the notations
\be
\phi(\vec{r}_0)=\sum_{b=1}^{N_2} u_b^*\,\phi_b(\vec{r}_0),
\label{eq:phi_of_r0}
\ee
and summation over $b$ runs over $N_2$ output channels.

To complete the derivation, we use Eq.~(\ref{EqSolExact}) and formally define optical sensitivity to a {\it local absorber} with $\delta \epsilon=i\epsilon''$ as
\be
{\cal S}(\vec{r}_0)\equiv\displaystyle\frac{dF(\vec{r}_0)}{d\epsilon''}=-k^2\,\delta V\,\text{Re}\left[ \phi(\vec{r}_0) E_0(\vec{r}_0)\right].
\label{EqSiBorn}
\ee 
This expression~\cite{2022_Bender_Coherent_Enhancement} serves as the foundation for the analysis that follows. 

It is illuminating to express the input state used to excite $E_0(\vec{r})$ in terms of $N_1$ flux-normalized input channels $\{\ket{\psi_a}\}$ as $\ket{\psi^{\text{in}}}=\sum_{a=1}^{N_1} v_a \ket{\psi_a}$. Here $v_a$ are amplitudes of each mode satisfying a constraint $\sum_{a=1}^{N_1}\vert v_a\vert^2=N_1$. Therefore, $E_0(\vec{r}_0)=\sum_{a=1}^{N_1} v_a\phi_a(\vec{r}_0)$,
similar to the output states introduced in Eq.~\eqref{eq:phi_of_r0}. We find
\be
{\cal{S}}(\vec{r_0})=-k^2\delta V\,\sum_{baa'}\text{Re}\left[
\phi_b(\vec{r_0})
{\cal{R}}^*_{ba'}
\phi_a(\vec{r_0})
v_{a'}^* v_a
\right],
\label{EqSensitivity_va}
\ee
where we took advantage of the relationship $u_b=\sum_{a=1}^{N_1}{\cal R}_{ba}v_a$. We note that Eqs.~(\ref{EqSiBorn},\ref{EqSensitivity_va}) are, in fact, equivalent. 

We end this section with several comments. First, Eqs.~(\ref{EqSiBorn},\ref{EqSensitivity_va}) are obtained in Born approximation (weak perturbation), which means they become exact in the limit  $k^{\cal D}\,\delta V\,\epsilon'' \ll 1$, where ${\cal D}=2,3$ is the dimensionality of the system. 
Secondly, sensitivity in Eqs.~(\ref{EqSiBorn},\ref{EqSensitivity_va}) is expressed in terms of complex-valued fields/coefficients $\{\phi(\vec{r}_0),E_0(\vec{r}_0)\}$ or $\{\phi_a(\vec{r}_0), \phi_b(\vec{r}_0), v_a, {\cal{R}}_{ba}\}$  respectively, all computed (or, in the case of $v_a$, specified) for the same realization of disorder. 
To compute the statistically averaged sensitivity from Eqs.~(\ref{EqSiBorn},\ref{EqSensitivity_va}), one must ensemble-average the full product of fields in the sensitivity expression, rather than averaging the fields themselves. 
This distinction preserves interference effects that contribute to coherent transport, even after disorder averaging. 
Lastly, we note that ${\cal S}(\vec{r}_0)$ has an implicit dependence on the location of the perturbation center that makes it {\it sensitivity map}: {\it external} measurement of the flux, contains the information about the {\it internal} location $\vec{r}_0$ of the perturbation. It is this key feature that enables of non-invasive approaches to tomographic reconstruction of optical properties of the medium.

\subsection{Phase conjugation and optical sensitivity\label{sec:phase_conjugation}}
\noindent Equation~\eqref{EqSensitivity_va} shows that the ability to enhance optical sensitivity at a target position $\vec{r}_0$ within a scattering medium hinges on a judicial choice of the input wavefront defined by $v_a$ coefficients. Among various wavefront shaping strategies, phase conjugation (PC) appears as a particularly intuitive and effective approach~\cite{2008_Yaqoob_Phase_conjugation,2010_Popoff_NatComm,2012_Mosk_NatPhoton,2013_Wang_NatPhoton,2022_Gigan_Roadmap}. Indeed, in a complex medium, a point source at $\vec{r}_0$ generates a wavefront that propagates outward, undergoing multiple scattering events. 
The phase-conjugated input, constructed by reversing the outgoing wavefront detected at the position of our extended source, excites a field which naturally retraces these scattering paths in time-reversed fashion. 
This is the consequence of the time-reversal symmetry~\cite{1997_Fink,2000_Fink} in CW propagation. 
Refocusing energy back at $\vec{r}_0$ is not perfect because time reversal of partial outgoing waves is incomplete. 
It is expected that this constructive interference should maximize the local field amplitude, for given finite control, thereby amplifying the response to small perturbations in the optical properties at $\vec{r}_0$ -- a concept that has been formalized using the generalized Wigner–Smith operator and related optimal input states~\cite{2017_Rotter_Focusing,2020_Rotter_Optimal,2021_Davy_GWS,2021_Bouchet_maximum_information}. 
The modal coefficients for the PC wavefront are given, in our notation, by
\be
v^\text{PC}_a=\left[\frac{N_1}{I_\text{in}(\vec{r}_0)}\right]^{1/2}\phi_a^*(\vec{r}_0),
\label{EqvaPC}
\ee
where $I_\text{in}(\vec{r_0})=\sum_{a=1}^{N_1} |\phi_{a}(\vec{r_0})|^2$, and $N_1^{1/2}$ factor ensures that the total input flux remains equal to $N_1$, the convention adopted earlier. By substituting Eq.~\eqref{EqvaPC} into Eq.~\eqref{EqSensitivity_va}, the sensitivity with PC input  becomes
\be
{\cal{S}}^\text{PC}(\vec{r_0})=-N_1 k^2 \delta V \sum_{ba}\text{Re}\left[ \phi_b(\vec{r_0}){\cal{R}}^*_{ba}\phi_a(\vec{r_0})\right].
\label{EqPCsensitivity}
\ee
To put this result in perspective, we obtain expression for sensitivity under two alternative excitation schemes. First, we consider a random combination of input channels, such that $\overline{v^*_{a'}v_a}=\delta_{a'a}$. Here, the averaging is performed only over different combinations of random inputs, in the particular realization of the system -- i.e. no disorder ensemble average is assumed. In this case Eq.~\eqref{EqSensitivity_va} yields
\begin{align}
{\cal S}^\text{RI}(\vec{r_0})
&=-k^2\delta V \sum_{ba}\text{Re}\left[ 
\phi_b(\vec{r_0})
{\cal{R}}^*_{ba}
\phi_a(\vec{r_0})
\right]\nonumber\\
&={\cal{S}}^\text{PC}(\vec{r_0})/N_1.
\label{EqRandomSensitivity}    
\end{align}
Here, superscript RI stands for `random input'. Remarkably, we observe that excitation with PC leads to enhancement of sensitivity by a factor of $N_1$ compared to random input. While such an enhancement might be expected, it is, in fact, a highly nontrivial result. Phase conjugation is well known to enhance the {\it intensity} at the target location by such a factor due to constructive interference. In the case optical sensitivity, however, the quantity of interest involves a product of two distinct fields, see Eqs.~(\ref{EqSiBorn},\ref{EqSensitivity_va}): one propagating from the injection site and the other from the detection port. An enhancement of the field originating from the source is expected, given that phase conjugation directly optimizes energy delivery to $\vec{r}_0$. Surprisingly, the second field, originating from the detection site, is also enhanced by the same factor. This outcome is not obvious, as the back-propagating field is not directly controlled by the phase-conjugated input. Instead, its enhancement arises due to the intrinsic reciprocity of wave propagation in complex media, which ensures that the time-reversed wavefront optimally reconstructs the field at the target location, thereby reinforcing the response to perturbations and leading to the obtained sensitivity enhancement.

In our previous work\cite{2022_Bender_Coherent_Enhancement}, we demonstrated enhancement of sensitivity when system was excited by singular vectors of the remission matrix ${\cal R}$. In this scheme, we use the singular value decomposition of ${\cal R}$,
\begin{align}
{\cal{R}}_{ba}=\sum^{\text{min}(N_1,N_2)}_{\alpha=1} U_{b\alpha}\rho^{1/2}_{\alpha} V^*_{\alpha a},
\label{EqSVD_operator}
\end{align}
and select a specific $\alpha$=1 -- the maximum remission eigenchannel (MRE). We chose $v_a=N_1^{1/2}\,V_{a1}$ -- the first column of the $V_{a\alpha}$ matrix, which corresponds to the maximum remission eigenvalue $\rho_1$. Here, by adding $N_1^{1/2}$ factor, we ensured that total flux used to excite the system is equal to $N_1$, i.e. $\sum_{a=1}^{N_1}|v_a|^2=N_1$, as in the previous two schemes Eqs.~(\ref{EqPCsensitivity},\ref{EqRandomSensitivity}). Substituting this input vector into Eq.~\eqref{EqSensitivity_va} and taking advantage of unitarity of $U$ and $V$ matrices in Eq.~\eqref{EqSVD_operator} we obtain
\begin{align}
&{\cal{S}}^\text{MRE}(\vec{r_0})=\label{EqRemissionEigenchannelSensitivity}\\
&-N_1 k^2 \delta V\sum_{ab}\text{Re}\left[ 
\phi_b(\vec{r_0})
\left(V_{a1}\rho^{1/2}_1 U^*_{1b}\right)
\phi_{a}(\vec{r_0})\right].\nonumber    
\end{align}
Comparing the above to Eqs.~(\ref{EqPCsensitivity},\ref{EqSVD_operator}), we observe that excitation of the remission eigenchannel amounted to the replacement ${\cal{R}}_{ba}=\sum_{\alpha} U_{b\alpha}\rho^{1/2}_{\alpha} V^*_{\alpha a}\rightarrow\left(V_{a1}\rho^{1/2}_1 U^*_{1b}\right)$ in the PC expression for sensitivity. In other words, in selecting MRE, we are singling out the largest singular value contribution the remission matrix and omitting the rest. 

Several remarks are in order. First of all, we would like to stress that similar to Eq.~\eqref{EqSensitivity_va}, all Eqs.~(\ref{EqPCsensitivity},\ref{EqRandomSensitivity},\ref{EqRemissionEigenchannelSensitivity}) hold in the single realization of the scattering system, i.e. no ensemble averaging is assumed. 

Secondly, our result above suggests that
\be
|{\bar{\cal{S}}}^\text{RI}(\vec{r_0})|<
|{\bar{\cal{S}}}^\text{MRE}(\vec{r_0})|<
|{\bar{\cal{S}}}^\text{PC}(\vec{r_0})|,
\label{EqSensitivityComparison}
\ee
with overline denoting statistical averaging. Hence, phase-conjugation approach indeed leads to the highest sensitivity for a given point $\vec{r}_0$. While the strict ordering $|{\cal{S}}^\text{RI}(\vec{r_0})|<|{\cal{S}}^\text{MRE}(\vec{r_0})|<|{\cal{S}}^\text{PC}(\vec{r_0})|$ probably holds in a typical sense, it is important to clarify that the relation between MRE and RI and MRE and PC might not be guaranteed to hold for a single realization of disorder. 
Unlike PC excitation, which deterministically enhances sensitivity by a factor of $N_1$ above the RI result, the MRE scheme involves interference effects determined by the eigenvectors of the remission matrix. 
As a result, ${\cal{S}}^\text{MRE}(\vec{r_0})$ is not guaranteed to even be negatively-defined and can exhibit significant fluctuations due to disorder-specific interference effects. However, in the average sense $|{\bar{\cal{S}}}^\text{MRE}(\vec{r_0})|$ exceeds $|{\bar{\cal{S}}}^\text{RI}(\vec{r_0})|$ by a factor approximately equal to the remission enhancement, as shown in Ref.~\cite{2022_Bender_Coherent_Enhancement} and later in this work. 

Third, {\it enhancement of sensitivity} $\eta_{\cal S}^\text{PC}(\vec{r}_0)\equiv{\cal{S}}^\text{PC}(\vec{r_0})/{\cal{S}}^\text{RI}(\vec{r_0})$ in the phase-conjugation scheme ($N_1$) exceeds that for the largest remission eigenchannel $\eta_{\cal S}^\text{MRE}(\vec{r}_0)\equiv{\cal{S}}^\text{MRE}(\vec{r_0})/{\cal{S}}^\text{RI}(\vec{r_0})$, see e.g. Ref.~\cite{2022_Bender_Coherent_Enhancement}. As it can be seen from Eq.~\eqref{EqvaPC}, achieving the largest sensitivity, the phase conjugation scheme requires input vector uniquely tailored to the specific observation point $\vec{r}_0$. In contrast, remission eigenchannels are computed based on ${\cal R}_{ba}$, which is a {\it global} quantity, independent of $\vec{r}_0$. As such, the largest remission eigenchannels leads to a global enhancement of sensitivity with position-independent input vector\cite{2022_Bender_Coherent_Enhancement}, see Sec.~\ref{sec:equivalence}. An ability to enhance sensitivity globally in a complex scattering system with a fixed input wavefront is highly non-trivial. This effect is captured by the filtered random matrix theory (see Appendix~\ref{sec:filtered_random_matrix_theory}) in diffusive systems, which we turn to next.

\subsection{Diffusion approximation for optical sensitivity\label{sec:sensitivity_diffusion_approximation}}
\noindent Having established a general formulation of optical sensitivity in arbitrary geometry, we now examine its implications in the diffusive regime, where light transport follows a statistical description. 
This section derives the corresponding sensitivity expressions within diffusion theory, providing a benchmark for comparison with the more general microscopic approach, see e.g., Ref.~\cite{2010_Durduran_DOT}. 
Seminal contributions to sensitivity analysis in the diffusive regime have been made by J. Schotland and collaborators, particularly in the context of inverse problems and optical tomography~\cite{1993_Schotland,1997_Schotland,1999_Arridge_DOT_inverse_problem,2001_Schotland_Markel,2009_Arridge_Schotland_DOT_inverse_problem}.
Unlike Sec.~\ref{sec:field_sensitivity}, the starting point of this consideration is the diffusion equation and not wave equation. We begin by discussing assumptions enabling such an approach. First, diffusion equation describes disorder-averaged quantities such as intensity $\overline{I}(\vec{r})$ denoted with an overline. Secondly, such a description is applicable at the scales much larger than transport mean free path $\ell$. Therefore, here it is assumed that relevant length scales, such as e.g. system size and separation distance between source and detector, are much larger than $\ell$~\cite{Akkermans_Montambaux_2007}. 

Let us first consider the case without any perturbation. The input state is expressed in terms of the $N_1$ flux-normalized input channels $\{\ket{\psi_a}\}$ and is assumed to be uncorrelated with the disordered system. The mean flux in the output port then reads:
\be
\overline{F} = \sum_{a,b}\,\overline{\left|{\cal R}_{ba}\right|^2}\,\vert v_a\vert^2.
\label{EqFluxRandom}
\ee
The elements of the remission matrix ${\cal R}_{ba}$ describing scattering from channel $a$ to channel $b$, see Fig.~\ref{fig:geometry_sketch_generic}c, can be expressed using the Fisher-Lee relation~\cite{1981_Fisher_Lee_Relation_between_conductivity_and_transmission_matrix}
\be
{\cal R}_{ba}=2i\sqrt{k_a^z k_b^z}\mel{\chi_b}{G_0}{\chi_a}.
\label{EqFisherLee}
\ee
The flux in Eq.~\eqref{EqFluxRandom} contains disorder-averaged product of two {\it field}\, Green's functions $G_0$. It can be evaluated in the diffusion framework, which amounts to keeping only so-called ladder diagrams in the Bethe-Salpeter equation~\cite{Akkermans_Montambaux_2007,2006_Sheng,1999_VanRossum}. We obtain
\begin{align}
&\overline{\left|{\cal R}_{ba}\right|^2}=\label{EqRab}\\
&4k_a^z k_b^z\int d\vec{r}_1\;d\vec{r}_2\,
\left|\mel{\chi_b}{\overline{G}_0}{\vec{r}_2}\right|^2\,
\Gamma(\vec{r}_2,\vec{r}_1)\,
\left|\mel{\vec{r}_1}{\overline{G}_0}{\chi_a}\right|^2,\nonumber
\end{align}
where the kernel $\Gamma(\vec{r}_2,\vec{r}_1)$ obeys the stationary diffusion equation
\be
-D\nabla^2\Gamma(\vec{r}_2,\vec{r}_1)=\frac{\gamma}{\tau}\delta(\vec{r}_2-\vec{r}_1).
\label{EqGamma_diffusion}
\ee
Here $D =\ell v_E/{\cal D}$ is the diffusion constant, with $v_E$ the energy velocity,  $\tau=\ell/v_E$ is the scattering time, and $\gamma$ is the scattering strength. The latter can be expressed in terms of $\ell$ and the mean density of states ${\rm DOS}(\omega)$ as
\be
\gamma=\frac{2k^2}{\pi c\ell\,{\rm DOS}(\omega)}.
\label{EqGamma}
\ee

The presence of the perturbation at $\vec{r}_0$ results in modification of the {\it mean}, disorder-averaged, flux at the output as
\begin{align}
&\delta\overline{\left|{\cal R}_{ba}\right|^2}=\label{EqdRba}\\
&4k_a^z k_b^z\int d\vec{r}_1\;d\vec{r}_2\,
\left|\mel{\chi_b}{\overline{G}_0}{\vec{r}_2}\right|^2\,
\delta\Gamma(\vec{r}_2,\vec{r}_1)\,
\left|\mel{\vec{r}_1}{\overline{G}_0}{\chi_a}\right|^2,\nonumber
\end{align}
The change in the diffusive kernel $\delta \Gamma(\vec{r}_2,\vec{r}_1)$ can be found using the perturbative approach within the diffusion framework described in Appendix~\ref{sec:diffusion_perturbation}. With the help of Eq.~\eqref{EqGamma_diffusion}, we obtain
\be
\delta\Gamma(\vec{r}_2,\vec{r}_1)\simeq\tilde{{\cal V}}^\text{I}\,\frac{\tau}{\gamma}\,\Gamma(\vec{r}_2,\vec{r}_0)\,\Gamma(\vec{r}_0,\vec{r}_1),\label{EqdGamma}
\ee
where $\tilde{{\cal V}}^\text{I}=-v_Ek\epsilon''\delta V$ is the strength of perturbation. Substituting Eq.~\eqref{EqdGamma} into Eq.~\eqref{EqdRba} we obtain the change in flux in the form
\be
\delta\overline{\left|{\cal R}_{ba}\right|^2}\simeq 4\tilde{{\cal V}}^\text{I}\frac{\,\tau}{\gamma}\,J_{b}(\vec{r_0})\,J_{a}(\vec{r_0}),
\label{EqdRba2}
\ee
where we introduced a diffusive flux
\be
J_{a}(\vec{r_0})=k_a^z\int d\vec{r}_1\,\Gamma(\vec{r}_0,\vec{r}_1)\,
\left|\mel{\vec{r}_1}{\overline{G}_0}{\chi_a}\right|^2.
\ee
This quantity is proportional to the mean intensity 
$\overline{I}_{a}(\vec{r_0})\equiv\overline{\left|\phi_a(\vec{r_0})\right|^2}=
\overline{\left|2i\sqrt{k_a^z}\mel{\vec{r}_0}{G_0}{\chi_a}\right|^2}$ 
excited at $\vec{r}_0$ by the input channel $\ket{\psi_a}$. Unlike $\overline{G}_0$, averaged modulus square of $G_0$ is described by diffusion and can be computed as
\begin{align}
\overline{I}_{a}(\vec{r_0})&=4k_a^z\!\int\!\! d\vec{r}_1d\vec{r}_2\!
\left|\!\mel{\vec{r}_0}{\overline{G}_0}{\vec{r}_2}\right|^2
\Gamma(\vec{r}_2,\vec{r}_1)
\left|\!\mel{\vec{r}_1}{\overline{G}_0}{\chi_a}\right|^2
\nonumber
\\
&\simeq 4k_a^z \!\int \!\! d\vec{r}_2\!\left|\! \mel{\vec{r}_0}{\overline{G}_0}{\vec{r}_2}\right|^2
\!\int\!\! d\vec{r}_1\Gamma(\vec{r}_0,\vec{r}_1)\left|\!\mel{\vec{r}_1}{\overline{G}_0}{\chi_a}\right|^2
\nonumber
\\
&=\frac{4}{\gamma}J_{a}(\vec{r_0})\label{EqIvsJ}.
\end{align}
Here we used the fact that $\int d\vec{r}_2\left| \mel{\vec{r}_0}{\overline{G}_0}{\vec{r}_2}\right|^2=-(\ell/k)\,\text{Im}\mel{\vec{r}_0}{\overline{G}_0}{\vec{r}_0}$ and, by definition of the mean density of states, $\text{Im}\mel{\vec{r}_0}{\overline{G}_0}{\vec{r}_0} = -\pi c{\rm DOS}(\omega)/2k$. Combining now Eqs.~\eqref{EqFluxRandom},\eqref{EqdRba2}, and \eqref{EqIvsJ}, we find the variation of the mean flux due the perturbation
\be
\delta  \bar{F}_\text{diff}\simeq\tilde{{\cal V}}^\text{I}\,\frac{\tau\gamma}{4}\,\overline{I}_{\text{in}}(\vec{r_0})\,\overline{I}_{\text{out}}(\vec{r_0}),
\label{EqdF}
\ee
in terms of the intensities at position $\vec{r}_0$ excited from input and output (i.e. detector) ports
\begin{align}
\overline{I}_{\text{in}}(\vec{r_0})&=\sum_{a=1}^{N_1}\,|v_a|^2\,\overline{I}_a(\vec{r_0}),\nonumber\\
\overline{I}_{\text{out}}(\vec{r_0})&=\sum_{b=1}^{N_2}\,\overline{I}_b(\vec{r_0}).\label{EqI_out}
\end{align}
We note that coefficients $|v_a|^2\rightarrow 1$, if we can assume random and statistically equivalent excitation. Therefore, we will refer to these intensities as $\overline{I}^{\text{RI}}_{\text{in}}(\vec{r_0}),\,\overline{I}^{\text{RI}}_{\text{out}}(\vec{r_0})$ below, where RI stands for `random input'.

To complete the derivation of the sensitivity $\overline{{\cal S}}_\text{diff}=d\bar{F}_\text{diff}/d\epsilon''$ in the diffusive approximation, we substitute all pre-factors into Eq.~\eqref{EqdF} and use the density of states ${\rm DOS}^\text{2D}(\omega)=\frac{k}{2\pi c}$ or ${\rm DOS}^\text{3D}(\omega)=\frac{k^2}{2\pi^2 c}$. We then obtain the main result of this section:
\be
\overline{{\cal S}}_\text{diff}(\vec{r_0}) = -k^2\delta V\left(\frac{\pi}{k}\right)^{{\cal D}-2}\,\overline{I}^{\text{RI}}_{\text{in}}(\vec{r_0})\,\overline{I}^{\text{RI}}_{\text{out}}(\vec{r_0}).
\label{EqDiffusiveSensitivity}
\ee
The coefficient in front of the expression depends on the dimensionality ${\cal D}=2,3$. This expression lends itself to a insightful physical interpretation. Indeed, $\overline{I}^{\text{RI}}_{\text{in}}(\vec{r_0})$ is proportional to the probability for a photon injected at the input port to reach the localized perturbation at $\vec{r_0}$. 
Likewise, $\overline{I}^{\text{RI}}_{\text{out}}(\vec{r_0})$ is proportional to the probability of a photon that passes through $\vec{r_0}$ to reach the output. 
Thus, Eq.~\eqref{EqDiffusiveSensitivity} is, up to a pre-factor, a conditional probability for a photon injected at the input to reach the output after having visited $\vec{r_0}$. 
For a half-space geometry, the spatial map described by the above expression is an arch, or `banana'~\cite{1991_Cui,1993_Schotland,2010_Durduran_DOT,2020_Fantini,2024_Blaney_sensitivity_review} as it is commonly referred to.

We now turn to a comparison between the expression for optical sensitivity derived within the diffusion approximation, Eq.~\eqref{EqDiffusiveSensitivity}, and the microscopic expression, Eq.~\eqref{EqSiBorn}, obtained in Sec.~\ref{sec:field_sensitivity}. 
The microscopic formulation applies to a single realization of disorder and does not rely on statistical averaging or assumptions about large, relative to the transport mean free path, system sizes or source-detector separations. 
This generality allows it to capture both sample-specific and ensemble-averaged behavior for arbitrary input state, whereas the diffusive result provides only the disorder-averaged sensitivity for random input state.
Therefore, we should expect that the macroscopic and diffusive descriptions both yield the same result when Eq.~\eqref{EqSiBorn} is applied to a diffusive system and appropriately averaged. In the following subsection, we present a formal proof of this equivalence.

\subsection{Microscopic sensitivity in the diffusive limit\label{sec:equivalence}}
\noindent In Sec.~\ref{sec:field_sensitivity}, we obtained the microscopic expression~\eqref{EqSensitivity_va} for the optical sensitivity. For systems exhibiting diffuse wave transport, this formula should be equivalent, in the statistical sense, to the diffuse sensitivity given in Eq.~\eqref{EqDiffusiveSensitivity}. To verify this, we need to take the proper limit and account for the underlying assumptions. These include: (1) the characteristic size parameters of the system are sufficiently larger than the transport mean free path, justifying diffusive treatment; (2) statistical average over an ensemble of disorder realizations; and (3) random excitation of the incident modes. 

\begin{figure}[b!]	
\centering
\includegraphics[width=3in]{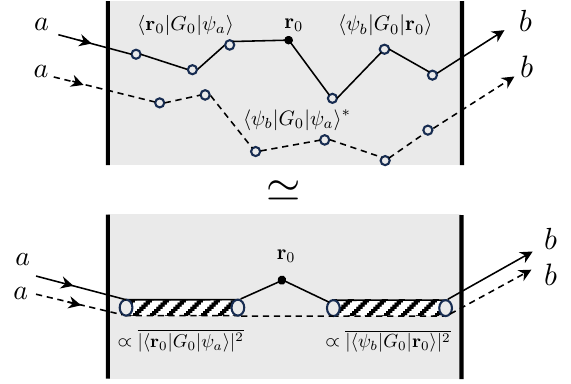}
\caption{Scattering representation of the identity~\eqref{EqGreenIdentity}. Solid and dashed lines represent the propagating field and its complex conjugate, while open circles represent scatterers. Shaded tubes represent diffusive paths where both field visit the same scatterers. A more formal representation of the lower diagram is given in Fig.~\ref{fig:diagram} in Appendix~\ref{sec:green_identity}.}
\label{fig:simplified_diagram}
\end{figure}

\begin{widetext}
As the first step, we express the sensitivity in Eq.~\eqref{EqSensitivity_va} in terms of the field Green's function. Using the Fisher-Lee relation~\eqref{EqFisherLee}, we obtain
\be
{\cal S}(\vec{r_0}) = -8k^2\delta V\sum_{baa'} v_a^*\,v_{a'}k_b^z\sqrt{k_a^zk_{a'}^z}\,\text{Im}\left[
\mel{\chi_b}{G_0}{\vec{r}_0}\mel{\vec{r}_0}{G_0}{\chi_{a'}}
\mel{\chi_b}{G_0}{\chi_a}^*
\right].
\label{EqFluxGreen1}
\ee

Likewise, the sensitivity in the diffusion approximation, Eq.~\eqref{EqDiffusiveSensitivity}, can be expressed in terms of the field Green's functions as
\be
\overline{{\cal S}}_\text{diff}(\vec{r_0})=-16k^2\delta V\,\left(\frac{\pi}{k}\right)^{{\cal D}-2}\,\sum_{ba} |v_a|^2 k_b^z k_a^z\,
\overline{\left|\mel{\chi_b}{G_0}{\vec{r}_0}\right|^2}\,\,
\overline{\left|\mel{\vec{r}_0}{G_0}{\chi_a}\right|^2}.
\label{EqFluxGreen2}
\ee
Therefore, we seek a connection between Eq.~\eqref{EqFluxGreen1} and Eq.~\eqref{EqFluxGreen2}, provided we take advantage of random excitation assumption (3), and statistical averaging (2). The former gives $\overline{v_a^*v_{a'}}=\delta_{aa'}$, whereas the latter leads to the condition
\be
\overline{\text{Im}\left[
\mel{\chi_b}{G_0}{\vec{r}_0}\!
\mel{\vec{r}_0}{G_0}{\chi_a}\!
\mel{\chi_b}{G_0}{\psi_a}^*
\right]}=2\left(\frac{\pi}{k}\right)^{{\cal D}-2}\,
\overline{\left|\mel{\chi_b}{G_0}{\vec{r}_0}\right|^2}\,\,
\overline{\left|\mel{\vec{r}_0}{G_0}{\chi_a}\right|^2} 
\label{EqGreenIdentity}
\ee
\end{widetext}
for the mean of Eq.~\eqref{EqFluxGreen1} to be equal to Eq.~\eqref{EqFluxGreen2}.
In Appendix~\ref{sec:green_identity}, we use a diagrammatic approach to prove that the equality~\eqref{EqGreenIdentity} indeed holds. This result can be qualitatively understood by considering the scattering paths that contribute the most after disorder averaging. These dominant scattering paths consist of diffusive trajectories, as shown in Fig.~\ref{fig:simplified_diagram}.
Consequently, we have established that under random illumination, the statistically averaged microscopic sensitivity, Eq.~\eqref{EqSiBorn}, and diffusive intensity sensitivity, Eq.~\eqref{EqDiffusiveSensitivity}, yield identical results.

We note that when the illumination is not random but remains independent of the disorder, this equivalence is still often preserved. 
For instance, it holds for a simple plane-wave input ($v_a = \delta_{a,a_0}$), since in this case $v_a^* v_{a'} = \delta_{a,a'}$ trivially. 
It also holds for input states with a large number of nonzero coefficients $v_a$ that are uncorrelated with the medium, 
because the cross terms ($a \neq a'$) in Eq.~\eqref{EqFluxGreen1} acquire random phase-dependent weights that are washed out upon summation.

This derivation above also exposes the mechanism via which the relationship between between Eq.~\eqref{EqSiBorn} and Eq.~\eqref{EqDiffusiveSensitivity} can be broken. Indeed, choosing the $v_a$ coefficients dependent on the specific realization of disorder would make them dependent on Green's functions matrix elements in Eq.~\eqref{EqFluxGreen1}, thereby preventing the factorization of disorder averages and rendering  Eq.~\eqref{EqDiffusiveSensitivity} no longer applicable. 
This is precisely the effect of  coherent excitation -- choosing incident wavefront uniquely tailored to the specific disorder realization explored in Secs.~\ref{sec:numerical_simulation_scalar} and \ref{sec:numerical_simulation_vector}. 

\subsection{From extensive to intensive optical sensitivity\label{sec:normalized_sensitivity}}
\noindent Earlier in this section we derived two expressions for computing sensitivity: a microscopic formulation, Eq.~\eqref{EqSiBorn}, and its diffusive approximation, Eq.~\eqref{EqDiffusiveSensitivity}.  
These expressions allow us to identify several key properties of sensitivity, which, by definition, is given by the ratio of flux variation to the magnitude of the perturbation in the imaginary part of the dielectric function.

{\it Volume of perturbation:}  
The spatial dependence of the absorbing center has been modeled by the delta-function, see Eq.~(\ref{EqBornField}). 
This assumption implies that the spatial extent of the perturbation $\delta V^{1/{\cal D}}$ is much smaller than the characteristic spatial variation scale of the field. 
In the microscopic formulation, Eq.~\eqref{EqSiBorn}, which applies to a single realization of disorder, the relevant length scale is the optical wavelength $\lambda=2\pi/k$, since interference effects govern the field structure. 
Although the structural correlation length (e.g., cell size) may be larger than $\lambda$, it does not limit the field resolution, and a conservative validity condition is still $\delta V^{1/{\cal D}}\ll\lambda$. 
In contrast, after statistical averaging -- as in the diffusive expression, Eq.~\eqref{EqDiffusiveSensitivity} -- both the wavelength and correlation length become irrelevant.
The spatial variations in the averaged fields are smoothed over the transport mean free path $\ell$, which sets the smallest relevant scale.
Thus, in the diffusive regime, the delta-function approximation remains valid so long as $\delta V\ll\ell^{\cal D}$, see Eq.~(\ref{EqPertI}) in Appendix~\ref{sec:diffusion_perturbation}. 
Therefore, weakness of perturbation amounts to the condition $k^{\cal D}\,\delta V\,\delta\epsilon\ll 1$, c.f. Eqs.~(\ref{EqBornField}) and (\ref{EqPertI}), making both Eq.~\eqref{EqSiBorn} and its diffusive approximation Eq.~(\ref{EqDiffusiveSensitivity}) trivially dependent on $\delta V$. 
As such sensitivity becomes an extensive quantity -- it scales with size of the perturbation. In contrast, intensive definition would normalize away this dependence making it `density'-like. Therefore, we propose normalization by the dimensionless quantity $k^{\cal D}\,\delta V$, see below.

{\it Number of the input spatial channels:} In Secs.~\ref{sec:field_sensitivity},~\ref{sec:sensitivity_diffusion_approximation} we adopted such a normalization that, on average, unit of flux is incident onto each of $N_1$ degrees of freedom in the input port. This is a `constant intensity' normalization -- it leads to the total flux being proportional to the {\it area} of the injection site and, thus, on $N_1$ itself. In other words, the sensitivity is trivially dependent on $N_1$, again, making it an extensive property and suggesting a natural normalization by $N_1$.

{\it Number of the output spatial channels:} The change in flux collected at the output should similarly increase with the increase of the collection area, or $N_2$ the number of degrees of freedom at the output port. In the diffusive approximation, Eq.~\eqref{EqDiffusiveSensitivity}, the proportionality should be nearly linear, as long as the size of port is much smaller than the transport mean free path. Similar to the two other dependencies discussed above, normalization by $N_2$ appears natural.

Based on the arguments above, we introduce a normalized optical sensitivity defined as
\be
{\cal s}(\vec{r}_{0})\equiv \frac{{\cal S}(\vec{r}_{0})}{k^{\cal D}\,\delta V\,N_1N_2}.
\label{EqInt_sensitivity}
\ee 
Such a normalization in both Eqs.~(\ref{EqSiBorn}) and (\ref{EqDiffusiveSensitivity}) defines sensitivity density. 
While this removes the dependence on the perturbation volume, it does not eliminate the dependence on other system parameters such as $N_1$, $N_2$, or geometry, c.f. Sec.~\ref{sec:remission_enhancement}. 
Nevertheless, the sensitivity density is particularly convenient when comparing different models or simulations -- the task we undertake in the following section. 

\section{Numerical analysis: scalar waves\label{sec:numerical_simulation_scalar}}
\begin{figure}
\includegraphics[width=3in]{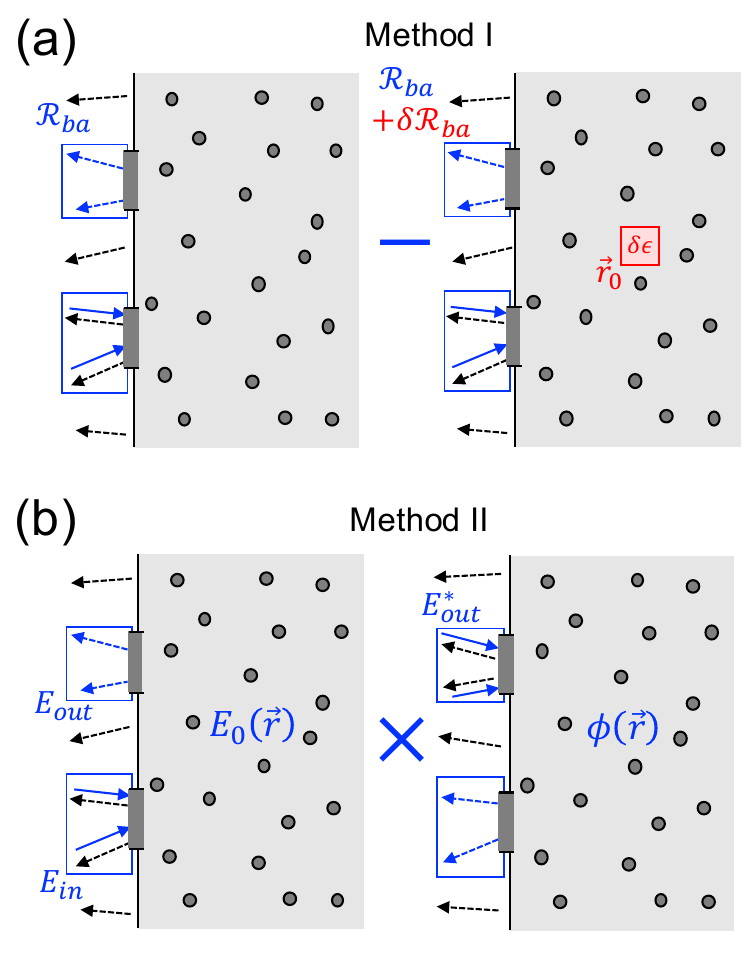}
\vskip -0.3cm
\caption{Two computational approaches for evaluating optical sensitivity.
(a)~Method I: Sensitivity is inferred from changes in remission coefficients ${\cal R}_{ba}$ due to the introduction of a {\it localized absorber} at position $\vec{r}_0$; see first part in Eq.~(\ref{EqSiBorn}). This method requires sequential simulations for each $\vec{r}_0$ and thus involves raster scanning. 
(b)~Method II: Sensitivity is computed using the product of two field distributions in the second part of Eq.~\eqref{EqSiBorn}. This approach is parallel and requires only two simulations to obtain the full spatial sensitivity map across all $\vec{r}_0$.
\label{fig:Sensitivity_methods}}
\end{figure}

\noindent In the previous section we introduced the microscopic expression for optical sensitivity Eq.~\eqref{EqSiBorn} and its diffusion approximation Eq.~\eqref{EqDiffusiveSensitivity}. 
Instead of sequential raster scanning the location of the absorbing center, see Fig.~\ref{fig:Sensitivity_methods}a, to map out the sensitivity, both formulations have the major advantage of being `parallel'. 
Indeed, Eq.~\eqref{EqSiBorn} expresses the entire sensitivity map in terms of two fields $E_0(\vec{r})$ and $\phi(\vec{r})$ requiring only two computations for each disorder configuration, see Fig.~\ref{fig:Sensitivity_methods}b, followed by ensemble averaging.
Eq.~\eqref{EqDiffusiveSensitivity} on the other hand, relates the sensitivity map to the ensemble-averaged intensity distributions for input and output ports.
In Sec.~\ref{sec:equivalence} we showed that both formulations should agree in diffusive system with random input excitation.
In Sec.~\ref{sec:scalar_random_input_results} we aim to verify that both Eq.~\eqref{EqSiBorn} and Eq.~\eqref{EqDiffusiveSensitivity} indeed give the same result in this limit in numerical simulation of scalar waves in 2D semi-infinite medium introduced in Sec.~\ref{sec:scalar_numerical_model}. 
This is not a trivial test because the quantities and the averaging procedure in two expressions are vastly different.
Next, in Sec.~\ref{sec:shaped_wavefront_excitation} we show that under excitation with disorder-specific wavefront, the maximum remission eigenchannel, the diffusive approximation fails to describe optical sensitivity computed using our microscopically exact Eq.~\eqref{EqSiBorn}. 
We also demonstrate that the sensitivity for the remission eigenchannel can be predicted by the filtered random matrix theory summarized in  Appendix~\ref{sec:filtered_random_matrix_theory}.
In Sec.~\ref{sec:remission_enhancement} we discuss relationship between enhancement of sensitivity and the remission enhancement.

\subsection{Numerical model\label{sec:scalar_numerical_model}}
\noindent We carry out our numerical simulations in a geometry chosen to represent the experimental sample from our previous work~\cite{2022_Bender_Coherent_Enhancement}.
The system is a two-dimensional (2D) disordered medium occupying a half-space, as shown in Fig.~\ref{fig:Sensitivity_methods}. 
This 2D configuration was not arbitrary: it models a planar photonic structures in the experiment, which enabled nearly non-invasive optical access to the internal field distribution~\cite{2020_Bender_Eigenchannels}, making it ideally suited for studying coherent wave transport {\it inside} multiple scattering medium.
The electromagnetic waves in such structures are effectively confined in the vertical (out-of-plane) direction~\cite{2015_Sarma,2016_Yamilov,2016_Sarma_Open_Channels,2017_Koirala}, with radiative leakage inhibited by total internal reflection due to the refractive index difference with surrounding media~\cite{2014_Yamilov,2017_Sarma} (air and substrate). 
This indeed allows a reduction to a 2D model, in which electromagnetic fields decouple into two independent polarizations: transverse magnetic (TM) and transverse electric (TE)~\cite{2005_Taflove_book,2018_Summers}. 
TM polarization corresponds to one out-of-plane electric field component and two in-plane magnetic field components, while TE polarization involves two in-plane electric field components and one out-of-plane magnetic field component. 
Waves with TM and TE polarizations propagate independently in 2D. 
Crucially, since scattering arises through the electric field, the interaction with disorder due to randomly varying dielectric permittivity differs: TM waves couple via a single electric field component, while TE waves involve two, leading to fundamentally different scattering behavior.

In this section, we consider TM polarization, which obeys a scalar wave equation~\cite{Akkermans_Montambaux_2007} matching Eq.~\eqref{EqWave1}. This choice is consistent with the modeling approach in Ref.~\cite{2022_Bender_Coherent_Enhancement}, where numerical simulations were performed using the KWANT package~\cite{Groth2014}. 
Here, we employ the MESTI (Maxwell’s Equations Solver with Thousands of Inputs) software package~\cite{2022_HoChun_MESTI}, which offers substantial computational speedup through a novel augmented partial factorization method. 
In this work, we do not consider spatially uniform loss, which would be required to simulate the effects of a weak out-of-plane scattering in experiments~\cite{2022_Bender_Coherent_Enhancement}.
In our previous work~\cite{2022_Jara_Simulation_Coherent_Remission}, MESTI was validated against KWANT and shown to produce quantitatively consistent results. 
In Sec.~\ref{sec:numerical_simulation_vector}, we extend our study to TE polarization using MESTI’s vector-wave capabilities, allowing further verification of the microscopic approach to optical sensitivity, thereby validating and extending the applicability of the microscopic sensitivity model to vector waves involving multiple polarization components.

Based on the above considerations, we conduct 2D simulations with the following parameters: 
vacuum wavelength $\lambda_{0}\!=$1.55 $\mu$m, 
slab dimensions $L \times W\!=$ 250 $\mu$m $\times$ 300 $\mu$m, and
background refractive index $n_\text{eff}\!=$ 2.85. 
Disorder is introduced by placing a low concentration of circular air holes -- filling fraction $f\!=$ 10 \%, radius $a\!=$ 100 nm, and refractive index $n_\text{air}\!=$ 1.
These parameters were chosen to ensure that diffusion approximation is satisfied, i.e. $\ell\ll W,L$. 
This condition also guarantees that the sample thickness greatly exceeds the input–output separation, so the medium can be regarded as effectively semi-infinite.
The actual value of $\ell\!=$ 6.4 $\mu$m was matched to the experiment in Ref.~\cite{2022_Bender_Coherent_Enhancement}, but is irrelevant in context of current discussion. 
The value of $\ell$ has been determined using procedure described in Ref.~\cite{2022_Jara_Simulation_Coherent_Remission}.
We will also present results of simulations where the transport mean free path is varied.
To model the effect of open boundaries, the numerical domain was surrounded by a Perfect Matching Layer (PML)~\cite{2005_Taflove_book}. 
Injection and detection ports of widths $W_1,\,W_2$, separated by distance $d\gg\ell$, are modeled by attaching two waveguides to the front interface, see Fig.~\ref{fig:Sensitivity_methods}. 
We will present simulations for different combinations of these parameters.

We finally note that while all two-dimensional disordered systems are expected to exhibit Anderson localization effects~\cite{2006_Sheng}, these only become relevant on lengths comparable to or exceeding the localization length $\xi$. In the weak-scattering regime considered here ($k\ell \gg 1$) this length is exponentially large, $\xi \propto \ell \exp(c\,k\ell)$ with $c \sim 1$~\cite{2006_Sheng}. For the parameters used in our simulations, as well as in typical optical experiments, $\xi$ is vastly larger than the system size and all other relevant scales, so localization effects can be safely neglected.

\subsection{Random input excitation\label{sec:scalar_random_input_results}}
\begin{figure*}
\begin{center}
\includegraphics[width=6in]{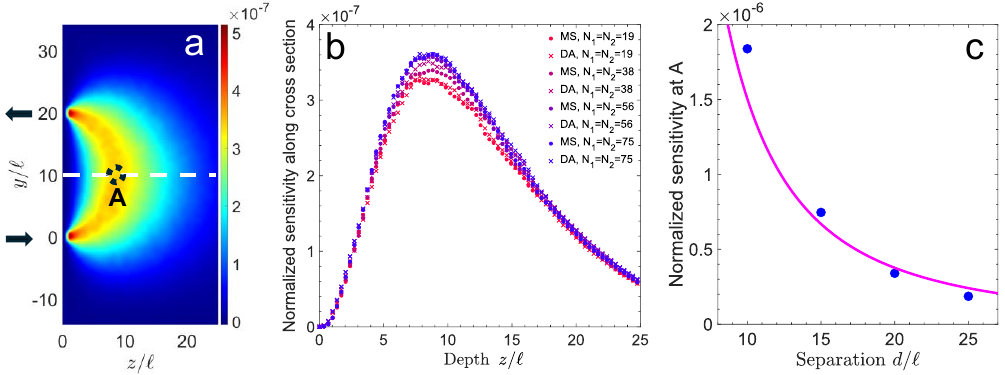}
\end{center}
\vskip -0.6cm
\caption{
Comparison of microscopic and diffusive models for optical sensitivity under random input excitation.  
All results are averaged over 1000 disorder realizations and normalized according to Eq.~\eqref{EqInt_sensitivity}. 
Transport mean free path is $\ell=~$6.4 $\mu$m.
(a)~Sensitivity map computed using the microscopic sensitivity (MS) expression, Eq.~\eqref{EqSiBorn}, exhibits the characteristic ‘banana’-shaped pattern.  
Input and output port widths are $W_1=W_2=10\,\mu$m; source-detector separation is $d=128\,\mu$m~$=20\times\ell$.
(b)~Depth cross-section of the sensitivity map along the dashed line in (a). 
Circles correspond to MS calculation, Eq.~\eqref{EqSiBorn}; crosses show results from the diffusion approximation (DA), Eq.~\eqref{EqDiffusiveSensitivity}.
Input and detector port widths are equal.  
Curves are shown for $5,\,10,\,15$ and $20\,\mu$m, with fixed separation $d=128\,\mu$m~$=20\times\ell$.
(c)~Sensitivity at the midpoint A of the banana (peak in panel b) as a function of source-detector separation. Symbols: MS results using Eq.~\eqref{EqSiBorn} for $W_1=W_2=10\,\mu$m; solid line: DA prediction based on analytic intensity profiles substituted into Eq.~\eqref{EqDiffusiveSensitivity}, as described in Appendix~\ref{sec:diffusion_in_remission_geometry}.
\label{fig:panel_random}}
\end{figure*}

\noindent Computing optical sensitivity using Eq.~\eqref{EqSiBorn} requires knowledge of two fields: $E_0(\vec{r}_0)$ and $\phi(\vec{r}_0)$. The former is the field inside the system obtained by illuminating it through the input port with a random combination of all modes. Computing the latter field requires a separate simulation, but for the same configuration of disorder. It is obtained by recording the field at the output port from the first simulation, conjugating it, and sending it back into the sample as input for the second simulation. Having computed internal fields, the quantity in Eq.~\eqref{EqSiBorn} is obtained for that specific disorder configuration. Subsequently, we compute this expression for 1000 realizations of disorder, requiring twice as many simulations, to achieve sufficient statistical average. Fig.~\ref{fig:panel_random}a depicts the result found for a system with $\ell\!=$ 6.4 $\mu$m, $W_1=W_2=10\,\mu$m, and input/output separation of $d=128\,\mu$m $=\!20\!\times\!\ell$. The characteristic `banana' shape can be seen. We would like to emphasize again the remarkable fact that statistical average of the product of two {\it fields} in Eq.~\eqref{EqSiBorn} results in a non-zero value. 

Computations based on Eq.~\eqref{EqDiffusiveSensitivity} require statistically averaged intensities throughout the volume of the system. These are obtained by illuminating it through input port with a random combination of all modes. Thus, simulations used to compute $E_0(\vec{r}_0)$ for Eq.~\eqref{EqSiBorn} can be reused for computing $\overline{I}^{\text{RI}}_{\text{out}}(\vec{r_0})$ -- it is obtained by a mirror reflection in the horizontal line through the middle of the system. 

In Fig.~\ref{fig:panel_random}b, we compare the sensitivities computed from Eqs.~(\ref{EqSiBorn}) and (\ref{EqDiffusiveSensitivity}) -- circles and crosses respectively -- and normalized via Eq.~\eqref{EqInt_sensitivity}. Here, the separation distance is fixed at $d=128\,\mu$m $= 20\!\times\!\ell$ and multiple values $W_1=W_2$ are tested. We can see that, as expected based on theoretical argument in Sec.~\ref{sec:equivalence}, the two methods give the same result. The fact that simulations for different values of $W_1=W_2$ agree among themselves, also supports our argument in Sec.~\ref{sec:normalized_sensitivity} for normalizing sensitivities. Furthermore, in Appendix~\ref{sec:sensitivity_asymmetric}, we demonstrate that the curves in Fig.~\ref{fig:panel_random}b agree with simulation performed for $W_1\neq W_2$ cases. Such an agreement is also expected for the normalized sensitivities, based on Eq.~\eqref{EqInt_sensitivity}. 

In Fig.~\ref{fig:panel_random}c, we demonstrate that dependence on the input/output separation $d$ can also be predicted by substituting analytical solutions for $\overline{I}^{\text{RI}}_{\text{in}}(\vec{r_0})$ and $\overline{I}^{\text{RI}}_{\text{out}}(\vec{r_0})$ found from diffusion equation. 
We derive such solution in Appendix~\ref{sec:diffusion_in_remission_geometry}. The decrease in the sensitivity $\overline{{\cal s}}_\text{diff}\simeq-1/(\pi^2 k^2 d^2)\propto 1/d^2$ with the increase of $d$ is the well-known trade-off between depth and sensitivity. 
Indeed, larger $d$ leads to increase of the radius of the banana pattern in the sensitivity map. 
Photons undergoing diffusive random walk along these long paths are less likely to reach the output port. 
As a consequence the remitted signal is less sensitive to the total flux variation due to local perturbation at larger depth inside the system, see Fig.~\ref{fig:panel_random}c.

For completeness, we computed sensitivity by using brute-force sequential approach -- Method I -- in Fig.~\ref{fig:Sensitivity_methods}a. Recall that sensitivity is defined as ${\cal S}(\vec{r}_0)=\delta F/\delta\epsilon''$ -- the difference between the average remitted flux detected at the output waveguide with and without perturbation center at $\vec{r}_0$. The agreement with Eq.~\eqref{EqSiBorn}, shown in Appendix~\ref{sec:diffusion_in_remission_geometry}, provides further evidence for robustness of our microscopical approach to optical sensitivity.

\subsection{Shaped wavefront excitation\label{sec:shaped_wavefront_excitation}}

\noindent We would like to motivate this section by revisiting assumptions used in obtaining Eq.~\eqref{EqDiffusiveSensitivity} -- diffusion approximation for optical sensitivity. 
In fact, it incorporates two separate assumptions that become particularly clear when we examine the process of averaging in the microscopically exact Eq.~\eqref{EqSensitivity_va}. 
The first assumption is the diffusion approximation itself, stating the the ensemble-averaged intensities can be described by diffusion equation and that the dimensions of the system are much larger than transport mean free path.
The second assumption has to do with averaging procedure of the input wavefront encoded by coefficients $v_a$ in Eq.~\eqref{EqSensitivity_va}.
The procedure involved in obtaining Eq.~\eqref{EqDiffusiveSensitivity} in Sec.~\ref{sec:sensitivity_diffusion_approximation} amounts to performing disorder averaging with no regard for $v_a$'s. 
Therefore, once diffusion approximation has been made, it is no longer possible to account for specific input wavefront.
This observation motivates the use of the microscopic expression for sensitivity Eq.~\eqref{EqSiBorn} [equivalent to Eq.~\eqref{EqSensitivity_va}], including the sample specific excitation coefficients, and only then performing statistical averaging.

Secondly, it has been shown in Ref.~\cite{2022_Bender_Coherent_Enhancement} that by exciting system with incident wavefront corresponding to  the maximum remission eigenchannel (MRE), see description before Eq.~\eqref{EqRemissionEigenchannelSensitivity}, we could enhance not only the signal in remission, but also the sensitivity. 
In our prior work, we used the proper microscopical expression Eq.~\eqref{EqSiBorn} to compute the sensitivity. 
However, it remained unclear whether Eq.~\eqref{EqDiffusiveSensitivity} could still apply with  $\overline{I}^{\text{MRE}}_{\text{in}}(\vec{r_0})$ and $\overline{I}^{\text{MRE}}_{\text{out}}(\vec{r_0})$ substituted instead of intensities for random input excitation.
Below, we clearly demonstrate that Eq.~\eqref{EqDiffusiveSensitivity} is {\it inadequate and cannot be used} in describing optical sensitivity of remission eigenchannels.

\begin{figure*}
\vskip -0.5in
\includegraphics[width=5in]{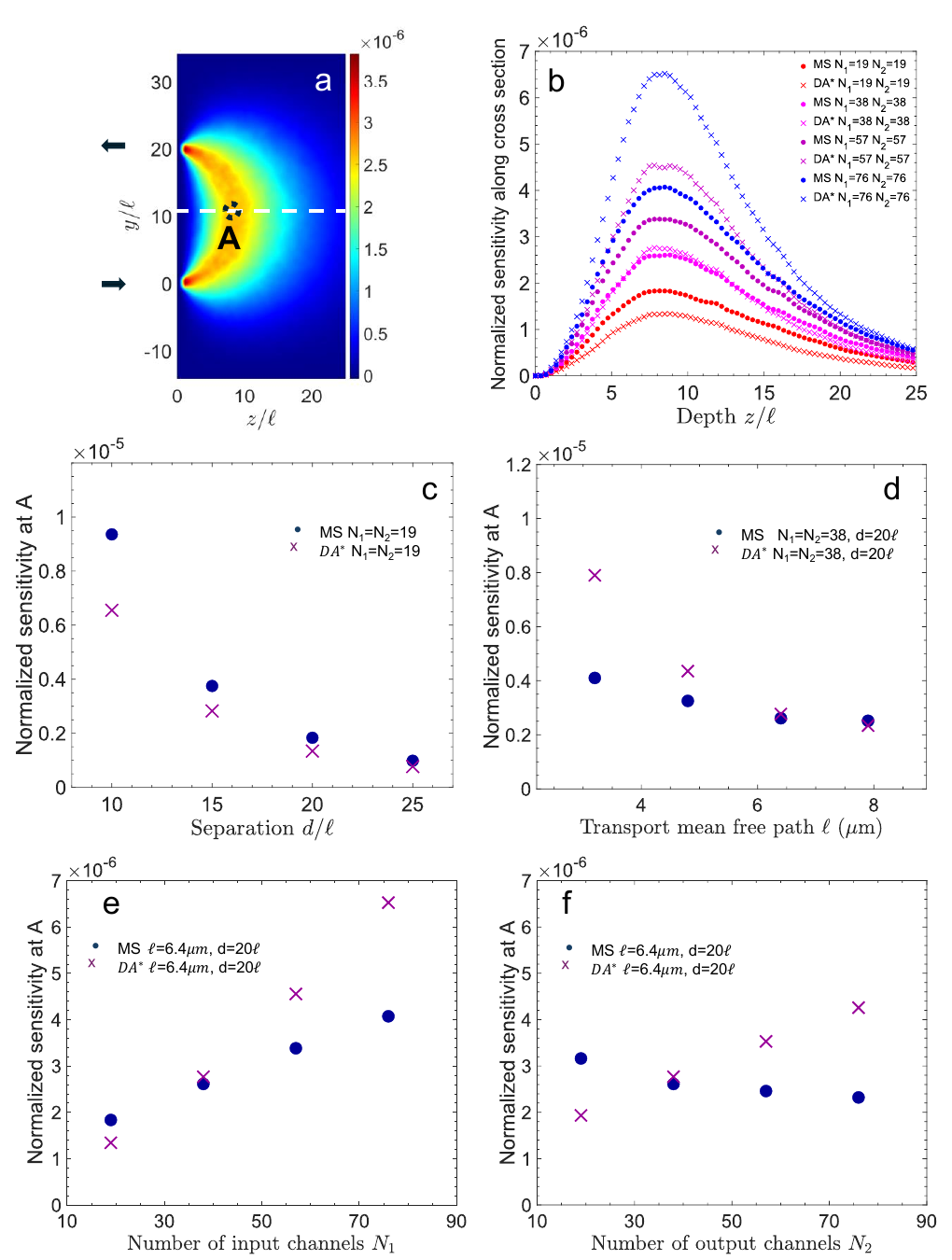}
\caption{
Sensitivity map for excitation with the maximum remission eigenchannel (MRE) and comparison with diffusion-based predictions. 
(a)~Sensitivity map computed using the microscopically exact expression, Eq.~\eqref{EqSiBorn}, for MRE excitation. The characteristic `banana'-shaped pattern is preserved, but overall sensitivity is enhanced compared to the random input case. Parameters: $W_1=W_2=10\,\mu$m, separation is $d=20\times\ell$, and transport mean free path $\ell=~$6.4 $\mu$m.
(b)~Depth cross-section along the mid-plane (dashed line in panel a), comparing sensitivity from the microscopic sensitivity (MS) formulation (Eq.~\eqref{EqSiBorn}, filled circles) and a heuristic application of Eq.~\eqref{EqDiffusiveSensitivity} using the disorder-averaged intensity of MREs (crosses, denoted as DA$^*$), see main text for details. 
(c)-(f)~Sensitivity at the central point A in panel (a), plotted as a function of: 
(c)~input-output separation $d$; 
(d)~transport mean free path $\ell$; 
(e)~number of input channels $N_1$, and 
(f)~number of output channels $N_2$. 
In all cases, results are averaged over 1000 disorder realizations and normalized according to Eq.~\eqref{EqInt_sensitivity}.
\label{fig:panel_eig_input}}
\end{figure*}

Here, we carry out comprehensive numerical tests of the results of Sec.~\ref{sec:sensitivity_diffusion_approximation}. 
Figure~\ref{fig:panel_eig_input}a shows disorder-averaged microscopic sensitivity (MS) map in Eq.~\eqref{EqSiBorn}, computed for MRE for system with $W_1=W_2=10\,\mu$m, separation is $d=20\times\ell$, and transport mean free path is $\ell=$6.4 $\mu$m.
We observe, in agreement with our earlier work~{\cite{2022_Bender_Coherent_Enhancement}}, that sensitivity map retains the banana shape, but is enhanced by a factor, which appears to be constant within accuracy of our simulation.
Remarkably, as we verify in Appendix~\ref{sec:sensitivity_asymmetric}, this conclusion holds even when $W_1\neq W_2$.
Therefore, below we present only the depth dependence of the sensitivity maps taken along the line $y=d/2$, shown as dashed line in Fig.~\ref{fig:panel_eig_input}a.

In Fig.~\ref{fig:panel_eig_input}b, we compare the sensitivity maps for MRE excitation obtained using two different methods. 
The filled circles represent numerical results computed using the microscopically exact Eq.~\eqref{EqSiBorn}, which remains valid for coherent excitation. 
The crosses, however, require careful interpretation. 
While Eq.~\eqref{EqDiffusiveSensitivity} (derived in the diffusion approximation) is rigorously justified for random input excitation, it does not strictly apply to the coherently excited MREs for two reasons. 
First, as noted above, averaging of intensities cannot be carried out separately from averaging the excitations coefficients $v_a$, which are unique to each disorder realization.  
Secondly, the diffusion equation does not accurately describe the disorder-averaged intensity distribution of MREs due to interference effects, which are neglected in the diffusion approach. 
Nevertheless, as an exploratory step, we substitute the numerically computed disorder-averaged intensity of MREs into Eq.~\eqref{EqDiffusiveSensitivity}, yielding the results represented by the crosses. 
While this substitution lacks theoretical justification, it provides an empirical comparison and helps assess the limitations of the diffusion approximation in this context. 
To avoid confusion, we denote this data as DA$^*$ rather than DA, emphasizing that it is not a direct application of the diffusion approximation but rather an ad hoc extension.

Among four datasets, corresponding to $W_1=W_2=5,\,10,\,15$ and $20\,\mu$m, DA$^*$ agrees with MS in only one case. Below, we rule this agreement accidental. 
To simplify the analysis, we will limit our analysis to the value of sensitivity in the middle of the banana, labeled with letter `B' in Fig.~\ref{fig:panel_eig_input}a. 
Since we have already determined that the sensitivity maps retain their shape, comparison of its value at a predetermined point is indeed sufficient. 
In Figs.~\ref{fig:panel_eig_input}c-f, we tested whether DA$^*$ is capable of describing MS when the parameters $d$, $\ell$, $N_1$, and $N_2$ are varied. 
We conclude that, in sharp contrast to applicability of DA to the random input case in Fig.~\ref{fig:panel_random}, DA$^*$ is inadequate in describing MS of MRE. 
In other words, the sensitivity in case of controlled wavefront excitation does not reduce to a product of two intensities as in Eq.~\eqref{EqDiffusiveSensitivity}.
This systematic deviation between MS and DA$^*$ is the direct evidence of existence of nontrivial correlation effects, neglected when statistical averaging over input field patterns is performed separately from intensities in Eq.~\eqref{EqSensitivity_va}.  

\subsection{Relationship between sensitivity enhancement and remission enhancement\label{sec:remission_enhancement}}
\noindent In our previous work~\cite{2022_Bender_Coherent_Enhancement}, we observed an intriguing result that the enhancement of sensitivity $\eta_{\cal S}^\text{MRE}$ was close to the enhancement of the remission $\eta_{\cal R}^\text{MRE}$ for MRE. 
If general, such a relationship could be extremely useful because enhancement of remission can be theoretically predicted. 
Indeed, theoretical technique~\cite{2013_Goetschy_FRM,Hsu2017,McIntosh2024}, called filtered random matrix (FRM) theory, has been shown to correctly reproduce the full distributions of the eigenvalues of a subset of the full scattering matrix of a complex system, in our case ${\cal R}$.  
In Ref.~\cite{2022_Bender_Coherent_Enhancement}, we used FRM theory to obtain an analytical expression for the remission enhancement -- the ratio $\rho_\text{max}/\bar{\rho}=\eta_{\cal R}^\text{MRE}$ between the maximum and the average eigenvalues of the remission matrix ${\cal{R}}^{\dagger}{\cal{R}}$. 
In Appendix~\ref{sec:filtered_random_matrix_theory} we have included a brief summary of the FRM theory. A less accurate but more intuitive model--the effective Marchenko-Pastur model--is also presented in Appendix~\ref{sec:Marcenko_Pastur}.

\begin{figure*}
\begin{center}
    \includegraphics[width=6in]{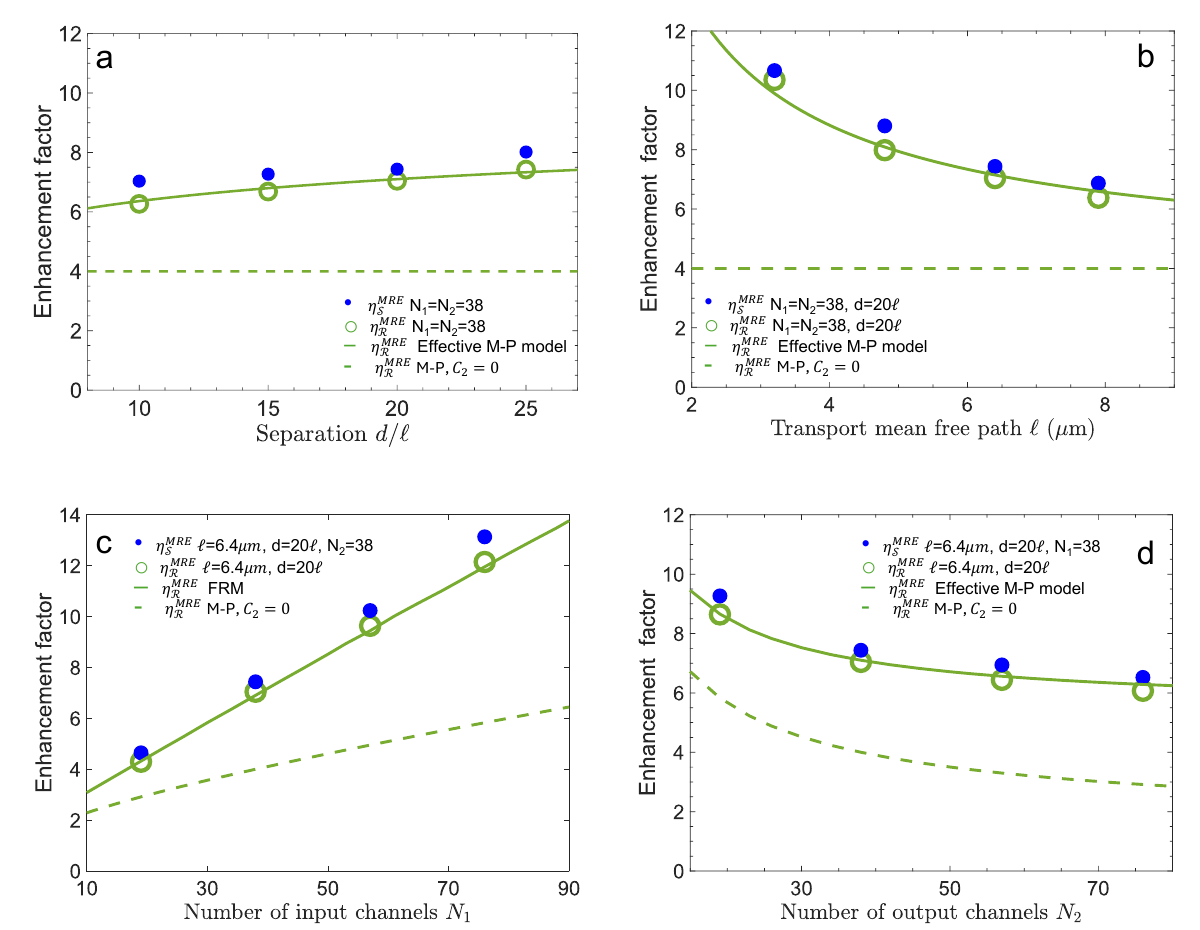}
\end{center}
\vskip -0.6cm
\caption{
Comparison between remission and sensitivity enhancement factors.
(a–d) Symbols show the enhancement factors for sensitivity $\eta_{\cal{S}}^\text{MRE}$ (filled circles) and remission $\eta_{\cal{R}}^\text{MRE}$ (open circles) as a function of 
(a)~source-detector separation $d$,
(b)~transport mean free path $\ell$,
(c)~input port width $W_1$, and 
(d)~output port width $W_2$; see text for definitions.
Solid lines in panels (a), (b), and (d) represent predictions from the effective Marchenko–Pastur (MP) model, see Eq.~\eqref{EqeffMP}in Appendix~\ref{sec:Marcenko_Pastur}. 
In panel (c), the solid line includes higher-order corrections based on filtered random matrix (FRM) theory, as described in Appendix~\ref{sec:filtered_random_matrix_theory}, Eq.~\eqref{Eqalpha}.
Dashed lines correspond to the MP model prediction when the contribution of long-range correlations $C_2$ is omitted, Eq.~\eqref{EqFRM3D}.
\label{fig:panel_enhan_factor}
}
\end{figure*}
In Figs.~\ref{fig:panel_enhan_factor}a-d, we compare theoretical predictions for the remission enhancement $\eta_{\cal R}^\text{MRE}$ to the sensitivity enhancement $\eta_{\cal S}^\text{MRE}$ and find that they indeed agree well. 
We now present an argument that this relationship is related to the invariance of the sensitivity map to the input field profile, observed in Sec.~\ref{sec:shaped_wavefront_excitation} above.

Let us recall that, by definition, the sensitivity is the response of the flux to the {\it localized} perturbation $\delta\epsilon(\vec{r})=i\epsilon''\delta(\vec{r}-\vec{r}_0)$: ${\cal S}(\vec{r}_0)=\delta F(\vec{r}_0)/\epsilon''$, see Eq.~(\ref{EqSiBorn}). 
As such, $\delta F(\vec{r}_0)$ implicitly depends on the location of the perturbation. 
By performing volume integration over $\vec{r}_0$ on both sides of the definition of sensitivity, we conclude that 
\be
\int{\cal S}(\vec{r}_0)d^3\vec{r}_0\propto
\int\delta F(\vec{r}_0)d^3\vec{r}_0.
\label{EqEtaProportionality0}
\ee
Volume integration of the localized perturbation is equivalent to estimating sensitivity to a global, i.e. spatially uniform, perturbation $\epsilon''$. 
Now we make an assumption that the volume-integrated change in flux, the right-hand side in Eq.~\eqref{EqEtaProportionality0}, is proportional to the total flux without perturbation
\be
\int{\cal S}(\vec{r}_0)d^3\vec{r}_0\propto
\left[\sum_{baa'}{\cal R}_{a'b}^*{\cal R}_{ba}v_{a'}^*v_a\right],
\label{EqEtaProportionality}
\ee
where we used explicit expression for the flux in terms of ${\cal R}$. 
The above amounts to assuming that a sufficiently small uniform perturbation should reduce all elements of remission matrix equally.
This assumption is supported by theoretical and numerical studies showing that, in the weak-absorption regime, uniform loss preserves the spatial profiles of individual quasimodes and consequently attenuates all scattering coefficients proportionally~\cite{2006_Yamilov,2014_Liew}.

Substituting appropriate input vectors $v_a$ and performing statistical averages we can find two versions of Eq.~\eqref{EqEtaProportionality}: one for MRE and one for random input excitation. 
Forming a ratio, we obtain an insightful constraint
\be
\frac{\int\overline{{\cal S}}^\text{MRE}(\vec{r_0})d^3\vec{r_0}}
     {\int\overline{{\cal S}}^\text{RI}(\vec{r_0})d^3\vec{r_0}}\approx
\frac{\bar{\rho}_\text{max}}{\bar{\rho}}\equiv
\eta_{\cal R}^\text{MRE}.
\label{Eqsen_enha1}
\ee
Here, we took note of the fact that random input excitation is statistically equivalent to exciting each eigenchannel with equal probability, $\rho_\text{RI}\equiv\bar{\rho}$, where $\bar{\rho}$ is the average remission eigenvalue. 
Such a relationship has profound implications -- it states that the maximum achievable {\it global} (i.e. volume integrated) enhancement of sensitivity is equal to the remission enhancement for the MRE:
\be
\frac{\int\overline{{\cal S}}(\vec{r_0})d^3\vec{r_0}}
     {\int\overline{{\cal S}}^\text{RI}(\vec{r_0})d^3\vec{r_0}}\leq
\eta_{\cal R}^\text{MRE}.
\label{Eqsen_enha3}
\ee
Wavefront shaping takes advantage of linearity of wave equation and is known to produce non-trivial constraints or sum rules~\cite{2022_Yamilov_Sum_rules}.

Furthermore, the fact that the spatial distribution (the map) of sensitivity for the MRE is
nearly the same as that for random input excitation $\overline{{\cal S}}^\text{MRE}(\vec{r_0})\propto\overline{{\cal S}}^\text{RI}(\vec{r_0})$, see Sec.~\ref{sec:shaped_wavefront_excitation}, allows us to take the argument in Eq.~\ref{Eqsen_enha1} a step further and conclude that 
\be
\eta_{\cal S}^\text{MRE}\simeq\eta_{\cal R}^\text{MRE}.
\label{Eqsen_enha2}
\ee
This is an interesting result in context of our discussion in Sec.~\ref{sec:phase_conjugation}, where we found ${\cal{S}}^\text{PC}$ yields highest {\it local} sensitivity enhancement. 
The argument presented above now allows us to conclude that MRE allows one to achieve the highest {\it global} enhancement of sensitivity.

In Ref.~\cite{2022_Bender_Coherent_Enhancement}, we observed that  discrepancy between $\eta_{\cal S}^\text{MRE}$ and $\eta_{\cal R}^\text{MRE}$ diminishes in the limit $d\gg W_1+W_2$, which is fully consistent with our extensive tests of this relationship reported in Fig.~\ref{fig:panel_enhan_factor}. In fact, we see good agreement between sensitivity and remission enhancements for $d$ as small as $\simeq 3\times(W_1+W_2)$, see panel (a) of the figure. In Appendix~\ref{sec:Marcenko_Pastur} we provide extensive numerical evidence relating $\eta_{\cal S}^\text{MRE}$ to $\eta_{\cal R}^\text{MRE}$.

The results presented in this section focus on 2D disordered systems, where the long-range intensity correlation  of speckle patterns, usually denoted as $C_2$, contributes significantly to the remission enhancement (see Appendix~\ref{sec:filtered_random_matrix_theory}). However, in most 3D diffusive media, particularly biological tissues, the impact of $C_2$ is expected to be considerably smaller, as 
its contribution to the enhancement, $N_1C_2$, scales as $\propto N_1/k\ell$ in 2D and as $\propto \sqrt{N_1}/k\ell$ in 3D. 
Consequently, in 3D media with $k\ell \gg 1$, the enhancement of optical sensitivity will primarily follow the prediction 
\be
\eta_{\cal R}^\text{MRE}\xrightarrow[C_2\rightarrow 0]{ }\left(1 + \sqrt{\frac{N_1}{N_2}}\right)^2, 
\label{EqFRM3D}
\ee
which stems from the Marcenko-Pastur limiting distribution for remission eigenvalues~\cite{2022_Bender_Coherent_Enhancement}. The dashed lines in Fig.~\ref{fig:panel_enhan_factor} illustrate this limiting behavior by omitting $C_2$ contribution from the sensitivity calculations. These results indicate that, although WFS in 2D can exploit mesoscopic correlations to enhance sensitivity beyond diffusion theory, enhancement of remission, and thus sensitivity, can be achieved in 3D, particularly in $N_1\gg N_2$ regime. The physical mechanism for enhancement of remission/sensitivity in absence of appreciable $C_2$ correlations is interference of multiple coherent contributions at the output port -- consequence of the determinism of wave transport.

\section{Optical sensitivity: vector waves\label{sec:sensitivity_analytical_vector}}
\noindent Our previous theoretical and numerical analyses in Secs.~\ref{sec:sensitivity_analytical_scalar} and \ref{sec:numerical_simulation_scalar} have been based on the scalar wave equation -- Eq.~\eqref{EqWave1}. 
Although scalar wave approximation is a common simplification in the study of wave transport in random media~\cite{Akkermans_Montambaux_2007}, its validity requires careful consideration in the context of coherent control via WFS.
Indeed, it has been demonstrated that polarization of the wave in strongly scattering media can reliably be manipulated~\cite{2012_Guan,2017_deAguiar}.
Notably, the first step towards the full 3D vector wave treatment can be readily made in 2D.
In this geometry, the electromagnetic wave equation decouples into TM and TE polarization states. 
The former is equivalent to 2D scalar wave equation, which we modeled numerically in Sec.~\ref{sec:numerical_simulation_scalar}.
TE polarization, on the other hand, consists of two components of the electric field.
Therefore, modeling TE waves incorporates an additional, compared to scalar or TM waves, scattering channel between two polarization components of electric field.
In Ref.~\cite{2022_Jara_Simulation_Coherent_Remission} we used numerical simulations to investigate whether wave polarization has any effect on the full distribution of the remission eigenvalues.
We found that the distribution remained independent of the polarization as long as two systems had the same macroscopical scattering parameters, i.e. the same transport mean free path. This is consistent with a recent numerical study~\cite{2024_Wade_Transmission_eigenvalues}, which found that the distribution of the transmission eigenvalues in fully vectorial 3D simulations coincides with that for scalar waves.

Generalizing this result to the sensitivity is far from obvious.
Clearly, the sensitivity for scalar waves in Eq.~\eqref{EqSiBorn} is determined by the interference between two fields -- one launched from the input port $E_0(\vec{r}_0)$ and the other from the output one $\phi(\vec{r}_0)$. 
Extending such a formula to TE waves should, in fact, include interference effects between two polarization states of each field. 
Such a generalization is nontrivial, motivating us to revisit the derivation of sensitivity for TE waves below.\\ 

\subsection{Mapping onto the scalar sensitivity model\label{sec:Helmholtz_equation_derivation}}
\noindent We begin by enumerating the underlying assumptions of our theoretical model.
First of all, we consider TE-polarized wave propagating into a 2D scattering system.
This corresponds to considering two in-plane components ($x,y$) of the electric field and one ($z$) component of the magnetic field.
Secondly, we consider non-magnetic media where $\mu\equiv \mu_{0}$.
Lastly, scattering originates from the spatially non-uniform dielectric permittivity $\epsilon(\vec{r})$, assumed to be real. Under these conditions, the equation describing the out-of-plane component of the magnetic field $H_z(\vec{r})$ becomes
\begin{equation}
\nabla^2 H_{z}(\vec{r})+
k^2\epsilon(\vec{r})\,H_{z}(\vec{r})-
\frac{1}{\epsilon(\vec{r})}
\nabla\epsilon(\vec{r})\cdot
\nabla\,H_{z}(\vec{r})=0.
\label{eq:magnetic_scalar_eq}
\end{equation}
In contrast to the scalar case (TM polarization) in Eq.~(\ref{EqWave1}), we gain an additional term $\sim \nabla\epsilon(\vec{r})\cdot\nabla H_z(\vec{r})$. Both vector quantities in this term lie in-plane. 

In the next step, we demonstrate that it is possible, with a judicial change of variables, to transform Eq.~\eqref{eq:magnetic_scalar_eq} into a Helmholtz equation similar to Eq.~\eqref{EqWave1} for scalar waves. 
Introducing the (scalar) function
\be
h(\vec{r})=\frac{H_{z}(\vec{r})}{\sqrt{\epsilon(\vec{r})}},
\label{eq:change_variable_eq}
\ee
we transform Eq.~\eqref{eq:magnetic_scalar_eq} into 
\be
\nabla^2h(\vec{r})+\kappa^2(\vec{r})\,h(\vec{r})=0,
\label{eq:magnetic_helmholtz_eq}
\ee
where $\kappa^2(\vec{r})$ differs markedly from $k^2\epsilon(\vec{r})$ in Eq.~\eqref{EqWave1}
\be
\kappa^2(\vec{r}) \equiv k^2\,\epsilon(\vec{r}) +
\frac{\nabla^2\epsilon(\vec{r})}{2\,\epsilon(\vec{r})}-\frac{3}{4}\frac{\nabla\epsilon(\vec{r})\cdot\nabla\epsilon(\vec{r})}{\epsilon^2(\vec{r})}.
\label{eq:kappa_0_reduced_eq}
\ee
Such a complex dependence on $\epsilon(\vec{r})$ causes further complications when we introduce a small perturbation $\delta\epsilon(\vec{r})$, needed for computing sensitivity. Unlike Eq.~\eqref{EqWave1}, the linear perturbation term in $\kappa^2(\vec{r})$ becomes
\begin{widetext}
\begin{equation}
\delta\kappa^2(\vec{r})\simeq
k^2\,\delta\epsilon(\vec{r})-
\frac{\nabla^2\epsilon(\vec{r})}{2\,\epsilon^2(\vec{r})}\,\delta\epsilon(\vec{r})+
\frac{3}{2}\frac{\nabla\epsilon(\vec{r})\cdot\nabla\epsilon(\vec{r})}{\epsilon^3(\vec{r})}\,\delta\epsilon(\vec{r})-
\frac{3}{2}\frac{\nabla\epsilon(\vec{r})\cdot\nabla\delta\epsilon(\vec{r})}{\epsilon^2(\vec{r})}.
\label{eq:approx_kappa_eq}
\end{equation}

Next, we apply the same theoretical approach as in Sec.~\ref{sec:field_sensitivity}, in order to derive the appropriate expression for the TE sensitivity, see Appendix~\ref{sec:diffusion_perturbation}. We arrive at 
\be
{\cal{S}}(\vec{r}_0)=-\delta V\left(
k^2-\frac{\nabla^2\epsilon(\vec{r}_0)}{2\,\epsilon^2(\vec{r}_0)}+
\frac{3}{2}\frac{\nabla\epsilon(\vec{r}_0)\cdot\nabla\epsilon(\vec{r}_0)}{\epsilon^3(\vec{r}_0)}-
\frac{3}{2}\nabla\!\left(\frac{\nabla\epsilon(\vec{r}_0)}{\epsilon^2(\vec{r}_0)}\right)
\right)
\text{Re}[\phi_h(\vec{r_0})h_0(\vec{r_0})],
\label{eq:final_sensitivity_expression_TE_fields_eq}
\ee
\end{widetext}
where $\phi_h(\vec{r_0})$ and $h_0(\vec{r_0})$ are defined analogously to $\phi(\vec{r}_0)$ and $E_0(\vec{r}_0)$ in Eq.~\eqref{EqSiBorn}. 
Also, just like Eq.~\eqref{EqSiBorn}, the above expression is exact in the limit of small perturbation. 
In contrast to the scalar case, Eq.~\eqref{eq:final_sensitivity_expression_TE_fields_eq} depends on derivatives of the dielectric function, making it too cumbersome to apply in numerical modeling. 

Here, we present arguments that the additional terms in the parenthesis in Eq.~\eqref{eq:final_sensitivity_expression_TE_fields_eq} can, in fact, be insignificant in many systems. Specifically, we consider two examples of such disordered media:

\noindent Case I: biological medium such as weakly scattering tissue. 
The dielectric function in this case varies rather weakly -- the characteristic fluctuation $\overline{\delta\epsilon(\vec{r}_0)^2}^{1/2}$ is much smaller than the mean value $\overline{\epsilon(\vec{r}_0)}$. 
Furthermore, the characteristic spatial extent of these variations $L_\epsilon$ is much larger than the wavelength of light. 
These two facts allow us to obtain a rough estimate for the gradient-terms in  Eq.~\eqref{eq:final_sensitivity_expression_TE_fields_eq} as $\sim\left(\overline{\delta\epsilon(\vec{r}_0)^2}/\overline{\epsilon(\vec{r}_0)}^2\right)/L_\epsilon^2$ and to conclude that they should be negligible compared to the first term $k^2$.

\noindent Case II: planar scattering media considered in Ref.~\cite{2022_Bender_Coherent_Enhancement}. 
It also used as a model in our scalar simulations in Sec.~\ref{sec:numerical_simulation_scalar}. 
Here, we consider a uniform (but large $\epsilon\sim 2.85^2$) dielectric function with a small concentration of air holes ($\epsilon_\text{air}=1$). 
In such a medium, the variations of $\epsilon(\vec{r}_0)$ are limited to the interfaces between the dielectric and air-holes, making them statistically rare. 
Therefore, although the non-zero contributions are on the order of unity, their statistical contribution (volume average) is small. 
Specifically in our numerical model we estimate that this contribution is of the order of $\sim (\Delta x/r)\times f\ll 1$. 
Here, $\Delta x$ is the pixel size, $r$ is the radius of the hole, and $f$ is the filling fraction of the air holes. 
The estimate shows that this contribution, limited to only interfaces between the dielectric and air, vanishes as $\Delta x/r\rightarrow 0$.
Consequently, in this case too, we are justified to omit the gradient terms, and we obtain the following simplified expression for the TE optical sensitivity:
\be
{\cal S}(\vec{r}_{0}) \simeq  -k^2\delta V\,
\text{Re}[\phi_h(\vec{r_0})h_0(\vec{r_0})].
\label{eq:Approximate_TE_fields_eq}
\ee
This is the expression we are going to investigate numerically in the next section.

\subsection{Numerical analysis\label{sec:numerical_simulation_vector}}
\begin{figure*}
\includegraphics[width=5in]{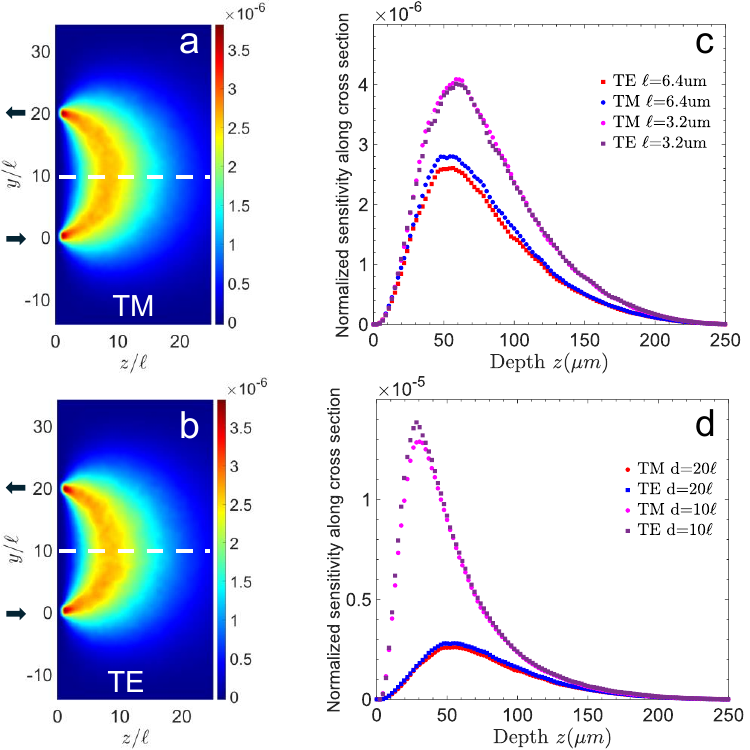}
\vskip -0.3cm
\caption{Effect of polarization on microscopic optical sensitivity.
(a,b)~Sensitivity maps computed using Eq.~\eqref{EqSiBorn} for TM polarization  and Eq.~\eqref{eq:Approximate_TE_fields_eq} for TE polarization, respectively. 
Results are averaged over 1000 disorder realizations. 
The microscopic disorder in (b) is adjusted to yield the same transport mean free path as in (a), $\ell=6.4\,\mu$m. 
The sensitivity distributions show strong agreement in both shape and magnitude.
(c)~Depth cross-section along the mid-plane (dashed line in panels (a) and (b)), comparing TE and TM polarizations for different values of $\ell$. Input-output separation is $d=20\ell$.
(d)~Further comparison of TE and TM sensitivity at fixed $\ell=6.4\,\mu$m, showing that the agreement persists across different geometrical parameters, such as the input-output port separation $d$.
\label{fig:polarization}}
\end{figure*}

\noindent We use the numerical model described in detail in Sec.~\ref{sec:scalar_numerical_model}, with the only exception that we now consider the TE polarization using MESTI software package~\cite{2022_HoChun_MESTI}.
Shaped wavefront used to excite the maximum remission eigenchannel is constructed from the remission matrix ${\cal R}$, which, in this case, represents the field remission matrix for the out-of-plane component of the magnetic field.
Our goal is to compare the vector-wave (TE -- two components of the electric field) sensitivity to the scalar (TM -- one component of the electric field) case considered in Sec.~\ref{sec:scalar_numerical_model}.
Fig.~\ref{fig:polarization}a,b shows 2D sensitivity map ${\cal s}(\vec{r}_0)$ in the TM and TE polarization, respectively. 
The data is statistically averaged over 1000 realizations. 
The input/output widths $W_1\!=\!W_2$ are 10 $\mu$m and input/output separation is $d\!=\!20\!\times\!\ell$.
We set the microscopical disorder parameters, air-holes size and density, for the TE case such that the macroscopical scattering parameter -- the transport mean free path -- is identical to that in the scalar simulation, $\ell\!=\!6.4\,\mu$m.
We observe the typical `banana' profile in both cases as the sensitivity maps agree quantitatively.
This can also be observed from the cross-section of each map along the mid-plane (dashed line in Fig.~\ref{fig:polarization}a,b) shown in Fig.~\ref{fig:polarization}c.
To confirm that the agreement between scalar and vector cases is not accidental, we varied the microscopic disorder in both cases to reduce the mean free path to $\ell=3.2\,\mu$m.
Again, we observe good agreement between simulations, see Fig.~\ref{fig:polarization}c.
We also verified that geometrical parameters, such as separation distance $d$ between input and output do not cause discrepancy between scalar and vector results.
Figure~\ref{fig:polarization}d shows that the agreement is indeed preserved irrespective of the value of $d$.

To provide insight into the polarization invariance of remission sensitivity, we make two complementary observations -- one formal and one physical. 
Formally, Eq.~(\ref{eq:Approximate_TE_fields_eq}), derived for TE-polarized vector waves in 2D, has the same mathematical structure as Eq.~\eqref{EqSiBorn} for scalar (TM) waves. As discussed in Sec.~\ref{sec:Helmholtz_equation_derivation}, such structural similarity is expected for a broad class of scattering systems. This supports the conclusion that optical sensitivity, defined via these equations, should be largely polarization-independent. 
Physically, it is difficult to identify any mechanism by which a particular component of the electric field would be systematically favored {\it inside} the diffusive scattering system, once scattering direction becomes completely randomized. 

\section{Conclusion\label{sec:conclusion}}
\noindent In this work, we developed a microscopic theory for optical sensitivity that remains valid under coherent wavefront control and verified its consistency with diffusion-based models in the appropriate limit. 
This formalism revealed how sensitivity can be systematically enhanced via phase conjugation and remission eigenchannels, and enabled quantitative predictions of the enhancement achieved with these input excitations. 
Although the numerical simulations presented in this work have been conducted in 2D, the theoretical description remains applicable to 3D systems. 

To summarize our results, we evaluated sensitivity map along the `banana' trajectory -- half-circle path of radius $d/2$ connecting input and output, see Table~\ref{tab:sensitivity_summary}. 
This shape corresponds to the spatial profile of sensitivity under random input. 
While this profile is preserved under remission eigenchannel (MRE) and phase-conjugated (PC) excitations, the required input wavefronts differ: MRE uses a fixed input wavefront found by maximizing remission, whereas PC requires a distinct inputs for each target position.
In the 2D case, sensitivity is nearly constant along the path, with PC input yielding a factor of $N_1$ enhancement $\eta^{\text{PC}}$ due to constructive interference of time-reversed paths. 
MRE yields a sensitivity enhancement $\eta^\text{MRE}$, which can be predicted either by the FRM theory or by the effective Marchenko-Pastur model reported in Table~\ref{tab:sensitivity_summary}, both incorporating the effect of  long-range correlation $C_2$; see Appendices~\ref{sec:filtered_random_matrix_theory} and~\ref{sec:Marcenko_Pastur}. 
In 3D, the sensitivity varies along the banana geometry as a function of polar angle $\theta$, see Appendix~\ref{sec:diffusion_in_remission_geometry}. 
However, both PC and MRE enhancement factors remain the same as in 2D. 
That said, the contribution of $C_2$ to $\eta^\text{MRE}$, is expected to be much smaller, if not insignificant, in most 3D systems where $\ell \gg \lambda$. 
This implies that, in the 3D case, the dominant mechanism for remission or sensitivity enhancement stems from  constructive interference, which is preserved due to the deterministic nature of coherent wave transport.
 
\begin{table*}[htbp]
\centering
\caption{Sensitivity ${\cal S}(r_0)$ evaluated along the banana trajectory for three excitation schemes: random input, phase-conjugated (PC) input, and maximum remission eigenchannel (MRE) excitation.}
\begin{tabularx}{\textwidth}{|>{\centering\arraybackslash}p{0.6cm}|>{\centering\arraybackslash}X|>{\centering\arraybackslash}p{3.3cm}|>{\centering\arraybackslash}p{3.9cm}|}
\hline
$\backslash$ &Random input sensitivity & PC enhancement $\eta^\text{PC}$ & MRE enhancement $\eta^\text{MRE}$ \\
\hline
\rule[-3ex]{0pt}{8ex} 2D:
& $-\delta V\times\displaystyle\frac{1}{\pi^2}\times\frac{N_1N_2}{d^2}$ 
& $\times N_1$ 
& $\times \displaystyle \left(1+\sqrt{\frac{N_1}{N_2} +N_1C_2 }\right)^2$ 
\\
\hline
\rule[-3ex]{0pt}{8ex} 3D:
& $-\delta V\times\displaystyle\frac{9}{8\pi k}\times\frac{N_1N_2}{ d^4\cos\theta}$  
& $\times N_1$ 
& $\times \displaystyle \left(1+\sqrt{\frac{N_1}{N_2} +N_1C_2 }\right)^2$ 
\\
\hline
\end{tabularx}
\label{tab:sensitivity_summary}
\end{table*}

We note two important considerations that motivate the practical implementation of coherent wavefront control strategies.
First, the decay of sensitivity with source-detector separation is more severe in three dimensions: while in 2D the sensitivity scales as ${\cal S}^\text{2D}\propto 1/d^2$, in 3D it follows ${\cal S}^\text{3D}\propto 1/d^4$, making the signal diminish even more rapidly with depth, c.f. Table~\ref{tab:sensitivity_summary} and Appendix~\ref{sec:diffusion_in_remission_geometry}. 
Second, numerical studies in Secs.~\ref{sec:numerical_simulation_scalar},~\ref{sec:sensitivity_analytical_vector} focused on a localized absorber used to define sensitivity, but did not account for attenuation due to absorption throughout the bulk medium, common in biological tissue samples. 
While the microscopic expression for sensitivity remains unchanged in the presence of absorption, the resulting spatial profile of sensitivity map does change. 
As we demonstrated in our earlier work~\cite{2022_Bender_Coherent_Enhancement}, bulk absorption makes the banana-shaped sensitivity map shallower. 
The above considerations motivate further study and demonstrate the promise for WFS to enhance signal strength via constructive interference in order to counteract the degradation of sensitivity due to both geometric spreading and attenuation.

From the practical standpoint,  MREs offer a unique opportunity -- they {\it maximize} the detection sensitivity throughout the entire imaging volume, while requiring only the knowledge of the remission matrix that can be obtained {\it non-invasively}. 
Furthermore, our analysis of fluctuation of sensitivity in Appendix \ref{sec:fluctuations} shows that, in addition to enhancing the mean sensitivity, MREs also exhibit reduced realization-to-realization variability compared to random input. This robustness ensures that standard inverse-problem techniques can still be applied with wavefront-shaped input, but now with significantly improved signal-to-noise and consequently greater penetration depth.

The next challenge is to translate these insights into experimental implementations that can overcome the constraints of the conventional DOT. The use of dynamic wavefront control in remission-based imaging needs to be explored in live biological systems, where temporal variations in tissue properties introduce additional complexity. Moreover, the question remains whether the principles demonstrated here can be extended to fluorescence and other optical contrast mechanisms that operate under highly scattering conditions.

Beyond imaging, these results point to a broader class of problems in wave physics where coherent control can be used to selectively enhance signal extraction in complex media. The ability to enhance and/or reshape sensitivity maps in a deterministic way suggests applications in adaptive sensing, optogenetics, and non-invasive diagnostics where penetration depth and resolution are fundamentally limited by scattering. The real impact of this approach will come from experimental validation and integration of computational wavefront control into next-generation imaging systems.

\begin{acknowledgments}
This work is supported partly by the US National Science Foundation (NSF) under Grants No. DMR-1905442 and No. DMR-1905465, and by the US Office of Naval Research (ONR) under Grant No. N00014-221-1-2026, and by the French Government under the program Investissements d’Avenir.
Extensive numerical simulations in this work were carried out on the Mill high-performance computing (HPC) cluster~\cite{2024_Cluster} at Missouri University of Science and Technology. PJ and AY gratefully acknowledge Predrag Lazic for technical assistance and IT support.
\end{acknowledgments}

\appendix

\section{Perturbative approach to diffuse intensity}\label{sec:diffusion_perturbation}

\noindent Intensity $\overline{I}_0(\vec{r})$ and $\overline{I}(\vec{r})$, before and after a perturbation is introduced, satisfy diffusion equation
\begin{align}
-D\nabla^2\,\overline{I}_0(\vec{r})=0,\label{EqDiff}\\
\left[-D\nabla^2+v_E\delta\mu(\vec{r})\right]\,\overline{I}(\vec{r})=0.
\label{EqDiff_with_pert}
\end{align}
Here $v_E$ is the energy velocity (equal to the the phase velocity $c/n$ in the absence of resonant scattering), $D$ is the diffusion coefficient, and $\delta\mu(\vec{r})$ a local perturbation in absorption. Although we assumed above that diffusion coefficient is spatially invariant and absorption is zero in absence of the perturbation, this description can be straightforwardly generalized to cover these effects. Inclusion of these effects would simply detract from our main goal to compare and contrast diffusive picture below with the microscopic treatment in Sec.~\ref{sec:field_sensitivity}. 

The leading correction to $\overline{I}_0(\vec{r})$ can be estimated perturbatively as
\begin{equation}
\overline{I}(\vec{r})\simeq\overline{I}_0(\vec{r})+\int d\vec{r}' d\vec{r}''\,G^\text{I}_0(\vec{r},\vec{r}')\,{\cal V}^\text{I}(\vec{r}',\vec{r}'')\,\overline{I}_0(\vec{r}''),
\label{EqLSI2}
\end{equation}
where $G^\text{I}_0(\vec{r},\vec{r}')$ is the Green's function of the diffusion equation
\begin{equation}
-D\nabla^2 G^\text{I}_0(\vec{r},\vec{r}')=\delta(\vec{r} - \vec{r}'),
\label{EqDiffGreen}
\end{equation}
and the perturbation, assumed to have linear size much smaller than the transport mean free path $\ell$, is represented by the delta potential
\begin{equation}
{\cal V}^\text{I}(\vec{r})=\tilde{{\cal V}}^\text{I}\,\delta(\vec{r}-\vec{r}_0), \;\; \text{with} \;\;\tilde{{\cal V}}^\text{I}=-v_E\,\delta\mu\,\delta V.
\label{EqPertI}
\end{equation}
Substituting Eq.~(\ref{EqPertI}) into Eq.~(\ref{EqLSI2}), gives the perturbation of intensity $\overline{\delta I}(\vec{r})=\overline{I}(\vec{r})-\overline{I}_0(\vec{r})$ similar to Eq.~(\ref{EqBornField}),
\begin{equation}
\overline{\delta I}(\vec{r}) = \tilde{{\cal V}}^\text{I}\,G^\text{I}_0(\vec{r},\vec{r}_0)\,\overline{I}_0(\vec{r}_0).
\label{dIDiff}
\end{equation}
Comparing the above to Eq.~(\ref{EqBornField}), we find $\delta \mu=k\epsilon''$.

Equation~\eqref{dIDiff} is used in Sec.~\ref{sec:sensitivity_diffusion_approximation} to evaluate the change in the flux at the detector in the response to the perturbation.  

\section{Proof of the Green's function identity Eq.~(\ref{EqGreenIdentity}) in diffusive approximation}\label{sec:green_identity}
\begin{figure}
\centering
\includegraphics[width=3in]{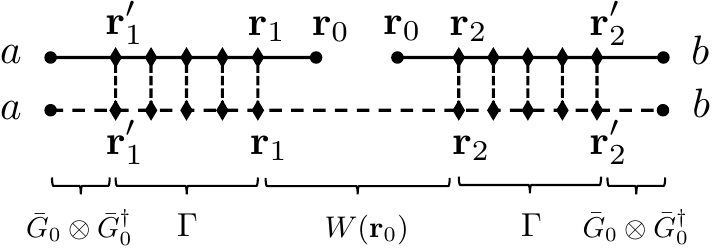}
\caption{Leading diagram in the expansion of $\overline{
\mel{\chi_b}{G_0}{\vec{r}_0}\!
\mel{\vec{r}_0}{G_0}{\chi_a}\!
\mel{\chi_b}{G_0}{\chi_a}^*}$. Solid and dashed horizontal lines represent the Green's operators $\bar{G}_0$ and $\bar{G}_0^\dagger$, respectively, while diamonds represent scatterers. Vertical dashed lines connect identical scatterers.}
\label{fig:diagram}
\end{figure}
\noindent To evaluate the left-hand-side in Eq.~(\ref{EqGreenIdentity}), we expand each Green's function over all possible scattering paths, selecting only those paths that do not accumulate phase terms dependent on the specific disorder configuration. These paths are represented by the ladder diagram shown in Fig.~\ref{fig:diagram}. We find
\begin{widetext}
\begin{align}
\overline{
\mel{\chi_b}{G_0}{\vec{r}_0}\!
\mel{\vec{r}_0}{G_0}{\chi_a}\!
\mel{\chi_b}{G_0}{\chi_a}^*}& =\label{EqMean3G}\\
\int d\vec{r}_1 d\vec{r}'_1 d\vec{r}_2 d\vec{r}'_2\,
\left|\!\mel{\chi_b}{\overline{G}_0}{\vec{r}'_2}\right|^2
\Gamma(\vec{r}'_2, \vec{r}_2) 
\mel{\vec{r}_2\vec{r}_2}{\hat{W}(\vec{r}_0)}{\vec{r}_1\vec{r}_1}\,
&\Gamma(\vec{r}_1,\vec{r}'_1) 
\left|\!\mel{\vec{r}'_1}{\overline{G}_0}{\chi_a}\right|^2,\nonumber
\end{align}
\end{widetext}
which involves the 4-rank tensor $\hat{W}(\vec{r}_0)=\left[ \bar{G}_0\ket{\vec{r}_0} \bra{\vec{r}_0}\bar{G}_0 \right]\otimes\bar{G}_0^\dagger$. The kernel $W(\vec{r}_0, \vec{r}_2, \vec{r}_1) \equiv \bra{\vec{r}_2 \vec{r}_2} \hat{W}(\vec{r}_0) \ket{\vec{r}_1 \vec{r}_1}$ can be expressed as
\begin{equation}
W(\vec{r}_0, \vec{r}_2, \vec{r}_1) =
\mel{\vec{r}_2}{\overline{G}_0}{\vec{r}_0}\!
\mel{\vec{r}_0}{\overline{G}_1}{\vec{r}_1}\!
\mel{\vec{r}_2}{\overline{G}_0}{\vec{r}_1}^*,
\end{equation}
as illustrated in Fig.~\ref{fig:diagram}. This  gives a non-negligible contribution to the integral Eq.~(\ref{EqMean3G}) for $\vec{r}_1$ and $\vec{r}_2$ in the vicinity of $\vec{r}_0$. Using $\Gamma(\vec{r}_1,\vec{r}'_1)\simeq\Gamma(\vec{r}_0,\vec{r}'_1)$ and $\Gamma(\vec{r}'_2,\vec{r}_2)\simeq\Gamma(\vec{r}'_2,\vec{r}_0)$, we get
\begin{equation}
\overline{
\mel{\chi_b}{G_0}{\vec{r}_0}\!
\mel{\vec{r}_0}{G_0}{\chi_a}\!
\mel{\chi_b}{G_0}{\chi_a}^*}=K_{b\vec{r}_0}W(\vec{r}_0)\,K_{\vec{r}_0 a},
\end{equation}
where $K_{\vec{r}_0 a}=K_{a\vec{r}_0 }$ is defined as 
\begin{equation}
K_{\vec{r}_0 a}=\int d\vec{r}_1 \Gamma(\vec{r}_0,\vec{r}_1)
\left|\!\mel{\vec{r}_1}{\overline{G}_0}{\chi_a}\right|^2,
\end{equation}
and $W(\vec{r}_0) = \int d\vec{r}_1d\vec{r}_2 W(\vec{r}_0, \vec{r}_2, \vec{r}_1)$. The integrated kernel  is independent of the position $\vec{r}_0$ of the perturbation, since
\begin{equation}
W(\vec{r}_0)=\bra{\vec{r}_0}\bar{G}_0\bar{G}_0^\dagger\bar{G}_0\ket{\vec{r}_0}
=\int\frac{d\vec{q}}{(2\pi)^d}\bar{G}_0(q)^2\bar{G}_0(q)^*.
\end{equation}
Using $\bar{G}_0(q)= (k^2-q^2-ik/\ell)^{-1}$, explicit calculation in the limit $k\ell \gg1$ gives
\begin{equation}
W(\vec{r}_0)\simeq \frac{i}{8}\frac{\ell^2}{k^2}\left( \frac{k}{\pi} \right)^{{\cal D}-2}.
\end{equation}
We conclude that the left hand-side of Eq.~(\ref{EqGreenIdentity}) can be expressed as
\begin{equation}
\overline{
\mel{\chi_b}{G_0}{\vec{r}_0}\!\!
\mel{\vec{r}_0}{G_0}{\chi_a}\!\!
\mel{\chi_b}{G_0}{\chi_a}^*}=\!\frac{i}{8}\,\frac{\ell^2}{k^2}\!\left( \frac{k}{\pi} \right)^{{\cal D}-2}\!\!\!\!\!\!\!K_{b\vec{r}_0}\,K_{\vec{r}_0 a}.
\label{EqMean3Gv2}
\end{equation}

The right-hand-side of Eq.~(\ref{EqGreenIdentity}) can also be expressed in terms of the product $K_{b\vec{r}_0} K_{\vec{r}_0 a}$. Indeed, in the diffusive limit, we have
\begin{align}
&\overline{\left|\mel{\vec{r}_0}{G_0}{\chi_a}\right|^2}
=\!\!\int \!\!d\vec{r}_1 d\vec{r}_2 \!
\left|\!\mel{\vec{r}_0}{\bar{G}_0}{\vec{r}_2}\right|^2
\Gamma(\vec{r}_2,\vec{r}_1)\!
\left|\!\mel{\vec{r}_1}{\bar{G}_0}{\chi_a}\right|^2 \nonumber\\
&\simeq\left[\int d\vec{r}_2
\left|\!\mel{\vec{r}_0}{\bar{G}_0}{\vec{r}_2}\right|^2
\right]\,\int d\vec{r}_1
\Gamma(\vec{r}_0,\vec{r}_1)
\left|\!\mel{\vec{r}_1}{\bar{G}_0}{\chi_a}\right|^2 \nonumber\\
&=\frac{\ell}{4k} \left(\frac{k}{\pi} \right)^{{\cal D}-2}K_{\vec{r_0}a}.
\end{align}
This result implies that Eq.~(\ref{EqMean3Gv2}) can be written as
\begin{align}
&\overline{
\mel{\chi_b}{G_0}{\vec{r}_0}\!
\mel{\vec{r}_0}{G_0}{\chi_a}\!
\mel{\chi_b}{G_0}{\chi_a}^*
}
\nonumber
\\
&=2i\left(\frac{\pi}{k}\right)^{{\cal D}-2}\,
\overline{\left|\mel{\chi_b}{G_0}{\vec{r}_0}\right|^2}\,\,
\overline{\left|\mel{\vec{r}_0}{G_0}{\chi_a}\right|^2},
\end{align}
and the identity Eq.~(\ref{EqGreenIdentity}) then follows naturally.

\section{ Asymmetric cases for random and coherent input sensitivities \label{sec:sensitivity_asymmetric}}

\begin{figure*}
\includegraphics[width=6in]{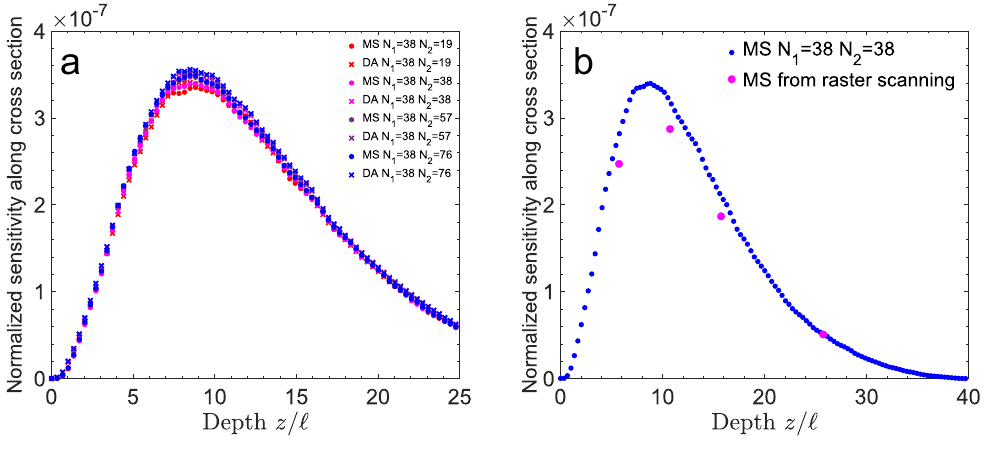}
\vskip -0.5cm
\caption{Comparison between microscopic sensitivity (MS) and diffusion approximation (DA) under random input excitation for asymmetric input-output port widths ($W_1\neq W_2$). 
(a)~Simulated sensitivity maps for systems with transport mean free path $\ell=$6.4 $\mu$m, using different input-output width combinations  (in $\mu$m) : 10$\times$5, 10$\times$10, 10$\times$15 and 10$\times$20. 
Excellent agreement is observed in all cases.
(b)~Validation of the microscopic sensitivity formula: results obtained using the second expression in Eq.~(\ref{EqSiBorn}) (blue circles) are compared with direct brute-force calculations based on the first part of Eq.~(\ref{EqSiBorn}) (red circles), confirming consistency.
\label{fig:panel_asymmetric_definition}}
\end{figure*}

\noindent In Sec.~\ref{sec:scalar_random_input_results} in Fig.~\ref{fig:panel_random} we showed the computation of the normalized sensitivity for the random input excitation for the symmetric configuration with $W_1=W_2$. In Fig.~\ref{fig:panel_asymmetric_definition}a we also include the numerical results for asymmetric input/output cases and observe perfect agreement irrespective of the ratio between $W_1$ and $W_2$. This provides further support for our normalization procedure in Sec.~\ref{sec:normalized_sensitivity}.

Furthermore, in Fig.~\ref{fig:panel_asymmetric_definition}b we have applied the first part of Eq.~(\ref{EqSiBorn}). This is the serial way to compute optical sensitivity by taking the ratio between change of the output flux and the strength of the perturbation, see Fig.~\ref{fig:Sensitivity_methods}a. Numerically, we added small amount of the absorption in a square area of $\ell\times\ell$, placed at several $z$'s along the center line. Good agreement with the sensitivity map calculation confirms the validity of the microscopic approach in Sec.~\ref{sec:field_sensitivity}.

\section{Diffusion approximation for sensitivity in remission geometry:
analytical result\label{sec:diffusion_in_remission_geometry}}
\noindent 
\noindent In the remission geometry considered in Sec.~\ref{sec:numerical_simulation_scalar}, a closed-form expression for optical sensitivity Eq.~(\ref{EqDiffusiveSensitivity}) can be obtained in 2D and 3D. To that end, we begin with the normalized sensitivity Eq.~(\ref{EqInt_sensitivity}) in terms of the random-input intensities 
\begin{equation}
\overline{{\cal s}}_\text{diff}(\vec{r_0}) = 
-\displaystyle\frac{1}{N_1 N_2}
\left(\frac{\pi}{k^2}\right)^{\mathcal{D}-2}
\overline{I}^{\text{RI}}_{\text{in}}(\vec{r_0})\,
\overline{I}^{\text{RI}}_{\text{out}}(\vec{r_0}).
\label{EqDiffusiveSensitivity2D}
\end{equation}
Here, input and output intensities can be found by solving 2D/3D diffusion equation in semi-infinite geometry $z>0$. At $z=0$, we apply open boundary condition with an extrapolation length $z_e$, see Ref.~\cite{Akkermans_Montambaux_2007}. We also assume that the widths of input and output, located at $y=0,d$, see e.g. Fig.~\ref{fig:panel_random}a, are much smaller than their separation $W_1,W_2\ll d$. 
Normalization adopted in Sec.~\ref{sec:sensitivity_diffusion_approximation}, assumes $N_1$ units of dimensionless flux is used to excite the system. 
The source of diffusive waves is $\nabla\cdot\vec{J}_\text{ball}$, where $\vec{J}_\text{ball}=c/(2n_\text{eff}k)\times e^{-z/\ell} \delta(y)$ is the ballistic (unscattered) flux. 
The normalization factor $[c/2kn_\text{eff}]$ is obtained from the microscopic wave equation~\cite{2022_HoChun_MESTI}.

In 2D, we find 
\begin{equation}
\overline{I}^{\text{RI,2D}}_{\text{in}}(\vec{r_0})=
\displaystyle\frac{c}{2kn_\text{eff}}\,
\displaystyle\frac{N_1}{\pi D}\,
\displaystyle\frac{\ell\,z_0}{(z_0+z_e)^2+y_0^2}.
\label{EqIinRI}
\end{equation}

\noindent Substituting Eq.~(\ref{EqIinRI}) into Eq.~(\ref{EqDiffusiveSensitivity2D}) we obtain the final result for the normalized sensitivity map
\begin{widetext}
\begin{equation}
\overline{{\cal s}}^\text{2D}_\text{diff}(y_0,z_0)= 
-\displaystyle\frac{1}{\pi^2k^2}
\displaystyle\frac{z_0^2}{\left[(z_0+z_e)^2+y_0^2\right]\left[(z_0+z_e)^2+(y_0-d)^2\right]}.
\label{EqBornDiff3}
\end{equation}
\end{widetext}
This is the desired analytical expression used to study dependence on the detector separation in Fig.~\ref{fig:panel_random} from Sec.~\ref{sec:remission_enhancement}. 
Interestingly, we find that Eq.~(\ref{EqBornDiff3}) is essentially independent on the transport mean free path. Indeed, at distances exceeding the extrapolation length $z_e\sim\ell$, the dependence on $\ell$ is insignificant. In fact, along the crescent banana trajectory, the sensitivity remains nearly constant
\begin{equation}
\overline{{\cal s}}^\text{2D}_\text{diff}\simeq
-\displaystyle\frac{1}{\pi^2}\times\frac{1}{k^2 d^2}.
\label{sdiff_dimensionless}
\end{equation}
It depends only on the vacuum wavenumber $k=2\pi/\lambda_0$ and source-detector separation $d$. 
For $\lambda_0=1.55\mu$m and $d=128\mu$m we get $\overline{{\cal s}}_\text{diff}(d/2,d/2)\simeq 3.8\times 10^{-7}$, which agrees with the microscopical numerical calculations reported in Fig.~\ref{fig:panel_random}.

The 3D result can be obtained analogously by using the the appropriate expression for Green's function. For $\overline{I}^{\text{RI}}_{\text{in}}(\vec{r_0})$ we get
\begin{equation}
\overline{I}^{\text{RI,3D}}_{\text{in}}(\vec{r_0})=
\displaystyle\frac{c}{2kn_\text{eff}}\,
\displaystyle\frac{N_1}{2\pi D}\,
\displaystyle\frac{\ell\,z_0}{[x_0^2+y_0^2+(z_0+z_e)^2]^{3/2}}.
\label{EqIinRI3D}
\end{equation}
where the normalization factor $c/2kn_\text{eff}$ remains the same as in 2D. Subsequently, we get the final dimensionless expression for 3D sensitivity map as
\begin{widetext}
\begin{equation}
\overline{{\cal s}}_\text{diff}^\text{3D}(\vec{r_0})= 
-\displaystyle\frac{9}{16\pi k^4}
\displaystyle\frac{z_0^2}{\left[x_0^2+y_0^2+(z_0+z_e)^2\right]^{3/2}\left[x_0^2+(y_0-d)^2+(z_0+z_e)^2\right]^{3/2}}.
\label{EqBornDiff3D}
\end{equation}
\end{widetext}

\noindent Similar to Eq.~(\ref{EqBornDiff3}), Eq.~(\ref{EqBornDiff3D}) does not significantly depend on the transport mean free path. Parameterizing the position along the banana trajectory of radius $d/2$ with a polar angle $\theta$, we get a simplified expression
\begin{equation}
\overline{{\cal s}}_\text{diff}^\text{3D}\simeq
-\displaystyle\frac{9}{8\pi}\times\frac{1}{k^4 d^4}\times\frac{1}{\cos\theta},
\label{sdiff_dimensionless_3D}
\end{equation}
which now, in 3D, depends on $k$, $d$ and $\theta$. Unlike the 2D expression, Eq.~(\ref{sdiff_dimensionless}), the sensitivity is no longer constant along the banana. The divergence at $\theta\sim\pm\pi/2$ is prevented by the surface-corrections at the level 
\begin{equation}
\overline{{\cal s}}_\text{diff}^\text{3D}\propto
-\displaystyle\frac{1}{k^4 d^3\ell}.
\label{sdiff_dimensionless_3D_BC}
\end{equation}

\section{Comparison between random input and the maximum remission eigenchannel profiles}\label{sec:intensity_profiles}
\begin{figure}
\includegraphics[width=\linewidth]
{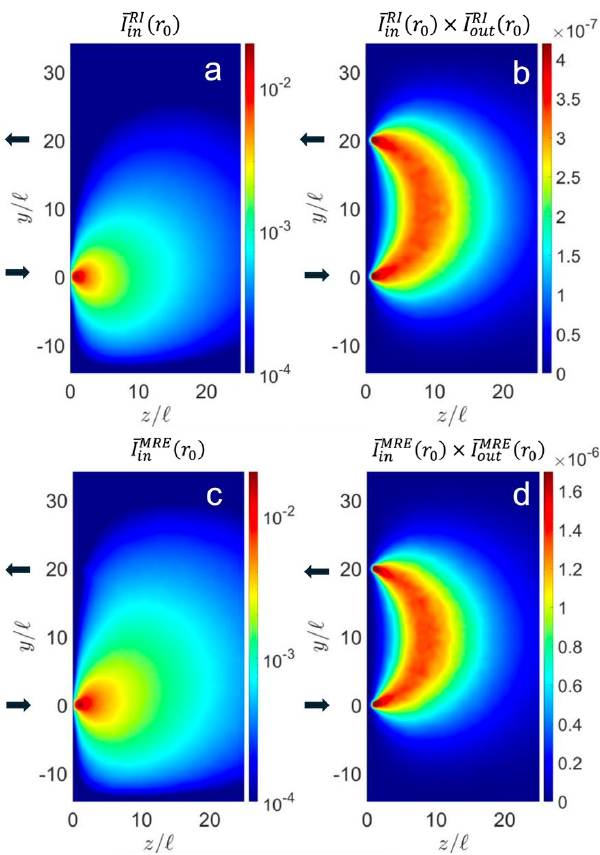}
\vskip -0.3cm
\caption{Comparison between random input and maximum remission eigenchannel (MRE) excitation.
(a)~Simulated input intensity $\overline{I}^{\text{RI}}_{\text{in}}(\vec{r_0})$ for random input in a system with $\ell\!=$6.4 $\mu$m, $d\!=\!20\,\ell$ and $W_1\! =\! W_2\!=\!5\,\mu$m.   
(b)~Product $\overline{I}^{\text{RI}}_{\text{in}}(\vec{r_0})\times\overline{I}^{\text{RI}}_{\text{out}}(\vec{r_0})$, representing the normalized sensitivity map for random input, as given by Eq.~(\ref{EqDiffusiveSensitivity}) under the diffusion approximation (DA). 
(c)~Input intensity $\overline{I}^{\text{MRE}}_{\text{in}}(\vec{r_0})$ corresponding to the maximal remission eigenchannel.
(d)~Product $\overline{I}^{\text{MRE}}_{\text{in}}(\vec{r_0})\times\overline{I}^{\text{MRE}}_{\text{out}}(\vec{r_0})$, constituting the DA$^*$ heuristic for the sensitivity map under MRE excitation, as discussed in Sec.~\ref{sec:shaped_wavefront_excitation}.
\label{fig:intensity_profiles}}
\end{figure}

\noindent Figure~\ref{fig:intensity_profiles} highlights a striking and unresolved feature in the behavior of optical sensitivity under coherent excitation.
According to the diffusion approximation (DA) in Sec.~\ref{sec:sensitivity_diffusion_approximation}, the sensitivity map $\bar{\cal s}(\vec{r}_0)$ is predicted to be proportional to the product of the input and output intensity profiles at position $\vec{r}_0$: 
$\bar{\cal s}_{\text{DA}}(\vec{r}_0) \propto \bar{I}_{\text{in}}(\vec{r}_0)\times\bar{I}_{\text{out}}(\vec{r}_0)$, see Eq.~(\ref{EqDiffusiveSensitivity}). 
This is illustrated in Figs.~\ref{fig:intensity_profiles}ab for the random input (RI) case, where the resulting sensitivity map exhibits the expected `banana'-shaped profile, consistent with established DOT theory. 

When the system is excited with the maximum remission eigenchannel (MRE), a coherent input state with a significantly different internal intensity distribution shown in Fig.~\ref{fig:intensity_profiles}c. 
Surprisingly, however, the product $\bar{I}^\text{MRE}_{\text{in}}(\vec{r}_0)\times \bar{I}^\text{MRE}_{\text{out}}(\vec{r}_0)$ still results in a  map with a similar shape, as shown in Fig.~\ref{fig:intensity_profiles}d. 
This result was discussed in Sec.~\ref{sec:shaped_wavefront_excitation} and referred to as a heuristic DA$^*$.
Resemblance of profiles in Figs.~\ref{fig:intensity_profiles}bd is unexpected, but it holds even quantitatively, see Sec.~\ref{sec:shaped_wavefront_excitation}. 
From the analysis in Sec.~\ref{sec:sensitivity_diffusion_approximation}, the DA prediction is not expected to be applicable for coherent input states such as MREs, since DA assumes statistical independence of input/output fields -- a condition clearly violated by the structured nature of MRE excitation. 
Nonetheless, Fig.~\ref{fig:intensity_profiles} suggests that the {\it spatial structure} of the sensitivity map remains robust, even when the underlying field intensities deviate significantly from those predicted by diffusion.

The origin of this robustness is not presently understood. The fact that the product $\bar{I}^\text{MRE}_{\text{in}}(\vec{r}_0) \times \bar{I}^\text{MRE}_{\text{out}}(\vec{r}_0)$ retains a similar shape despite strong modification of each factor individually implies the presence of subtle constraints in the underlying wave transport that are not captured by existing theory. This observation highlights an important open question: why does the DA-inspired structure of the sensitivity map persist even for coherent input states far outside the DA regime? Resolving this will require deeper theoretical insight into why the the banana-like profile of sensitivity is preserved even under coherent illumination. One promising direction  could be sum rules~\cite{2022_Yamilov_Sum_rules}, which are known to produce non-trivial constraints in mesoscopic wave transport.

\section{Effect of fluctuations on sensitivity\label{sec:fluctuations}}

\begin{figure*}
  \centering
  \includegraphics[width=7in]{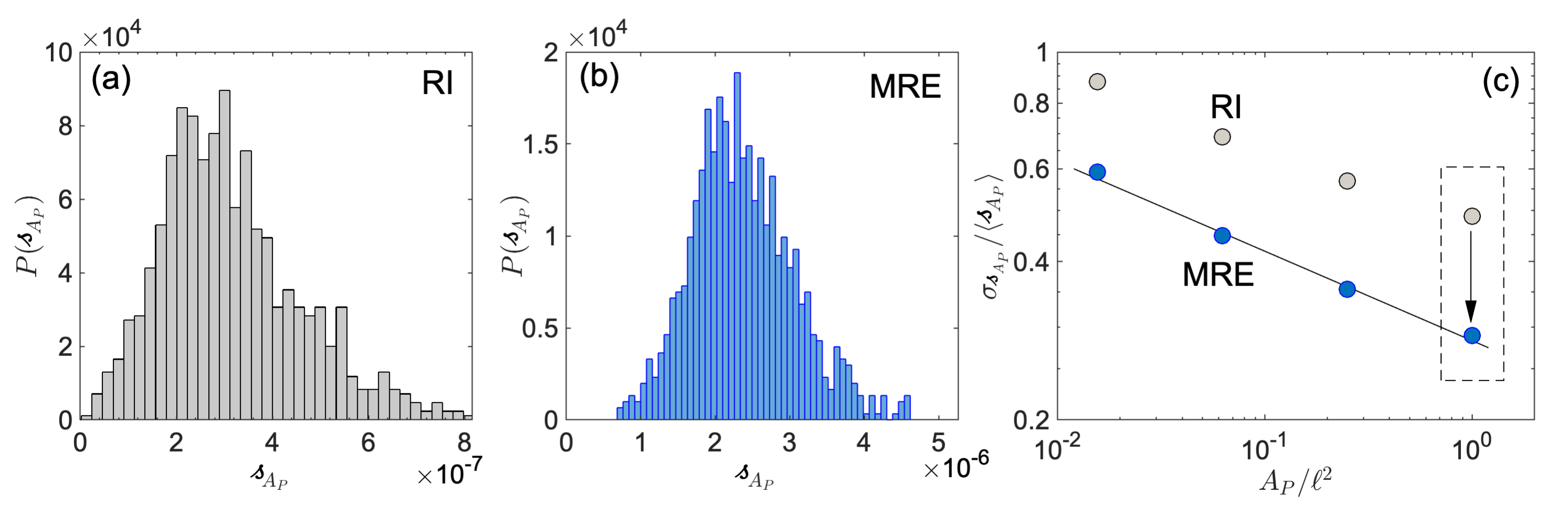}
  \vskip -0.5cm
  \caption{
  (a) Distribution of ${\cal S}_{A_P}$ sensitivity averaged over a perturbation area $A_{P}=\ell \times \ell$ under random input (RI) excitation. $A_{P}$ is centered in the middle of the banana region. The histogram is compiled from 1000 disorder realizations.  
  (b) Same as in (a) but for maximum remission eigenchannel (MRE) excitation. 
  (c) Log–log plot of relative fluctuations (standard deviation normalized by the mean) computed from $P({\cal S}_{A_P})$ versus perturbation area $A_P=\{1/64,1/16,1/4,1\}\times\ell^2$. Both RI and MRE exhibit similar scaling $\sigma_{{\cal S}_{A_P}}/\langle {\cal S}_{A_P} \rangle \propto 1/A_P^\alpha$ with $\alpha\simeq 0.17$ (solid line). This confirms that fluctuations are suppressed by volume averaging in the $A_P\rightarrow\infty$ limit. Importantly, the fluctuations for MRE excitation are significantly suppressed compared to those for random input.
  \label{fig:fluctuation}}
\end{figure*}

\noindent The results presented in the main text correspond to disorder-averaged quantities. In practice, however, both extrinsic (sample-to-sample) and intrinsic (point-to-point within the same sample) fluctuations play an important role~\cite{2018_Fang} in the interpretation of experimental measurements of sensitivity. Here we quantify these fluctuations and analyze their relevance for applications such as diffuse optical imaging.

For each disorder realization, we computed the sensitivity averaged over a perturbation area $A_P$ centered in the middle of the banana region. Treating this volume-averaged sensitivity as a single data point, we compiled distributions from 1000 disorder realizations. The results are shown in Fig.~\ref{fig:fluctuation}. Panel (a) displays the distribution of ${\cal s}_{A_P}$ for random input (RI), which is broad with a finite variance. Panel (b) shows the distribution for the maximum remission eigenchannel (MRE) excitation: the mean sensitivity is increased, while the relative fluctuations are smaller. Thus, MRE achieves both {\it enhanced mean sensitivity and reduced fluctuations} compared to RI.

Panel (c) quantifies the relative fluctuations, $\sigma_{{\cal s}_{A_P}}/{\bar {\cal s}}_{A_P}$, as a function of perturbation area. Both RI and MRE show a decrease of fluctuations with increasing $A_P$, but with a weaker scaling than the ideal $1/\sqrt{A_P}$ dependence expected for uncorrelated random contributions. Importantly, across all $A_P$, the fluctuations of sensitivity under MRE excitation are consistently smaller than those for RI. This robustness resembles the behavior for the intensity of high-transmission eigenchannels in the waveguide geometry, which also exhibit suppressed realization-to-realization fluctuations compared to random inputs~\cite{2020_Bender_Eigenchannels}. 

The above results demonstrate that remission eigenchannels are not only advantageous in terms of mean sensitivity, but also in terms of statistical stability. Their reduced fluctuations make them particularly promising for diffuse optical imaging applications, where this property should result in an increase in robustness of the inverse-problem algorithms reconstructing the locations of the perturbations from remission measurements.\\

\section{Filtered random matrix theory  for predicting coherent remission enhancement}\label{sec:filtered_random_matrix_theory}
\noindent The detailed description of the analytical procedure used for computing remission enhancement by applying filtered random matrix theory can be found in Ref.~\cite{2022_Bender_Coherent_Enhancement}. However, for the sake of completeness we will sum up here the overall scheme of the method. In Fig.~\ref{fig:panel_enhan_factor} in Sec.~\ref{sec:remission_enhancement} we included solid lines which represent the analytical solutions when computing sensitivity as a function of $\ell$ and the geometrical parameters $d$, $N_{1}$ and $N_{2}$.

As is well known, the scattering matrix fully encodes  multiple scattering within the medium. It is a powerful tool that relates arbitrary input fields to their corresponding outputs, and in principle, allows for the reconstruction or prediction of either, see Fig.~\ref{fig:geometry_sketch_generic} in Sec.~\ref{sec:field_sensitivity}. Thus, to compute the theoretical enhancement, we rely on full knowledge of the scattering matrix. Specifically, access to the scattering matrix provides information about the remission matrix $\cal R$, from which we can compute the eigenvalue probability density function of ${\cal R}^\dagger{\cal R}$ and its statistical moments. In our setup, light propagates through an open geometry, where input and output channels cover only a small fraction of the total surface area. Therefore, we must also account for incomplete channel control in both injection and detection~\cite{2022_Bender_Coherent_Enhancement}.
A novel method based on random matrix theory called `Filtered Random Matrix Theory'  (FRM), see Ref.~\cite{2013_Goetschy_FRM}, allow us to model the remission matrix ${\cal R}$ as a filtered matrix of dimension $N_{2}\times N_{1}$, drawn from a virtual $M_{0}\times M_{0}$ matrix ${\cal R}_{0}$ characterized by a bimodal distribution of remission eigenvalues with a specific mean value. This model captures the idea that only a fraction of the scattering channels are effectively excited and detected at the injection and remission sites~\cite{2022_Bender_Coherent_Enhancement}. Both $M_{0}$ and the mean $\overline \rho_{0}$ of the bimodal eigenvalue PDF of ${\cal R}^\dagger_{0}{\cal R}_{0}$ are effective parameters of the model, which are unknown a priori. A theoretical
calculation of the flux and its fluctuations detected in the output port will be used to express
$M_{0}$ and $\overline \rho_{0}$ in terms of the parameters involved in the remission experiment: the mean free path, the injection-remission distance $d$, and the numbers of input and output spatial channels $N_{1}$ and $N_{2}$.

According to Refs.~\cite{2022_Bender_Coherent_Enhancement} and \cite{2013_Goetschy_FRM}, the FRM theory allows one to establish the mathematical relation between the PDFs of ${\cal R}^\dagger{\cal R}$ and ${\cal R}^\dagger_{0}{\cal R}_{0}$. Below we give explicit equations for the case $N_1\ge N_2$, as in Ref.~\cite{2022_Bender_Coherent_Enhancement}. The case $N_1\le N_2$ can be obtained by inverting the role of $N_1$ and $N_2$ in the equations for the eigenvalues $\rho$. We note, however, that the prediction for the largest eigenvalue $\rho_\text{max}$ remains the same in both cases. For $N_1\ge N_2$ the PDF of the non-zero eigenvalues of ${\cal R}^\dagger{\cal R}$ is identical to the PDF $P(\rho)$ of the eigenvalues of ${\cal R}{\cal R}^\dagger$. It is given by $P(\rho)=-\lim_{\eta\to 0^{+}} \Im [g(\rho+i\eta)]/\pi$, where $g(w)$ is the solution to the following implicit equation~\cite{2013_Goetschy_FRM}:
\begin{widetext}
\begin{equation}
 \frac{w m_{2}g(w)+1-m_{2}}{m_{2}g(w)[w m_{2}g(w)+m_{1}-m_{2}]}g\bigg(\frac{[w m_{2}g(w)+1-m_{2}]^2}{m_{2}g(w)[w m_{2}g(w)+m_{1}-m_{2}]}\bigg)=1,\label{Eq1FRMT}
\end{equation}
\end{widetext}
where $m_{1}=N_{1}/M_{0}$, $m_{2}=N_{2}/M_{0}$ and
\begin{equation}
 g(w)=\frac{1}{w}-\frac{\overline\rho_{0}}{w\sqrt{1-w}}\arctanh\left[ \frac{\tanh(1/\overline\rho_{0})}{\sqrt(1-w)}\right].
\end{equation}

 Based on Eq.~(\ref{Eq1FRMT}), it is possible to express the moments $P(\rho)$ in terms of the eigenvalue PDF of ${\cal R}^\dagger_{0}{\cal R}_{0}$. Thus, for the first two cumulants, we find~\cite{2013_Goetschy_FRM}:
\begin{equation}
\langle\rho\rangle=m_{1}\langle\rho_{0}\rangle=m_{1}\overline\rho_{0},
\end{equation}

\begin{equation}
\Var\left(\frac{\rho}{\langle\rho\rangle}\right)=m_{2}\bigg[ \frac{2}{3\overline\rho_{0}}+\frac{1}{m_{1}}-2\bigg].
\end{equation}

We now solve these equations to express the unknown parameters $m_{1}$, $m_{2}$ and $\overline\rho_{0}$ of Eq.~(\ref{Eq1FRMT}) in terms of the first two moments, $\langle\rho\rangle$ and $\langle\rho^2\rangle$ to get,
\begin{equation}
m_{1}=\sqrt{\frac{3\langle\rho\rangle}{2}\left[ \beta \Var\bigg(\frac{\rho}{\langle\rho\rangle}\bigg)-1\right]+\bigg( \frac{3\langle\rho\rangle}{2}\bigg)^2}+\frac{3\langle\rho\rangle}{2}\label{Eqm1},
\end{equation}
\begin{equation}
m_{2}=\frac{m_{1}}{\beta}\label{Eqm2},
\end{equation}
\begin{equation}
\overline\rho_{0}=\frac{\langle\rho\rangle}{\rho}\label{Eqro},
\end{equation}
where $\beta=N_{1}/N_{2}$. The above expressions together with Eq.~(\ref{Eq1FRMT}) allow us to predict the full PDF $P(\rho)$  from the first two moments. The results of this prediction are shown in Ref.~\cite{2022_Bender_Coherent_Enhancement}. 

The two first moment can also be evaluated analytically. First, we note that $\langle\rho\rangle=(N_{1}/N_{2})\rho_\text{RI}$, where $\rho_\text{RI}$ is the mean diffusive flux measured in the output port, and it is given by the solution to the Fick's law of diffusion,
\begin{equation}
\rho_\text{RI} \approx W_{2} D \partial_{z}I(y=d,z=0),
\end{equation}
 where $I(y,z)$ is the solution to the stationary diffusion equation,
\begin{equation}
-D\nabla^2 I(y,z)=\delta(y)\delta(z-z_{e}).
\end{equation} 
In the previous equations, $W_{2}$ is the width of the output, $D=\ell c/2$ is the diffusion coefficient for light, and $z_{e}$ in extrapolation length defined in Sec.~\ref{sec:diffusion_in_remission_geometry}. An approximate solution is $I(y,z) \approx z\ell/(\pi D y^2)$ in the limit $y \gg z,\ell$. We also know that $N_{2}=k W_{2}/\pi$, we get $\rho_\text{RI} \approx N_{2}k \ell/(k d)^2$ and
\begin{equation}
\langle\rho\rangle \approx N_{1}\frac{k\ell}{(k d)^2}.
\end{equation}

 Next, we need to find the expression for the normalized variance $\Var(\rho/\langle\rho\rangle)$. This variance can be expressed as a function of the full eigenvalues $\tilde{\rho}$ of ${\cal R}^\dagger{\cal R}$ as,
 \begin{equation}
\Var\bigg( \frac{\rho}{\langle\rho\rangle}\bigg)=\frac{N_{2}}{N_{1}}\left[\Var\bigg(\frac{\tilde{\rho}}{\langle\tilde{\rho}\rangle}\bigg)+1\right]-1\label{EqVar1},
\end{equation}
where we have $\tilde{\rho}=\rho_\text{RI}$. To find an expression for $\Var( \tilde{\rho}/{\langle\tilde{\rho}\rangle})$ we relate it to the fluctuations in the intensity $I_{a}$ measured in the output waveguide after exciting the channel $a$ at the input. This is done by applying a singular value decomposition to $\cal R$. By this, we find that $I_{a}=\sum_{n}|V_{an}|^2\tilde{\rho_{n}}$. As a result the fluctuations in $I_{a}$ depend on the fluctuations in $\tilde{\rho}$ and the statistical properties of the matrix $V$.  By assuming that $V$ is randomly distributed in the unitary group, see Ref.~\cite{2022_Bender_Depth_Targeted_Energy_Deposition}, we can show that in the limit $M_{1} \gg 1$,
\begin{equation}
\Var\left( \frac{\tilde{\rho}}{\langle \tilde{\rho}\rangle}\right) \approx N_{1} \Var\bigg( \frac{I_{a}}{\langle I_{a}\rangle}\bigg)\label{EqVar2}.
\end{equation}
As explained in Ref.~\cite{2022_Bender_Coherent_Enhancement}, the average in the unitary group makes the result independent of the channel a. It is thus also equal to the fluctuations of intensity $I = |E|^2$ in the output waveguide resulting from a uniform excitation of all modes in the input waveguide. We evaluate the latter by decomposing the field $E$ as a sum of all possible scattering contributions reaching the output waveguide. Using standard diagrammatic techniques, we find
\begin{equation}
\frac{\Var I}{\langle I\rangle^2}=\frac{1}{N_{2}}+C_{2}\label{EqVar3},
\end{equation}
where $C_{2}$ is the intensity-intensity correlation function given by,
\begin{equation}
C_{2}=\frac{1}{4 k\ell\langle I(d,0)\rangle^2}\int\int dydz\langle I(y,z)\rangle^2[\nabla_{y,z}K(d,0;y,z)]^2\label{EqC2theoretical}.
\end{equation}

It can be shown that in the limit $d\gg W_{1},\ell$, the calculation of $C_{2}$ gives,
\begin{equation}
C_{2} \approx \frac{1}{k \ell}\left[ \frac{4}{\pi}\ln{\bigg(\frac{d}{W_{1}}\bigg)}+\gamma\right]\label{EqC2},
\end{equation}
where numerical simulations show a $\gamma \approx 0.6$. Finally, Eqs.~(\ref{EqVar1},\ref{EqVar2}) and Eq.~(\ref{EqVar3}) together give us,
\begin{equation}
\Var\left[\frac{\rho}{\langle\rho\rangle}\right]=\frac{N_{2}}{N_{1}}+N_{2}C_{2}\label{EqVariance}.
\end{equation}

In order to get to the final step towards the sensitivity enhancement expression, we need to determine the upper edge $\rho_\text{max}$ of the distribution $P(\rho)$. An equation for this quantity can be derived from~\eqref{Eq1FRMT}; see Eqs.~[8] and~[9] in the Supporting Information of Ref.~\cite{2022_Bender_Coherent_Enhancement}. While this equation is exact, it does not yield an explicit analytical expression of $\rho_\text{max}$ in terms of $N_1$, $N_2$, and $C_2$. 
However, an approximate explicit solution can be obtained under the conditions $N_1, N_2 \gg 1$ and $d \gg W_1$, which are satisfied for most of our simulations. The upper edge of $P(\rho)$, as derived in Ref.~\cite{2022_Bender_Depth_Targeted_Energy_Deposition}, takes the form,
\begin{widetext}
\begin{equation}
\frac{\rho_\text{max}}{\langle\rho\rangle} \approx \frac{[(\alpha -1)^{2/3}+(\pi/2)^{2/3}]^2[\alpha - 1+(\pi/2)^{2/3}(\alpha - 1)^{1/3}]}{\alpha(\alpha - 1)^{1/3}}+{\cal{O}}(m)\label{Eqalpha},
\end{equation}
\end{widetext}
where $\alpha =m/\overline\rho_{0}$. By Eq.~(\ref{Eqm1}) and (\ref{Eqro}) we get,
\begin{equation}
\alpha=\frac{m^2}{\langle\rho\rangle}\approx \frac{3}{2}\left[\Var\bigg(\frac{\rho}{\langle\rho\rangle}\bigg)-1 \right]=\frac{3}{2}\bigg(N_{2}C_{2}+\frac{N_{2}}{N_{1}}-1\bigg).\label{EqC_2_enhance}
\end{equation}
Because of the values of the parameters in our simulations and experiment, we can consider the limit $\alpha \gg 1$ in Eq.~(\ref{Eqalpha}),
\begin{equation}
\frac{\rho_\text{max}}{\langle\rho\rangle} \approx \alpha +3\bigg(\frac{\pi}{2}\bigg)^{2/3}\alpha^{1/3}-2+{\cal O}(\alpha^{-1/3}).
\end{equation}
This expression allows to compute the enhancement in the remitted signal, $\rho_\text{max}/ \rho_\text{RI}=(N_1/N_2)\rho_\text{max}/\langle\rho\rangle$. We can even make an additional assumption to capture the behavior in the limit $N_{2}C_{2} \gg 1$ to obtain,
\begin{equation}
\frac{\rho_\text{max}}{\rho_\text{RI}} \approx \frac{3 W_{1}}{2\pi\ell}\left[ \frac{4}{\pi}\ln{}\bigg( \frac{d}{W_{1}}\bigg)+\gamma\right].\label{Eqaproxxenhan}
\end{equation}
This last expression shows explicitly the strong dependence of the enhancement on the number of input channels and the transport mean free path, as well as a logarithmic dependence of the distance $d$. However, we remark that the complete dependence on these parameters and the number of output channels is captured by Eq.~(\ref{Eq1FRMT}).

\section{Effective Marchenko-Pastur model}\label{sec:Marcenko_Pastur}

\begin{table}
\centering
\caption{Comparison of remission enhancement $\eta_{\cal{R}}$ obtained from numerical simulations with analytical predictions based on the effective Marchenko-Pastur model and the idealized Marchenko-Pastur model (i.e. $C_2=0$), for various input and output port widths. All results correspond to a system with transport mean free path $\ell=6.4\,\mu$m and source-detector separation $d=128\,\mu$m.}
\begin{tabular}{|c|c|c|c|c|c|c|c|c|}
\hline
$N_1$ & $N_2$ & $N1 /N2$ & $C_2$ & $1 / C_2$ & $\eta_{\cal{R}}$ & $\eta_{\cal{S}}$ & \normalfont Effective MP &MP \\ \hline
3     & 57 & 0.05 & 0.08 & 12.7 & 1.6 & 1.7 & 2.4 & 1.5 \\ \hline
3  & 38 & 0.08 & 0.08 & 12.7 & 1.7 & 1.7 & 2.4 & 1.6 \\ \hline
3  & 19 & 0.16 & 0.08 & 12.7 & 1.7 & 1.8 & 2.6 & 2.0 \\ \hline
19 & 38 & 0.50 & 0.05 & 19.4 & 4.3 & 4.6 & 4.9 & 2.9 \\ \hline
38 & 76 & 0.50 & 0.05 & 21.7 & 6.1 & 6.5 & 6.2 & 2.9 \\ \hline
38 & 57 & 0.67 & 0.05 & 21.7 & 6.4 & 6.9 & 6.5 & 3.3 \\ \hline
19 & 19 & 1.00 & 0.05 & 19.4 & 5.1 & 5.5 & 5.8 & 4.0 \\ \hline
38 & 38 & 1.00 & 0.05 & 21.7 & 7.0 & 7.4 & 7.1 & 4.0 \\ \hline
57 & 57 & 1.00 & 0.05 & 21.8 & 8.7 & 9.4 & 8.4 & 4.0 \\ \hline
76 & 76 & 1.00 & 0.05 & 20.8 & 10.1 & 11.2 & 10.0 & 4.0 \\ \hline
57 & 38 & 1.50 & 0.05 & 21.8 & 9.6 & 10.2 & 9.2 & 4.9 \\ \hline
38 & 19 & 2.00 & 0.05 & 21.7 & 8.6 & 9.3 & 8.6 & 5.8 \\ \hline
57 & 19 & 3.00 & 0.05 & 21.8 & 12.1 & 12.8 & 11.3 & 7.5 \\ \hline
19 & 3  & 6.33 & 0.05 & 19.4 & 11.3 & 12.0 & 13.7 & 12.4 \\ \hline
38 & 3  & 12.67 & 0.05 & 21.7 & 21.7 & 22.0 & 23.0 & 20.8 \\ \hline
57 & 3  & 19.00 & 0.05 & 21.8 & 31.8 & 32.2 & 31.9 & 28.7 \\ \hline
\end{tabular}
\label{tab:table_effective_MP_model}
\end{table}

As explained in the previous section, while the FRM model is accurate, it does not provide an explicit analytical expression for the remission enhancement for arbitrary aspect ratio $N_1/N_2$. An alternative model for the remission enhancement is the effective Marchenko-Pastur model, which is less accurate but has the advantages of being very simple, intuitive, and providing an explicit result in terms of $N_1$, $N_2$, and $C_2$. This model was first introduced in Ref.~\cite{Hsu2017} as a simple alternative to the FRM approach in transmission. It was later shown to yield good results for maximizing energy deposition with broadband light~\cite{McIntosh2024}.
In the present context of remission, this model amounts to assuming that the non-Gaussian remission matrix $\mathcal{R}$ of size $N_2 \times N_1$ can be replaced by an effective Gaussian matrix of size $N_\text{eff} \times N_1$, where $N_\text{eff}$ is the number of effectively independent channels at the output. This number is smaller than $N_2$ due to long-range correlations at the output, and is given by $1/N_\text{eff} = 1/N_2 + C_2$~\cite{Hsu2017,McIntosh2024}.
In this model, the matrix $\mathcal{R}^\dagger \mathcal{R}$ behaves as a Wishart matrix of aspect ratio $N_1/N_\text{eff}$, whose eigenvalue distribution is given by the Marchenko-Pastur law, parameterized solely by $N_1/N_\text{eff}$. In particular, in the limit of large matrix size, the upper edge of the distribution, normalized by the mean eigenvalue, yields
\begin{equation}
\frac{\rho_\text{max}}{\rho_\text{RI}} = \left(1 + \sqrt{\frac{N_1}{N_\text{eff}}}\right)^2 = \left(1 + \sqrt{\frac{N_1}{N_2} + N_1 C_2}\right)^2.\label{EqeffMP}
\end{equation}

To assess the accuracy of the effective Marchenko–Pastur (MP) model in predicting remission enhancement, we consider various combinations of $N_1$, $1/C_2$ and  $N_2$, looking for systematic deviations between remission/sensitivity enhancements obtained numerically and theoretical predictions. In simulations, we consider the system with the source-detector separation to $d = 20\,\ell$ with $\ell = 6.4~\mu$m, and vary the number of degrees of freedom at the input and output to scan the relevant interval of the MP parameter $C_2$, see Table~\ref{tab:table_effective_MP_model}. The value of $C_2$ is computed using Eq.~(\ref{EqC2}), originally derived in Ref.~\cite{2022_Bender_Coherent_Enhancement}, which provides an analytical expression valid in the limit $d \gg W_1, L$.

As shown, the sensitivity enhancement $\eta_{\cal S}$ consistently exceeds the remission enhancement $\eta_{\cal R}$, yet their magnitudes remain close to the effective MP model. 
This is in agreement with the prediction from Sec.~\ref{sec:remission_enhancement} that remission enhancement serves as a proxy for global sensitivity enhancement. 
Moreover, the data corroborate the hypothesis that MP model with $C_2 = 0$ indeed sets a lower bound on enhancement: any increase beyond this baseline stems from the long-range correlations, whose effect is captured by the $C_2$ contribution. 
The observed agreement with theory, despite various finite-size effects, highlights the robustness of the effective MP model in predicting remission enhancement in diffusive media. We also note that the disagreement observed for small $N_1$ is due to the limitation of the Marchenko-Pastur law, which strictly applies only to matrices of large dimensions.


\begin{thebibliography}{74}%
\makeatletter
\providecommand \@ifxundefined [1]{%
 \@ifx{#1\undefined}
}%
\providecommand \@ifnum [1]{%
 \ifnum #1\expandafter \@firstoftwo
 \else \expandafter \@secondoftwo
 \fi
}%
\providecommand \@ifx [1]{%
 \ifx #1\expandafter \@firstoftwo
 \else \expandafter \@secondoftwo
 \fi
}%
\providecommand \natexlab [1]{#1}%
\providecommand \enquote  [1]{``#1''}%
\providecommand \bibnamefont  [1]{#1}%
\providecommand \bibfnamefont [1]{#1}%
\providecommand \citenamefont [1]{#1}%
\providecommand \href@noop [0]{\@secondoftwo}%
\providecommand \href [0]{\begingroup \@sanitize@url \@href}%
\providecommand \@href[1]{\@@startlink{#1}\@@href}%
\providecommand \@@href[1]{\endgroup#1\@@endlink}%
\providecommand \@sanitize@url [0]{\catcode `\\12\catcode `\$12\catcode `\&12\catcode `\#12\catcode `\^12\catcode `\_12\catcode `\%12\relax}%
\providecommand \@@startlink[1]{}%
\providecommand \@@endlink[0]{}%
\providecommand \url  [0]{\begingroup\@sanitize@url \@url }%
\providecommand \@url [1]{\endgroup\@href {#1}{\urlprefix }}%
\providecommand \urlprefix  [0]{URL }%
\providecommand \Eprint [0]{\href }%
\providecommand \doibase [0]{https://doi.org/}%
\providecommand \selectlanguage [0]{\@gobble}%
\providecommand \bibinfo  [0]{\@secondoftwo}%
\providecommand \bibfield  [0]{\@secondoftwo}%
\providecommand \translation [1]{[#1]}%
\providecommand \BibitemOpen [0]{}%
\providecommand \bibitemStop [0]{}%
\providecommand \bibitemNoStop [0]{.\EOS\space}%
\providecommand \EOS [0]{\spacefactor3000\relax}%
\providecommand \BibitemShut  [1]{\csname bibitem#1\endcsname}%
\let\auto@bib@innerbib\@empty
\bibitem [{\citenamefont {Ishimaru}(1978)}]{1978_Ishimaru}%
  \BibitemOpen
  \bibfield  {author} {\bibinfo {author} {\bibfnamefont {A.}~\bibnamefont {Ishimaru}},\ }\href@noop {} {\emph {\bibinfo {title} {Wave Propagation and Scattering in Random Media}}},\ Vol.\ \bibinfo {volume} {1-2}\ (\bibinfo  {publisher} {Academic Press},\ \bibinfo {address} {New York},\ \bibinfo {year} {1978})\BibitemShut {NoStop}%
\bibitem [{\citenamefont {Aki}\ and\ \citenamefont {Richards}(1980)}]{1980_Aki}%
  \BibitemOpen
  \bibfield  {author} {\bibinfo {author} {\bibfnamefont {K.}~\bibnamefont {Aki}}\ and\ \bibinfo {author} {\bibfnamefont {P.~G.}\ \bibnamefont {Richards}},\ }\href@noop {} {\emph {\bibinfo {title} {Quantitative Seismology: Theory and Methods}}}\ (\bibinfo  {publisher} {W.H. Freeman and Company},\ \bibinfo {address} {San Francisco},\ \bibinfo {year} {1980})\BibitemShut {NoStop}%
\bibitem [{\citenamefont {Pike}\ and\ \citenamefont {Sabatier}(2002)}]{2002_Pike_Sabatier}%
  \BibitemOpen
  \bibinfo {editor} {\bibfnamefont {E.~R.}\ \bibnamefont {Pike}}\ and\ \bibinfo {editor} {\bibfnamefont {P.~C.}\ \bibnamefont {Sabatier}},\ eds.,\ \href@noop {} {\emph {\bibinfo {title} {Scattering: Scattering and Inverse Scattering in Pure and Applied Science}}}\ (\bibinfo  {publisher} {Academic Press},\ \bibinfo {address} {San Diego},\ \bibinfo {year} {2002})\BibitemShut {NoStop}%
\bibitem [{\citenamefont {Chew}(1995)}]{1995_Chew_Scattering_book}%
  \BibitemOpen
  \bibfield  {author} {\bibinfo {author} {\bibfnamefont {W.~C.}\ \bibnamefont {Chew}},\ }\href@noop {} {\emph {\bibinfo {title} {Waves and Fields in Inhomogeneous Media}}}\ (\bibinfo  {publisher} {IEEE Press},\ \bibinfo {address} {New York},\ \bibinfo {year} {1995})\ \bibinfo {note} {reprinted by Wiley-IEEE Press, 1999}\BibitemShut {NoStop}%
\bibitem [{\citenamefont {Bertolotti}\ and\ \citenamefont {Katz}(2022)}]{2022_Bertolotti_Imaging_review}%
  \BibitemOpen
  \bibfield  {author} {\bibinfo {author} {\bibfnamefont {J.}~\bibnamefont {Bertolotti}}\ and\ \bibinfo {author} {\bibfnamefont {O.}~\bibnamefont {Katz}},\ }\bibfield  {title} {\bibinfo {title} {Imaging in complex media},\ }\href {https://doi.org/10.1038/s41567-022-01723-8} {\bibfield  {journal} {\bibinfo  {journal} {Nature Physics}\ }\textbf {\bibinfo {volume} {18}},\ \bibinfo {pages} {1008} (\bibinfo {year} {2022})}\BibitemShut {NoStop}%
\bibitem [{\citenamefont {Wang}\ and\ \citenamefont {Wu}(2007)}]{2007_Wang}%
  \BibitemOpen
  \bibfield  {author} {\bibinfo {author} {\bibfnamefont {L.}~\bibnamefont {Wang}}\ and\ \bibinfo {author} {\bibfnamefont {H.-i.}\ \bibnamefont {Wu}},\ }\href@noop {} {\emph {\bibinfo {title} {Biomedical optics: principles and imaging}}}\ (\bibinfo  {publisher} {John Wiley \& Sons},\ \bibinfo {address} {Hoboken, NJ},\ \bibinfo {year} {2007})\BibitemShut {NoStop}%
\bibitem [{\citenamefont {Bouma}\ and\ \citenamefont {Tearney}(2001)}]{2001_Bouma_OCT}%
  \BibitemOpen
  \bibfield  {author} {\bibinfo {author} {\bibfnamefont {B.~E.}\ \bibnamefont {Bouma}}\ and\ \bibinfo {author} {\bibfnamefont {G.~J.}\ \bibnamefont {Tearney}},\ }\href@noop {} {\emph {\bibinfo {title} {Handbook of optical coherence tomography}}}\ (\bibinfo  {publisher} {CRC Press},\ \bibinfo {address} {Boca Raton},\ \bibinfo {year} {2001})\BibitemShut {NoStop}%
\bibitem [{\citenamefont {Yodh}\ and\ \citenamefont {Chance}(1995)}]{1995_Yodh_DOT}%
  \BibitemOpen
  \bibfield  {author} {\bibinfo {author} {\bibfnamefont {A.}~\bibnamefont {Yodh}}\ and\ \bibinfo {author} {\bibfnamefont {B.}~\bibnamefont {Chance}},\ }\bibfield  {title} {\bibinfo {title} {Spectroscopy and imaging with diffusing light},\ }\href {https://doi.org/10.1063/1.881445} {\bibfield  {journal} {\bibinfo  {journal} {Physics Today}\ }\textbf {\bibinfo {volume} {48}},\ \bibinfo {pages} {34} (\bibinfo {year} {1995})}\BibitemShut {NoStop}%
\bibitem [{\citenamefont {Boas}\ \emph {et~al.}(2001)\citenamefont {Boas}, \citenamefont {Brooks}, \citenamefont {Miller}, \citenamefont {DiMarzio}, \citenamefont {Kilmer}, \citenamefont {Gaudette},\ and\ \citenamefont {Zhang}}]{2001_Boas_DOT}%
  \BibitemOpen
  \bibfield  {author} {\bibinfo {author} {\bibfnamefont {D.}~\bibnamefont {Boas}}, \bibinfo {author} {\bibfnamefont {D.}~\bibnamefont {Brooks}}, \bibinfo {author} {\bibfnamefont {E.}~\bibnamefont {Miller}}, \bibinfo {author} {\bibfnamefont {C.}~\bibnamefont {DiMarzio}}, \bibinfo {author} {\bibfnamefont {M.}~\bibnamefont {Kilmer}}, \bibinfo {author} {\bibfnamefont {R.}~\bibnamefont {Gaudette}},\ and\ \bibinfo {author} {\bibfnamefont {Q.}~\bibnamefont {Zhang}},\ }\bibfield  {title} {\bibinfo {title} {Imaging the body with diffuse optical tomography},\ }\href {https://doi.org/10.1109/79.962278} {\bibfield  {journal} {\bibinfo  {journal} {IEEE Signal Processing Magazine}\ }\textbf {\bibinfo {volume} {18}},\ \bibinfo {pages} {57} (\bibinfo {year} {2001})}\BibitemShut {NoStop}%
\bibitem [{\citenamefont {Durduran}\ \emph {et~al.}(2010)\citenamefont {Durduran}, \citenamefont {Choe}, \citenamefont {Baker},\ and\ \citenamefont {Yodh}}]{2010_Durduran_DOT}%
  \BibitemOpen
  \bibfield  {author} {\bibinfo {author} {\bibfnamefont {T.}~\bibnamefont {Durduran}}, \bibinfo {author} {\bibfnamefont {R.}~\bibnamefont {Choe}}, \bibinfo {author} {\bibfnamefont {W.~B.}\ \bibnamefont {Baker}},\ and\ \bibinfo {author} {\bibfnamefont {A.~G.}\ \bibnamefont {Yodh}},\ }\bibfield  {title} {\bibinfo {title} {Diffuse optics for tissue monitoring and tomography},\ }\href {https://doi.org/10.1088/0034-4885/73/7/076701} {\bibfield  {journal} {\bibinfo  {journal} {Reports on Progress in Physics}\ }\textbf {\bibinfo {volume} {73}},\ \bibinfo {pages} {076701} (\bibinfo {year} {2010})}\BibitemShut {NoStop}%
\bibitem [{\citenamefont {Scholkmann}\ \emph {et~al.}(2014)\citenamefont {Scholkmann}, \citenamefont {Kleiser}, \citenamefont {Metz}, \citenamefont {Zimmermann}, \citenamefont {{Mata Pavia}}, \citenamefont {Wolf},\ and\ \citenamefont {Wolf}}]{2014_Scholkmann_fNIRS}%
  \BibitemOpen
  \bibfield  {author} {\bibinfo {author} {\bibfnamefont {F.}~\bibnamefont {Scholkmann}}, \bibinfo {author} {\bibfnamefont {S.}~\bibnamefont {Kleiser}}, \bibinfo {author} {\bibfnamefont {A.~J.}\ \bibnamefont {Metz}}, \bibinfo {author} {\bibfnamefont {R.}~\bibnamefont {Zimmermann}}, \bibinfo {author} {\bibfnamefont {J.}~\bibnamefont {{Mata Pavia}}}, \bibinfo {author} {\bibfnamefont {U.}~\bibnamefont {Wolf}},\ and\ \bibinfo {author} {\bibfnamefont {M.}~\bibnamefont {Wolf}},\ }\bibfield  {title} {\bibinfo {title} {A review on continuous wave functional near-infrared spectroscopy and imaging instrumentation and methodology},\ }\href {https://doi.org/https://doi.org/10.1016/j.neuroimage.2013.05.004} {\bibfield  {journal} {\bibinfo  {journal} {NeuroImage}\ }\textbf {\bibinfo {volume} {85}},\ \bibinfo {pages} {6} (\bibinfo {year} {2014})}\BibitemShut {NoStop}%
\bibitem [{\citenamefont {Vidal-Rosas}\ \emph {et~al.}(2023)\citenamefont {Vidal-Rosas}, \citenamefont {von L{\"u}hmann}, \citenamefont {Pinti},\ and\ \citenamefont {Cooper}}]{2023_Vidal_fNIRS_DOT_perspectives}%
  \BibitemOpen
  \bibfield  {author} {\bibinfo {author} {\bibfnamefont {E.~E.}\ \bibnamefont {Vidal-Rosas}}, \bibinfo {author} {\bibfnamefont {A.}~\bibnamefont {von L{\"u}hmann}}, \bibinfo {author} {\bibfnamefont {P.}~\bibnamefont {Pinti}},\ and\ \bibinfo {author} {\bibfnamefont {R.~J.}\ \bibnamefont {Cooper}},\ }\bibfield  {title} {\bibinfo {title} {{Wearable, high-density fNIRS and diffuse optical tomography technologies: a perspective}},\ }\href {https://doi.org/10.1117/1.NPh.10.2.023513} {\bibfield  {journal} {\bibinfo  {journal} {Neurophotonics}\ }\textbf {\bibinfo {volume} {10}},\ \bibinfo {pages} {023513} (\bibinfo {year} {2023})}\BibitemShut {NoStop}%
\bibitem [{\citenamefont {J{\"o}bsis}(1977)}]{1977_Jobsis}%
  \BibitemOpen
  \bibfield  {author} {\bibinfo {author} {\bibfnamefont {F.~F.}\ \bibnamefont {J{\"o}bsis}},\ }\bibfield  {title} {\bibinfo {title} {Noninvasive, infrared monitoring of cerebral and myocardial oxygen sufficiency and circulatory parameters},\ }\href {https://doi.org/10.1126/science.929199} {\bibfield  {journal} {\bibinfo  {journal} {Science}\ }\textbf {\bibinfo {volume} {198}},\ \bibinfo {pages} {1264} (\bibinfo {year} {1977})}\BibitemShut {NoStop}%
\bibitem [{\citenamefont {Arridge}(1999)}]{1999_Arridge_DOT_inverse_problem}%
  \BibitemOpen
  \bibfield  {author} {\bibinfo {author} {\bibfnamefont {S.~R.}\ \bibnamefont {Arridge}},\ }\bibfield  {title} {\bibinfo {title} {Optical tomography in medical imaging},\ }\href {https://doi.org/10.1088/0266-5611/15/2/022} {\bibfield  {journal} {\bibinfo  {journal} {Inverse Problems}\ }\textbf {\bibinfo {volume} {15}},\ \bibinfo {pages} {R41} (\bibinfo {year} {1999})}\BibitemShut {NoStop}%
\bibitem [{\citenamefont {Arridge}\ and\ \citenamefont {Schotland}(2009)}]{2009_Arridge_Schotland_DOT_inverse_problem}%
  \BibitemOpen
  \bibfield  {author} {\bibinfo {author} {\bibfnamefont {S.~R.}\ \bibnamefont {Arridge}}\ and\ \bibinfo {author} {\bibfnamefont {J.~C.}\ \bibnamefont {Schotland}},\ }\bibfield  {title} {\bibinfo {title} {Optical tomography: forward and inverse problems},\ }\href {https://doi.org/10.1088/0266-5611/25/12/123010} {\bibfield  {journal} {\bibinfo  {journal} {Inverse Problems}\ }\textbf {\bibinfo {volume} {25}},\ \bibinfo {pages} {123010} (\bibinfo {year} {2009})}\BibitemShut {NoStop}%
\bibitem [{\citenamefont {Blaney}\ \emph {et~al.}(2024)\citenamefont {Blaney}, \citenamefont {Sassaroli},\ and\ \citenamefont {Fantini}}]{2024_Blaney_sensitivity_review}%
  \BibitemOpen
  \bibfield  {author} {\bibinfo {author} {\bibfnamefont {G.}~\bibnamefont {Blaney}}, \bibinfo {author} {\bibfnamefont {A.}~\bibnamefont {Sassaroli}},\ and\ \bibinfo {author} {\bibfnamefont {S.}~\bibnamefont {Fantini}},\ }\bibfield  {title} {\bibinfo {title} {Spatial sensitivity to absorption changes for various near-infrared spectroscopy methods: a compendium review},\ }\href {https://doi.org/10.1142/S1793545824300015} {\bibfield  {journal} {\bibinfo  {journal} {Journal of Innovative Optical Health Sciences}\ }\textbf {\bibinfo {volume} {17}},\ \bibinfo {pages} {2430001} (\bibinfo {year} {2024})}\BibitemShut {NoStop}%
\bibitem [{\citenamefont {Tsai}\ and\ \citenamefont {Hamblin}(2017)}]{2017_Tsai_Damage_threshold}%
  \BibitemOpen
  \bibfield  {author} {\bibinfo {author} {\bibfnamefont {S.-R.}\ \bibnamefont {Tsai}}\ and\ \bibinfo {author} {\bibfnamefont {M.~R.}\ \bibnamefont {Hamblin}},\ }\bibfield  {title} {\bibinfo {title} {Biological effects and medical applications of infrared radiation},\ }\href {https://doi.org/https://doi.org/10.1016/j.jphotobiol.2017.04.014} {\bibfield  {journal} {\bibinfo  {journal} {Journal of Photochemistry and Photobiology B: Biology}\ }\textbf {\bibinfo {volume} {170}},\ \bibinfo {pages} {197} (\bibinfo {year} {2017})}\BibitemShut {NoStop}%
\bibitem [{\citenamefont {Mora}\ \emph {et~al.}(2015)\citenamefont {Mora}, \citenamefont {Contini}, \citenamefont {Arridge}, \citenamefont {Martelli}, \citenamefont {Tosi}, \citenamefont {Boso}, \citenamefont {Farina}, \citenamefont {Durduran}, \citenamefont {Martinenghi}, \citenamefont {Torricelli},\ and\ \citenamefont {Pifferi}}]{2015_Mora_Depth_limit}%
  \BibitemOpen
  \bibfield  {author} {\bibinfo {author} {\bibfnamefont {A.~D.}\ \bibnamefont {Mora}}, \bibinfo {author} {\bibfnamefont {D.}~\bibnamefont {Contini}}, \bibinfo {author} {\bibfnamefont {S.}~\bibnamefont {Arridge}}, \bibinfo {author} {\bibfnamefont {F.}~\bibnamefont {Martelli}}, \bibinfo {author} {\bibfnamefont {A.}~\bibnamefont {Tosi}}, \bibinfo {author} {\bibfnamefont {G.}~\bibnamefont {Boso}}, \bibinfo {author} {\bibfnamefont {A.}~\bibnamefont {Farina}}, \bibinfo {author} {\bibfnamefont {T.}~\bibnamefont {Durduran}}, \bibinfo {author} {\bibfnamefont {E.}~\bibnamefont {Martinenghi}}, \bibinfo {author} {\bibfnamefont {A.}~\bibnamefont {Torricelli}},\ and\ \bibinfo {author} {\bibfnamefont {A.}~\bibnamefont {Pifferi}},\ }\bibfield  {title} {\bibinfo {title} {Towards next-generation time-domain diffuse optics for extreme depth penetration and sensitivity},\ }\href {https://doi.org/10.1364/BOE.6.001749} {\bibfield  {journal} {\bibinfo  {journal} {Biomed. Opt. Express}\ }\textbf {\bibinfo {volume} {6}},\ \bibinfo
  {pages} {1749} (\bibinfo {year} {2015})}\BibitemShut {NoStop}%
\bibitem [{\citenamefont {Pifferi}\ \emph {et~al.}(2016)\citenamefont {Pifferi}, \citenamefont {Contini}, \citenamefont {Mora}, \citenamefont {Farina}, \citenamefont {Spinelli},\ and\ \citenamefont {Torricelli}}]{2016_Pifferi_Depth_limit}%
  \BibitemOpen
  \bibfield  {author} {\bibinfo {author} {\bibfnamefont {A.}~\bibnamefont {Pifferi}}, \bibinfo {author} {\bibfnamefont {D.}~\bibnamefont {Contini}}, \bibinfo {author} {\bibfnamefont {A.~D.}\ \bibnamefont {Mora}}, \bibinfo {author} {\bibfnamefont {A.}~\bibnamefont {Farina}}, \bibinfo {author} {\bibfnamefont {L.}~\bibnamefont {Spinelli}},\ and\ \bibinfo {author} {\bibfnamefont {A.}~\bibnamefont {Torricelli}},\ }\bibfield  {title} {\bibinfo {title} {New frontiers in time-domain diffuse optics, a review},\ }\href {https://doi.org/10.1117/1.JBO.21.9.091310} {\bibfield  {journal} {\bibinfo  {journal} {Journal of Biomedical Optics}\ }\textbf {\bibinfo {volume} {21}},\ \bibinfo {pages} {091310} (\bibinfo {year} {2016})}\BibitemShut {NoStop}%
\bibitem [{\citenamefont {A.P.Mosk}\ \emph {et~al.}(2012)\citenamefont {A.P.Mosk}, \citenamefont {A.Lagendijk}, \citenamefont {G.Lerosey},\ and\ \citenamefont {M.Fink}}]{2012_Mosk_Wavefront_shaping_review}%
  \BibitemOpen
  \bibfield  {author} {\bibinfo {author} {\bibnamefont {A.P.Mosk}}, \bibinfo {author} {\bibnamefont {A.Lagendijk}}, \bibinfo {author} {\bibnamefont {G.Lerosey}},\ and\ \bibinfo {author} {\bibnamefont {M.Fink}},\ }\bibfield  {title} {\bibinfo {title} {Controlling waves in space and time for imaging and focusing in complex media},\ }\href {https://doi.org/10.1038/nphoton.2012.88} {\bibfield  {journal} {\bibinfo  {journal} {Nature Photonics}\ }\textbf {\bibinfo {volume} {6}},\ \bibinfo {pages} {283} (\bibinfo {year} {2012})}\BibitemShut {NoStop}%
\bibitem [{\citenamefont {Yu}\ \emph {et~al.}(2015)\citenamefont {Yu}, \citenamefont {Park}, \citenamefont {Lee}, \citenamefont {Yoon}, \citenamefont {Kim}, \citenamefont {Lee},\ and\ \citenamefont {Park}}]{2015_Yu_Wavefront_Shaping_Review}%
  \BibitemOpen
  \bibfield  {author} {\bibinfo {author} {\bibfnamefont {H.}~\bibnamefont {Yu}}, \bibinfo {author} {\bibfnamefont {J.}~\bibnamefont {Park}}, \bibinfo {author} {\bibfnamefont {K.}~\bibnamefont {Lee}}, \bibinfo {author} {\bibfnamefont {J.}~\bibnamefont {Yoon}}, \bibinfo {author} {\bibfnamefont {K.}~\bibnamefont {Kim}}, \bibinfo {author} {\bibfnamefont {S.}~\bibnamefont {Lee}},\ and\ \bibinfo {author} {\bibfnamefont {Y.}~\bibnamefont {Park}},\ }\bibfield  {title} {\bibinfo {title} {Recent advances in wavefront shaping techniques for biomedical applications},\ }\href {https://doi.org/10.1016/j.cap.2015.02.015} {\bibfield  {journal} {\bibinfo  {journal} {Current Applied Physics}\ }\textbf {\bibinfo {volume} {15}},\ \bibinfo {pages} {632} (\bibinfo {year} {2015})}\BibitemShut {NoStop}%
\bibitem [{\citenamefont {S.Rotter}\ and\ \citenamefont {S.Gigan}(2017)}]{2017_Rotter_Gigan_review}%
  \BibitemOpen
  \bibfield  {author} {\bibinfo {author} {\bibnamefont {S.Rotter}}\ and\ \bibinfo {author} {\bibnamefont {S.Gigan}},\ }\bibfield  {title} {\bibinfo {title} {Light fields in complex media: Mesoscopic scattering meets wave control},\ }\href {https://doi.org/10.1103/RevModPhys.89.015005} {\bibfield  {journal} {\bibinfo  {journal} {Reviews of Modern Physics}\ }\textbf {\bibinfo {volume} {89}},\ \bibinfo {pages} {1} (\bibinfo {year} {2017})}\BibitemShut {NoStop}%
\bibitem [{\citenamefont {Gigan}\ \emph {et~al.}(2022)\citenamefont {Gigan}, \citenamefont {Katz}, \citenamefont {de~Aguiar}, \citenamefont {Andresen}, \citenamefont {Aubry}, \citenamefont {Bertolotti}, \citenamefont {Bossy}, \citenamefont {Bouchet}, \citenamefont {Brake}, \citenamefont {Brasselet}, \citenamefont {Bromberg}, \citenamefont {Cao}, \citenamefont {Chaigne}, \citenamefont {Cheng}, \citenamefont {Choi}, \citenamefont {{\v{C}}i{\v{z}}m{\'a}r}, \citenamefont {Cui}, \citenamefont {Curtis}, \citenamefont {Defienne}, \citenamefont {Hofer}, \citenamefont {Horisaki}, \citenamefont {Horstmeyer}, \citenamefont {Ji}, \citenamefont {LaViolette}, \citenamefont {Mertz}, \citenamefont {Moser}, \citenamefont {Mosk}, \citenamefont {P{\'e}gard}, \citenamefont {Piestun}, \citenamefont {Popoff}, \citenamefont {Phillips}, \citenamefont {Psaltis}, \citenamefont {Rahmani}, \citenamefont {Rigneault}, \citenamefont {Rotter}, \citenamefont {Tian}, \citenamefont {Vellekoop}, \citenamefont {Waller}, \citenamefont {Wang},
  \citenamefont {Weber}, \citenamefont {Xiao}, \citenamefont {Xu}, \citenamefont {Yamilov}, \citenamefont {Yang},\ and\ \citenamefont {Y{\i}lmaz}}]{2022_Gigan_Roadmap}%
  \BibitemOpen
  \bibfield  {author} {\bibinfo {author} {\bibfnamefont {S.}~\bibnamefont {Gigan}}, \bibinfo {author} {\bibfnamefont {O.}~\bibnamefont {Katz}}, \bibinfo {author} {\bibfnamefont {H.~B.}\ \bibnamefont {de~Aguiar}}, \bibinfo {author} {\bibfnamefont {E.~R.}\ \bibnamefont {Andresen}}, \bibinfo {author} {\bibfnamefont {A.}~\bibnamefont {Aubry}}, \bibinfo {author} {\bibfnamefont {J.}~\bibnamefont {Bertolotti}}, \bibinfo {author} {\bibfnamefont {E.}~\bibnamefont {Bossy}}, \bibinfo {author} {\bibfnamefont {D.}~\bibnamefont {Bouchet}}, \bibinfo {author} {\bibfnamefont {J.}~\bibnamefont {Brake}}, \bibinfo {author} {\bibfnamefont {S.}~\bibnamefont {Brasselet}}, \bibinfo {author} {\bibfnamefont {Y.}~\bibnamefont {Bromberg}}, \bibinfo {author} {\bibfnamefont {H.}~\bibnamefont {Cao}}, \bibinfo {author} {\bibfnamefont {T.}~\bibnamefont {Chaigne}}, \bibinfo {author} {\bibfnamefont {Z.}~\bibnamefont {Cheng}}, \bibinfo {author} {\bibfnamefont {W.}~\bibnamefont {Choi}}, \bibinfo {author} {\bibfnamefont {T.}~\bibnamefont
  {{\v{C}}i{\v{z}}m{\'a}r}}, \bibinfo {author} {\bibfnamefont {M.}~\bibnamefont {Cui}}, \bibinfo {author} {\bibfnamefont {V.~R.}\ \bibnamefont {Curtis}}, \bibinfo {author} {\bibfnamefont {H.}~\bibnamefont {Defienne}}, \bibinfo {author} {\bibfnamefont {M.}~\bibnamefont {Hofer}}, \bibinfo {author} {\bibfnamefont {R.}~\bibnamefont {Horisaki}}, \bibinfo {author} {\bibfnamefont {R.}~\bibnamefont {Horstmeyer}}, \bibinfo {author} {\bibfnamefont {N.}~\bibnamefont {Ji}}, \bibinfo {author} {\bibfnamefont {A.~K.}\ \bibnamefont {LaViolette}}, \bibinfo {author} {\bibfnamefont {J.}~\bibnamefont {Mertz}}, \bibinfo {author} {\bibfnamefont {C.}~\bibnamefont {Moser}}, \bibinfo {author} {\bibfnamefont {A.~P.}\ \bibnamefont {Mosk}}, \bibinfo {author} {\bibfnamefont {N.~C.}\ \bibnamefont {P{\'e}gard}}, \bibinfo {author} {\bibfnamefont {R.}~\bibnamefont {Piestun}}, \bibinfo {author} {\bibfnamefont {S.}~\bibnamefont {Popoff}}, \bibinfo {author} {\bibfnamefont {D.~B.}\ \bibnamefont {Phillips}}, \bibinfo {author} {\bibfnamefont
  {D.}~\bibnamefont {Psaltis}}, \bibinfo {author} {\bibfnamefont {B.}~\bibnamefont {Rahmani}}, \bibinfo {author} {\bibfnamefont {H.}~\bibnamefont {Rigneault}}, \bibinfo {author} {\bibfnamefont {S.}~\bibnamefont {Rotter}}, \bibinfo {author} {\bibfnamefont {L.}~\bibnamefont {Tian}}, \bibinfo {author} {\bibfnamefont {I.~M.}\ \bibnamefont {Vellekoop}}, \bibinfo {author} {\bibfnamefont {L.}~\bibnamefont {Waller}}, \bibinfo {author} {\bibfnamefont {L.}~\bibnamefont {Wang}}, \bibinfo {author} {\bibfnamefont {T.}~\bibnamefont {Weber}}, \bibinfo {author} {\bibfnamefont {S.}~\bibnamefont {Xiao}}, \bibinfo {author} {\bibfnamefont {C.}~\bibnamefont {Xu}}, \bibinfo {author} {\bibfnamefont {A.}~\bibnamefont {Yamilov}}, \bibinfo {author} {\bibfnamefont {C.}~\bibnamefont {Yang}},\ and\ \bibinfo {author} {\bibfnamefont {H.}~\bibnamefont {Y{\i}lmaz}},\ }\bibfield  {title} {\bibinfo {title} {Roadmap on wavefront shaping and deep imaging in complex media},\ }\href {https://doi.org/10.1088/2515-7647/ac76f9} {\bibfield  {journal}
  {\bibinfo  {journal} {Journal of Physics: Photonics}\ }\textbf {\bibinfo {volume} {4}},\ \bibinfo {pages} {042501} (\bibinfo {year} {2022})}\BibitemShut {NoStop}%
\bibitem [{\citenamefont {Cao}\ \emph {et~al.}(2022)\citenamefont {Cao}, \citenamefont {Mosk},\ and\ \citenamefont {Rotter}}]{2022_Cao_Mosk_Rotter_review}%
  \BibitemOpen
  \bibfield  {author} {\bibinfo {author} {\bibfnamefont {H.}~\bibnamefont {Cao}}, \bibinfo {author} {\bibfnamefont {A.~P.}\ \bibnamefont {Mosk}},\ and\ \bibinfo {author} {\bibfnamefont {S.}~\bibnamefont {Rotter}},\ }\bibfield  {title} {\bibinfo {title} {Shaping the propagation of light in complex media},\ }\href {https://doi.org/10.1038/s41567-022-01677-x} {\bibfield  {journal} {\bibinfo  {journal} {Nature Physics}\ }\textbf {\bibinfo {volume} {18}},\ \bibinfo {pages} {994} (\bibinfo {year} {2022})}\BibitemShut {NoStop}%
\bibitem [{\citenamefont {Yoon}\ \emph {et~al.}(2020)\citenamefont {Yoon}, \citenamefont {Kim}, \citenamefont {Lee},\ and\ \citenamefont {Choi}}]{2020_Yoon_Review}%
  \BibitemOpen
  \bibfield  {author} {\bibinfo {author} {\bibfnamefont {S.}~\bibnamefont {Yoon}}, \bibinfo {author} {\bibfnamefont {C.}~\bibnamefont {Kim}}, \bibinfo {author} {\bibfnamefont {J.}~\bibnamefont {Lee}},\ and\ \bibinfo {author} {\bibfnamefont {W.}~\bibnamefont {Choi}},\ }\bibfield  {title} {\bibinfo {title} {Deep optical imaging within complex scattering media},\ }\href {https://doi.org/10.1038/s42254-019-0143-2} {\bibfield  {journal} {\bibinfo  {journal} {Nature Reviews Physics}\ }\textbf {\bibinfo {volume} {2}},\ \bibinfo {pages} {141} (\bibinfo {year} {2020})}\BibitemShut {NoStop}%
\bibitem [{\citenamefont {Bender}\ \emph {et~al.}(2022{\natexlab{a}})\citenamefont {Bender}, \citenamefont {Yamilov}, \citenamefont {Goetschy}, \citenamefont {Yilmaz}, \citenamefont {Hsu},\ and\ \citenamefont {Cao}}]{2022_Bender_Depth_Targeted_Energy_Deposition}%
  \BibitemOpen
  \bibfield  {author} {\bibinfo {author} {\bibfnamefont {N.}~\bibnamefont {Bender}}, \bibinfo {author} {\bibfnamefont {A.}~\bibnamefont {Yamilov}}, \bibinfo {author} {\bibfnamefont {A.}~\bibnamefont {Goetschy}}, \bibinfo {author} {\bibfnamefont {H.}~\bibnamefont {Yilmaz}}, \bibinfo {author} {\bibfnamefont {C.~W.}\ \bibnamefont {Hsu}},\ and\ \bibinfo {author} {\bibfnamefont {H.}~\bibnamefont {Cao}},\ }\bibfield  {title} {\bibinfo {title} {Depth-targeted energy deposition deep inside scattering media},\ }\href {https://doi.org/10.1038/s41567-021-01475-x} {\bibfield  {journal} {\bibinfo  {journal} {Nature Physics}\ }\textbf {\bibinfo {volume} {18}},\ \bibinfo {pages} {309} (\bibinfo {year} {2022}{\natexlab{a}})}\BibitemShut {NoStop}%
\bibitem [{\citenamefont {Choi}\ \emph {et~al.}(2011)\citenamefont {Choi}, \citenamefont {Mosk}, \citenamefont {Park},\ and\ \citenamefont {Choi}}]{2011_Mosk}%
  \BibitemOpen
  \bibfield  {author} {\bibinfo {author} {\bibfnamefont {W.}~\bibnamefont {Choi}}, \bibinfo {author} {\bibfnamefont {A.~P.}\ \bibnamefont {Mosk}}, \bibinfo {author} {\bibfnamefont {Q.~H.}\ \bibnamefont {Park}},\ and\ \bibinfo {author} {\bibfnamefont {W.}~\bibnamefont {Choi}},\ }\bibfield  {title} {\bibinfo {title} {Transmission eigenchannels in a disordered medium},\ }\href {https://doi.org/10.1103/PhysRevB.83.134207} {\bibfield  {journal} {\bibinfo  {journal} {Phys.~Rev.~B}\ }\textbf {\bibinfo {volume} {83}},\ \bibinfo {eid} {134207} (\bibinfo {year} {2011})}\BibitemShut {NoStop}%
\bibitem [{\citenamefont {Davy}\ \emph {et~al.}(2015)\citenamefont {Davy}, \citenamefont {Shi}, \citenamefont {Park}, \citenamefont {Tian},\ and\ \citenamefont {Genack}}]{2015_Genack_Eigenchannels_Inside}%
  \BibitemOpen
  \bibfield  {author} {\bibinfo {author} {\bibfnamefont {M.}~\bibnamefont {Davy}}, \bibinfo {author} {\bibfnamefont {Z.}~\bibnamefont {Shi}}, \bibinfo {author} {\bibfnamefont {J.}~\bibnamefont {Park}}, \bibinfo {author} {\bibfnamefont {C.}~\bibnamefont {Tian}},\ and\ \bibinfo {author} {\bibfnamefont {A.~Z.}\ \bibnamefont {Genack}},\ }\bibfield  {title} {\bibinfo {title} {Universal structure of transmission eigenchannels inside opaque media},\ }\href {https://doi.org/10.1038/ncomms7893} {\bibfield  {journal} {\bibinfo  {journal} {Nature Communications}\ }\textbf {\bibinfo {volume} {6}},\ \bibinfo {pages} {6893} (\bibinfo {year} {2015})}\BibitemShut {NoStop}%
\bibitem [{\citenamefont {Sarma}\ \emph {et~al.}(2016)\citenamefont {Sarma}, \citenamefont {Yamilov}, \citenamefont {Petrenko}, \citenamefont {Bromberg},\ and\ \citenamefont {Cao}}]{2016_Sarma_Open_Channels}%
  \BibitemOpen
  \bibfield  {author} {\bibinfo {author} {\bibfnamefont {R.}~\bibnamefont {Sarma}}, \bibinfo {author} {\bibfnamefont {A.~G.}\ \bibnamefont {Yamilov}}, \bibinfo {author} {\bibfnamefont {S.}~\bibnamefont {Petrenko}}, \bibinfo {author} {\bibfnamefont {Y.}~\bibnamefont {Bromberg}},\ and\ \bibinfo {author} {\bibfnamefont {H.}~\bibnamefont {Cao}},\ }\bibfield  {title} {\bibinfo {title} {Control of energy density inside a disordered medium by coupling to open or closed channels},\ }\href {https://doi.org/10.1103/PhysRevLett.117.086803} {\bibfield  {journal} {\bibinfo  {journal} {Physical Review Letters}\ }\textbf {\bibinfo {volume} {117}},\ \bibinfo {pages} {086803} (\bibinfo {year} {2016})}\BibitemShut {NoStop}%
\bibitem [{\citenamefont {Ojambati}\ \emph {et~al.}(2016)\citenamefont {Ojambati}, \citenamefont {Y{\i}lmaz}, \citenamefont {Lagendijk}, \citenamefont {Mosk},\ and\ \citenamefont {Vos}}]{2016_Ojambati_Fundamental_Mode_Experiment}%
  \BibitemOpen
  \bibfield  {author} {\bibinfo {author} {\bibfnamefont {O.~S.}\ \bibnamefont {Ojambati}}, \bibinfo {author} {\bibfnamefont {H.}~\bibnamefont {Y{\i}lmaz}}, \bibinfo {author} {\bibfnamefont {A.}~\bibnamefont {Lagendijk}}, \bibinfo {author} {\bibfnamefont {A.~P.}\ \bibnamefont {Mosk}},\ and\ \bibinfo {author} {\bibfnamefont {W.~L.}\ \bibnamefont {Vos}},\ }\bibfield  {title} {\bibinfo {title} {Coupling of energy into the fundamental diffusion mode of a complex nanophotonic medium},\ }\href {https://doi.org/10.1088/1367-2630/18/4/043032} {\bibfield  {journal} {\bibinfo  {journal} {New Journal of Physics}\ }\textbf {\bibinfo {volume} {18}},\ \bibinfo {pages} {043032} (\bibinfo {year} {2016})}\BibitemShut {NoStop}%
\bibitem [{\citenamefont {Carminati}\ and\ \citenamefont {Schotland}(2021)}]{2021_Carminati_Schotland}%
  \BibitemOpen
  \bibfield  {author} {\bibinfo {author} {\bibfnamefont {R.}~\bibnamefont {Carminati}}\ and\ \bibinfo {author} {\bibfnamefont {J.~C.}\ \bibnamefont {Schotland}},\ }\href {https://doi.org/10.1017/9781316544693} {\emph {\bibinfo {title} {Principles of Scattering and Transport of Light}}}\ (\bibinfo  {publisher} {Cambridge University Press},\ \bibinfo {address} {Cambridge},\ \bibinfo {year} {2021})\BibitemShut {NoStop}%
\bibitem [{\citenamefont {Bender}\ \emph {et~al.}(2022{\natexlab{b}})\citenamefont {Bender}, \citenamefont {Goetschy}, \citenamefont {Hsu}, \citenamefont {Yilmaz}, \citenamefont {Jara}, \citenamefont {Yamilov},\ and\ \citenamefont {Cao}}]{2022_Bender_Coherent_Enhancement}%
  \BibitemOpen
  \bibfield  {author} {\bibinfo {author} {\bibfnamefont {N.}~\bibnamefont {Bender}}, \bibinfo {author} {\bibfnamefont {A.}~\bibnamefont {Goetschy}}, \bibinfo {author} {\bibfnamefont {C.~W.}\ \bibnamefont {Hsu}}, \bibinfo {author} {\bibfnamefont {H.}~\bibnamefont {Yilmaz}}, \bibinfo {author} {\bibfnamefont {P.}~\bibnamefont {Jara}}, \bibinfo {author} {\bibfnamefont {A.}~\bibnamefont {Yamilov}},\ and\ \bibinfo {author} {\bibfnamefont {H.}~\bibnamefont {Cao}},\ }\bibfield  {title} {\bibinfo {title} {Coherent enhancement of optical remission in diffusive media},\ }\href {https://doi.org/10.1073/pnas.2207089119} {\bibfield  {journal} {\bibinfo  {journal} {Proceedings of the National Academy of Sciences}\ }\textbf {\bibinfo {volume} {119}},\ \bibinfo {pages} {1} (\bibinfo {year} {2022}{\natexlab{b}})}\BibitemShut {NoStop}%
\bibitem [{\citenamefont {Sheng}(2006)}]{2006_Sheng}%
  \BibitemOpen
  \bibfield  {author} {\bibinfo {author} {\bibfnamefont {P.}~\bibnamefont {Sheng}},\ }\href {https://doi.org/10.1007/3-540-29156-3} {\emph {\bibinfo {title} {Introduction to Wave Scattering, Localization, and Mesoscopic Phenomena}}},\ \bibinfo {edition} {2nd}\ ed.,\ \bibinfo {series} {Springer Series in Materials Science}, Vol.~\bibinfo {volume} {88}\ (\bibinfo  {publisher} {Springer},\ \bibinfo {year} {2006})\BibitemShut {NoStop}%
\bibitem [{\citenamefont {Beenakker}(1997)}]{1997_Beenakker}%
  \BibitemOpen
  \bibfield  {author} {\bibinfo {author} {\bibfnamefont {C.~W.~J.}\ \bibnamefont {Beenakker}},\ }\bibfield  {title} {\bibinfo {title} {Random-matrix theory of quantum transport},\ }\href {https://doi.org/10.1103/RevModPhys.69.731} {\bibfield  {journal} {\bibinfo  {journal} {Reviews of Modern Physics}\ }\textbf {\bibinfo {volume} {69}},\ \bibinfo {pages} {731} (\bibinfo {year} {1997})}\BibitemShut {NoStop}%
\bibitem [{\citenamefont {Schotland}(1997)}]{1997_Schotland}%
  \BibitemOpen
  \bibfield  {author} {\bibinfo {author} {\bibfnamefont {J.~C.}\ \bibnamefont {Schotland}},\ }\bibfield  {title} {\bibinfo {title} {Inverse scattering methods for diffuse optical tomography},\ }\href {https://doi.org/10.1088/0266-5611/13/6/001} {\bibfield  {journal} {\bibinfo  {journal} {Inverse Problems}\ }\textbf {\bibinfo {volume} {13}},\ \bibinfo {pages} {R1} (\bibinfo {year} {1997})}\BibitemShut {NoStop}%
\bibitem [{\citenamefont {Yaqoob}\ \emph {et~al.}(2008)\citenamefont {Yaqoob}, \citenamefont {Psaltis}, \citenamefont {Feld},\ and\ \citenamefont {Yang}}]{2008_Yaqoob_Phase_conjugation}%
  \BibitemOpen
  \bibfield  {author} {\bibinfo {author} {\bibfnamefont {Z.}~\bibnamefont {Yaqoob}}, \bibinfo {author} {\bibfnamefont {D.}~\bibnamefont {Psaltis}}, \bibinfo {author} {\bibfnamefont {M.~S.}\ \bibnamefont {Feld}},\ and\ \bibinfo {author} {\bibfnamefont {C.}~\bibnamefont {Yang}},\ }\bibfield  {title} {\bibinfo {title} {Optical phase conjugation for turbidity suppression in biological samples},\ }\href {https://doi.org/10.1038/nphoton.2007.297} {\bibfield  {journal} {\bibinfo  {journal} {Nature Photonics}\ }\textbf {\bibinfo {volume} {2}},\ \bibinfo {pages} {110} (\bibinfo {year} {2008})}\BibitemShut {NoStop}%
\bibitem [{\citenamefont {Fink}\ \emph {et~al.}(2000)\citenamefont {Fink}, \citenamefont {Cassereau}, \citenamefont {Derode}, \citenamefont {Prada}, \citenamefont {Roux}, \citenamefont {Tanter}, \citenamefont {Thomas},\ and\ \citenamefont {Wu}}]{2000_Fink}%
  \BibitemOpen
  \bibfield  {author} {\bibinfo {author} {\bibfnamefont {M.}~\bibnamefont {Fink}}, \bibinfo {author} {\bibfnamefont {D.}~\bibnamefont {Cassereau}}, \bibinfo {author} {\bibfnamefont {A.}~\bibnamefont {Derode}}, \bibinfo {author} {\bibfnamefont {C.}~\bibnamefont {Prada}}, \bibinfo {author} {\bibfnamefont {P.}~\bibnamefont {Roux}}, \bibinfo {author} {\bibfnamefont {M.}~\bibnamefont {Tanter}}, \bibinfo {author} {\bibfnamefont {J.-L.}\ \bibnamefont {Thomas}},\ and\ \bibinfo {author} {\bibfnamefont {F.}~\bibnamefont {Wu}},\ }\bibfield  {title} {\bibinfo {title} {Time-reversed acoustics},\ }\href {https://doi.org/10.1088/0034-4885/63/12/202} {\bibfield  {journal} {\bibinfo  {journal} {Reports on Progress in Physics}\ }\textbf {\bibinfo {volume} {63}},\ \bibinfo {pages} {1933} (\bibinfo {year} {2000})}\BibitemShut {NoStop}%
\bibitem [{\citenamefont {D.Fisher}\ and\ \citenamefont {P.Lee}(1981)}]{1981_Fisher_Lee_Relation_between_conductivity_and_transmission_matrix}%
  \BibitemOpen
  \bibfield  {author} {\bibinfo {author} {\bibnamefont {D.Fisher}}\ and\ \bibinfo {author} {\bibnamefont {P.Lee}},\ }\bibfield  {title} {\bibinfo {title} {Relation between conductivity and transmission matrix},\ }\href {https://doi.org/10.1103/PhysRevB.23.6851} {\bibfield  {journal} {\bibinfo  {journal} {Physical Review B}\ }\textbf {\bibinfo {volume} {23}},\ \bibinfo {pages} {6851} (\bibinfo {year} {1981})}\BibitemShut {NoStop}%
\bibitem [{\citenamefont {Popoff}\ \emph {et~al.}(2010)\citenamefont {Popoff}, \citenamefont {Lerosey}, \citenamefont {Carminati}, \citenamefont {Fink}, \citenamefont {Boccara},\ and\ \citenamefont {Gigan}}]{2010_Popoff_NatComm}%
  \BibitemOpen
  \bibfield  {author} {\bibinfo {author} {\bibfnamefont {S.~M.}\ \bibnamefont {Popoff}}, \bibinfo {author} {\bibfnamefont {G.}~\bibnamefont {Lerosey}}, \bibinfo {author} {\bibfnamefont {R.}~\bibnamefont {Carminati}}, \bibinfo {author} {\bibfnamefont {M.}~\bibnamefont {Fink}}, \bibinfo {author} {\bibfnamefont {A.~C.}\ \bibnamefont {Boccara}},\ and\ \bibinfo {author} {\bibfnamefont {S.}~\bibnamefont {Gigan}},\ }\bibfield  {title} {\bibinfo {title} {Image transmission through an opaque material},\ }\href {https://doi.org/10.1038/ncomms1078} {\bibfield  {journal} {\bibinfo  {journal} {Nature Communications}\ }\textbf {\bibinfo {volume} {1}},\ \bibinfo {pages} {81} (\bibinfo {year} {2010})}\BibitemShut {NoStop}%
\bibitem [{\citenamefont {Mosk}\ \emph {et~al.}(2012)\citenamefont {Mosk}, \citenamefont {Lagendijk}, \citenamefont {Lerosey},\ and\ \citenamefont {Fink}}]{2012_Mosk_NatPhoton}%
  \BibitemOpen
  \bibfield  {author} {\bibinfo {author} {\bibfnamefont {A.~P.}\ \bibnamefont {Mosk}}, \bibinfo {author} {\bibfnamefont {A.}~\bibnamefont {Lagendijk}}, \bibinfo {author} {\bibfnamefont {G.}~\bibnamefont {Lerosey}},\ and\ \bibinfo {author} {\bibfnamefont {M.}~\bibnamefont {Fink}},\ }\bibfield  {title} {\bibinfo {title} {Controlling waves in space and time for imaging and focusing in complex media},\ }\href {https://doi.org/10.1038/nphoton.2012.88} {\bibfield  {journal} {\bibinfo  {journal} {Nature Photonics}\ }\textbf {\bibinfo {volume} {6}},\ \bibinfo {pages} {283} (\bibinfo {year} {2012})}\BibitemShut {NoStop}%
\bibitem [{\citenamefont {Wang}\ and\ \citenamefont {Zhao}(2013)}]{2013_Wang_NatPhoton}%
  \BibitemOpen
  \bibfield  {author} {\bibinfo {author} {\bibfnamefont {L.~V.}\ \bibnamefont {Wang}}\ and\ \bibinfo {author} {\bibfnamefont {J.}~\bibnamefont {Zhao}},\ }\bibfield  {title} {\bibinfo {title} {Photoacoustic tomography: principles and advances},\ }\href {https://doi.org/10.1038/nphoton.2013.191} {\bibfield  {journal} {\bibinfo  {journal} {Nature Photonics}\ }\textbf {\bibinfo {volume} {7}},\ \bibinfo {pages} {503} (\bibinfo {year} {2013})}\BibitemShut {NoStop}%
\bibitem [{\citenamefont {Fink}(1997)}]{1997_Fink}%
  \BibitemOpen
  \bibfield  {author} {\bibinfo {author} {\bibfnamefont {M.}~\bibnamefont {Fink}},\ }\bibfield  {title} {\bibinfo {title} {Time reversed acoustics},\ }\href {https://doi.org/10.1063/1.881692} {\bibfield  {journal} {\bibinfo  {journal} {Physics Today}\ }\textbf {\bibinfo {volume} {50}},\ \bibinfo {pages} {34} (\bibinfo {year} {1997})}\BibitemShut {NoStop}%
\bibitem [{\citenamefont {Ambichl}\ \emph {et~al.}(2017)\citenamefont {Ambichl}, \citenamefont {Brandst{\"o}tter}, \citenamefont {B{\"o}hm}, \citenamefont {K{\"u}hmayer}, \citenamefont {Kuhl},\ and\ \citenamefont {Rotter}}]{2017_Rotter_Focusing}%
  \BibitemOpen
  \bibfield  {author} {\bibinfo {author} {\bibfnamefont {P.}~\bibnamefont {Ambichl}}, \bibinfo {author} {\bibfnamefont {A.}~\bibnamefont {Brandst{\"o}tter}}, \bibinfo {author} {\bibfnamefont {J.}~\bibnamefont {B{\"o}hm}}, \bibinfo {author} {\bibfnamefont {M.}~\bibnamefont {K{\"u}hmayer}}, \bibinfo {author} {\bibfnamefont {U.}~\bibnamefont {Kuhl}},\ and\ \bibinfo {author} {\bibfnamefont {S.}~\bibnamefont {Rotter}},\ }\bibfield  {title} {\bibinfo {title} {Focusing inside disordered media with the generalized wigner–smith operator},\ }\href {https://doi.org/10.1103/PhysRevLett.119.033903} {\bibfield  {journal} {\bibinfo  {journal} {Physical Review Letters}\ }\textbf {\bibinfo {volume} {119}},\ \bibinfo {pages} {033903} (\bibinfo {year} {2017})}\BibitemShut {NoStop}%
\bibitem [{\citenamefont {Horodynski}\ \emph {et~al.}(2020)\citenamefont {Horodynski}, \citenamefont {K{\"u}hmayer}, \citenamefont {Brandst{\"o}tter}, \citenamefont {Pichler}, \citenamefont {Fyodorov}, \citenamefont {Kuhl},\ and\ \citenamefont {Rotter}}]{2020_Rotter_Optimal}%
  \BibitemOpen
  \bibfield  {author} {\bibinfo {author} {\bibfnamefont {M.}~\bibnamefont {Horodynski}}, \bibinfo {author} {\bibfnamefont {M.}~\bibnamefont {K{\"u}hmayer}}, \bibinfo {author} {\bibfnamefont {A.}~\bibnamefont {Brandst{\"o}tter}}, \bibinfo {author} {\bibfnamefont {K.}~\bibnamefont {Pichler}}, \bibinfo {author} {\bibfnamefont {Y.~V.}\ \bibnamefont {Fyodorov}}, \bibinfo {author} {\bibfnamefont {U.}~\bibnamefont {Kuhl}},\ and\ \bibinfo {author} {\bibfnamefont {S.}~\bibnamefont {Rotter}},\ }\bibfield  {title} {\bibinfo {title} {Optimal wave fields for micro‑manipulation in complex scattering environments},\ }\href {https://doi.org/10.1038/s41566-019-0550-z} {\bibfield  {journal} {\bibinfo  {journal} {Nature Photonics}\ }\textbf {\bibinfo {volume} {14}},\ \bibinfo {pages} {149} (\bibinfo {year} {2020})}\BibitemShut {NoStop}%
\bibitem [{\citenamefont {del Hougne}\ \emph {et~al.}(2021)\citenamefont {del Hougne}, \citenamefont {Yeo}, \citenamefont {Besnier},\ and\ \citenamefont {Davy}}]{2021_Davy_GWS}%
  \BibitemOpen
  \bibfield  {author} {\bibinfo {author} {\bibfnamefont {P.}~\bibnamefont {del Hougne}}, \bibinfo {author} {\bibfnamefont {K.~B.}\ \bibnamefont {Yeo}}, \bibinfo {author} {\bibfnamefont {P.}~\bibnamefont {Besnier}},\ and\ \bibinfo {author} {\bibfnamefont {M.}~\bibnamefont {Davy}},\ }\bibfield  {title} {\bibinfo {title} {Coherent wave control in complex media with arbitrary wavefronts},\ }\href {https://doi.org/10.1103/PhysRevLett.126.193903} {\bibfield  {journal} {\bibinfo  {journal} {Physical Review Letters}\ }\textbf {\bibinfo {volume} {126}},\ \bibinfo {pages} {193903} (\bibinfo {year} {2021})}\BibitemShut {NoStop}%
\bibitem [{\citenamefont {Bouchet}\ \emph {et~al.}(2021)\citenamefont {Bouchet}, \citenamefont {Rotter},\ and\ \citenamefont {Mosk}}]{2021_Bouchet_maximum_information}%
  \BibitemOpen
  \bibfield  {author} {\bibinfo {author} {\bibfnamefont {D.}~\bibnamefont {Bouchet}}, \bibinfo {author} {\bibfnamefont {S.}~\bibnamefont {Rotter}},\ and\ \bibinfo {author} {\bibfnamefont {A.~P.}\ \bibnamefont {Mosk}},\ }\bibfield  {title} {\bibinfo {title} {Maximum information states for coherent scattering measurements},\ }\href {https://doi.org/doi.org/10.1038/s41567-020-01137-4} {\bibfield  {journal} {\bibinfo  {journal} {Nature Physics}\ }\textbf {\bibinfo {volume} {17}},\ \bibinfo {pages} {564} (\bibinfo {year} {2021})}\BibitemShut {NoStop}%
\bibitem [{\citenamefont {Schotland}\ \emph {et~al.}(1993)\citenamefont {Schotland}, \citenamefont {Haselgrove},\ and\ \citenamefont {Leigh}}]{1993_Schotland}%
  \BibitemOpen
  \bibfield  {author} {\bibinfo {author} {\bibfnamefont {J.~C.}\ \bibnamefont {Schotland}}, \bibinfo {author} {\bibfnamefont {J.~C.}\ \bibnamefont {Haselgrove}},\ and\ \bibinfo {author} {\bibfnamefont {J.~S.}\ \bibnamefont {Leigh}},\ }\bibfield  {title} {\bibinfo {title} {Photon hitting density},\ }\href {https://doi.org/10.1364/AO.32.000448} {\bibfield  {journal} {\bibinfo  {journal} {Applied Optics}\ }\textbf {\bibinfo {volume} {32}},\ \bibinfo {pages} {448} (\bibinfo {year} {1993})}\BibitemShut {NoStop}%
\bibitem [{\citenamefont {Schotland}\ and\ \citenamefont {Markel}(2001)}]{2001_Schotland_Markel}%
  \BibitemOpen
  \bibfield  {author} {\bibinfo {author} {\bibfnamefont {J.~C.}\ \bibnamefont {Schotland}}\ and\ \bibinfo {author} {\bibfnamefont {V.~A.}\ \bibnamefont {Markel}},\ }\bibfield  {title} {\bibinfo {title} {Inverse scattering with diffusing waves},\ }\href {https://doi.org/10.1364/JOSAA.18.002767} {\bibfield  {journal} {\bibinfo  {journal} {J. Opt. Soc. Am. A}\ }\textbf {\bibinfo {volume} {18}},\ \bibinfo {pages} {2767} (\bibinfo {year} {2001})}\BibitemShut {NoStop}%
\bibitem [{\citenamefont {Akkermans}\ and\ \citenamefont {Montambaux}(2007)}]{Akkermans_Montambaux_2007}%
  \BibitemOpen
  \bibfield  {author} {\bibinfo {author} {\bibfnamefont {E.}~\bibnamefont {Akkermans}}\ and\ \bibinfo {author} {\bibfnamefont {G.}~\bibnamefont {Montambaux}},\ }\href@noop {} {\emph {\bibinfo {title} {Mesoscopic Physics of Electrons and Photons}}}\ (\bibinfo  {publisher} {Cambridge University Press},\ \bibinfo {year} {2007})\BibitemShut {NoStop}%
\bibitem [{\citenamefont {van Rossum}\ and\ \citenamefont {Nieuwenhuizen}(1999)}]{1999_VanRossum}%
  \BibitemOpen
  \bibfield  {author} {\bibinfo {author} {\bibfnamefont {M.~C.~W.}\ \bibnamefont {van Rossum}}\ and\ \bibinfo {author} {\bibfnamefont {T.~M.}\ \bibnamefont {Nieuwenhuizen}},\ }\bibfield  {title} {\bibinfo {title} {Multiple scattering of classical waves: microscopy, mesoscopy, and diffusion},\ }\href {https://doi.org/10.1103/RevModPhys.71.313} {\bibfield  {journal} {\bibinfo  {journal} {Rev. Mod. Phys.}\ }\textbf {\bibinfo {volume} {71}},\ \bibinfo {pages} {313} (\bibinfo {year} {1999})}\BibitemShut {NoStop}%
\bibitem [{\citenamefont {Cui}\ \emph {et~al.}(1991)\citenamefont {Cui}, \citenamefont {Kumar},\ and\ \citenamefont {Chance}}]{1991_Cui}%
  \BibitemOpen
  \bibfield  {author} {\bibinfo {author} {\bibfnamefont {W.}~\bibnamefont {Cui}}, \bibinfo {author} {\bibfnamefont {C.}~\bibnamefont {Kumar}},\ and\ \bibinfo {author} {\bibfnamefont {B.}~\bibnamefont {Chance}},\ }\bibfield  {title} {\bibinfo {title} {Experimental study of migration depth for the photons measured at sample surface},\ }in\ \href {https://doi.org/10.1117/12.44189} {\emph {\bibinfo {booktitle} {Proceedings of SPIE - Time-Resolved Spectroscopy and Imaging of Tissues}}},\ Vol.\ \bibinfo {volume} {1431}\ (\bibinfo  {publisher} {SPIE},\ \bibinfo {year} {1991})\ pp.\ \bibinfo {pages} {180--191}\BibitemShut {NoStop}%
\bibitem [{\citenamefont {Fantini}\ \emph {et~al.}(2020)\citenamefont {Fantini}, \citenamefont {Blaney},\ and\ \citenamefont {Sassaroli}}]{2020_Fantini}%
  \BibitemOpen
  \bibfield  {author} {\bibinfo {author} {\bibfnamefont {S.}~\bibnamefont {Fantini}}, \bibinfo {author} {\bibfnamefont {G.}~\bibnamefont {Blaney}},\ and\ \bibinfo {author} {\bibfnamefont {A.}~\bibnamefont {Sassaroli}},\ }\bibfield  {title} {\bibinfo {title} {Transformational change in the field of diffuse optics: From going bananas to going nuts},\ }\href {https://doi.org/10.1142/S1793545819300131} {\bibfield  {journal} {\bibinfo  {journal} {Journal of Innovative Optical Health Sciences}\ }\textbf {\bibinfo {volume} {13}},\ \bibinfo {pages} {1930013} (\bibinfo {year} {2020})}\BibitemShut {NoStop}%
\bibitem [{\citenamefont {Bender}\ \emph {et~al.}(2020)\citenamefont {Bender}, \citenamefont {Yamilov}, \citenamefont {Y{\i}lmaz},\ and\ \citenamefont {Cao}}]{2020_Bender_Eigenchannels}%
  \BibitemOpen
  \bibfield  {author} {\bibinfo {author} {\bibfnamefont {N.}~\bibnamefont {Bender}}, \bibinfo {author} {\bibfnamefont {A.}~\bibnamefont {Yamilov}}, \bibinfo {author} {\bibfnamefont {H.}~\bibnamefont {Y{\i}lmaz}},\ and\ \bibinfo {author} {\bibfnamefont {H.}~\bibnamefont {Cao}},\ }\bibfield  {title} {\bibinfo {title} {Fluctuations and correlations of transmission eigenchannels in diffusive media},\ }\href {https://doi.org/10.1103/PhysRevLett.125.165901} {\bibfield  {journal} {\bibinfo  {journal} {Physical Review Letters}\ }\textbf {\bibinfo {volume} {125}},\ \bibinfo {pages} {165901} (\bibinfo {year} {2020})}\BibitemShut {NoStop}%
\bibitem [{\citenamefont {Sarma}\ \emph {et~al.}(2015)\citenamefont {Sarma}, \citenamefont {Yamilov}, \citenamefont {Liew}, \citenamefont {Guy},\ and\ \citenamefont {Cao}}]{2015_Sarma}%
  \BibitemOpen
  \bibfield  {author} {\bibinfo {author} {\bibfnamefont {R.}~\bibnamefont {Sarma}}, \bibinfo {author} {\bibfnamefont {A.}~\bibnamefont {Yamilov}}, \bibinfo {author} {\bibfnamefont {S.~F.}\ \bibnamefont {Liew}}, \bibinfo {author} {\bibfnamefont {M.}~\bibnamefont {Guy}},\ and\ \bibinfo {author} {\bibfnamefont {H.}~\bibnamefont {Cao}},\ }\bibfield  {title} {\bibinfo {title} {Control of mesoscopic transport by modifying transmission channels in opaque media},\ }\href {https://doi.org/10.1103/PhysRevB.92.214206} {\bibfield  {journal} {\bibinfo  {journal} {Physical Review B}\ }\textbf {\bibinfo {volume} {92}},\ \bibinfo {pages} {214206} (\bibinfo {year} {2015})}\BibitemShut {NoStop}%
\bibitem [{\citenamefont {Yamilov}\ \emph {et~al.}(2016)\citenamefont {Yamilov}, \citenamefont {Petrenko}, \citenamefont {Sarma},\ and\ \citenamefont {Cao}}]{2016_Yamilov}%
  \BibitemOpen
  \bibfield  {author} {\bibinfo {author} {\bibfnamefont {A.}~\bibnamefont {Yamilov}}, \bibinfo {author} {\bibfnamefont {S.}~\bibnamefont {Petrenko}}, \bibinfo {author} {\bibfnamefont {R.}~\bibnamefont {Sarma}},\ and\ \bibinfo {author} {\bibfnamefont {H.}~\bibnamefont {Cao}},\ }\bibfield  {title} {\bibinfo {title} {Shape-dependence of transmission, reflection and absorption eigenvalue densities in disordered waveguides with dissipation},\ }\href {https://doi.org/10.1103/PhysRevB.93.100201} {\bibfield  {journal} {\bibinfo  {journal} {Physical Review B}\ }\textbf {\bibinfo {volume} {93}},\ \bibinfo {pages} {100201} (\bibinfo {year} {2016})}\BibitemShut {NoStop}%
\bibitem [{\citenamefont {Koirala}\ \emph {et~al.}(2017)\citenamefont {Koirala}, \citenamefont {Sarma}, \citenamefont {Cao},\ and\ \citenamefont {Yamilov}}]{2017_Koirala}%
  \BibitemOpen
  \bibfield  {author} {\bibinfo {author} {\bibfnamefont {M.}~\bibnamefont {Koirala}}, \bibinfo {author} {\bibfnamefont {R.}~\bibnamefont {Sarma}}, \bibinfo {author} {\bibfnamefont {H.}~\bibnamefont {Cao}},\ and\ \bibinfo {author} {\bibfnamefont {A.}~\bibnamefont {Yamilov}},\ }\bibfield  {title} {\bibinfo {title} {Inverse design of perfectly transmitting eigenchannels in scattering media},\ }\href {https://doi.org/10.1103/PhysRevB.96.054209} {\bibfield  {journal} {\bibinfo  {journal} {Physical Review B}\ }\textbf {\bibinfo {volume} {96}},\ \bibinfo {pages} {054209} (\bibinfo {year} {2017})}\BibitemShut {NoStop}%
\bibitem [{\citenamefont {Yamilov}\ \emph {et~al.}(2014)\citenamefont {Yamilov}, \citenamefont {Sarma}, \citenamefont {Redding}, \citenamefont {Payne}, \citenamefont {Noh},\ and\ \citenamefont {Cao}}]{2014_Yamilov}%
  \BibitemOpen
  \bibfield  {author} {\bibinfo {author} {\bibfnamefont {A.~G.}\ \bibnamefont {Yamilov}}, \bibinfo {author} {\bibfnamefont {R.}~\bibnamefont {Sarma}}, \bibinfo {author} {\bibfnamefont {B.}~\bibnamefont {Redding}}, \bibinfo {author} {\bibfnamefont {B.}~\bibnamefont {Payne}}, \bibinfo {author} {\bibfnamefont {H.}~\bibnamefont {Noh}},\ and\ \bibinfo {author} {\bibfnamefont {H.}~\bibnamefont {Cao}},\ }\bibfield  {title} {\bibinfo {title} {Position-dependent diffusion of light in disordered waveguides},\ }\href {https://doi.org/10.1103/PhysRevLett.112.023904} {\bibfield  {journal} {\bibinfo  {journal} {Physical Review Letters}\ }\textbf {\bibinfo {volume} {112}},\ \bibinfo {pages} {023904} (\bibinfo {year} {2014})}\BibitemShut {NoStop}%
\bibitem [{\citenamefont {Sarma}\ \emph {et~al.}(2017)\citenamefont {Sarma}, \citenamefont {Yamilov},\ and\ \citenamefont {Cao}}]{2017_Sarma}%
  \BibitemOpen
  \bibfield  {author} {\bibinfo {author} {\bibfnamefont {R.}~\bibnamefont {Sarma}}, \bibinfo {author} {\bibfnamefont {A.}~\bibnamefont {Yamilov}},\ and\ \bibinfo {author} {\bibfnamefont {H.}~\bibnamefont {Cao}},\ }\bibfield  {title} {\bibinfo {title} {Enhancing light transmission through a random medium with inhomogeneous scattering and loss},\ }\href {https://doi.org/10.1063/1.4973459} {\bibfield  {journal} {\bibinfo  {journal} {Applied Physics Letters}\ }\textbf {\bibinfo {volume} {110}},\ \bibinfo {pages} {021103} (\bibinfo {year} {2017})}\BibitemShut {NoStop}%
\bibitem [{\citenamefont {Taflove}\ and\ \citenamefont {Hagness}(2005)}]{2005_Taflove_book}%
  \BibitemOpen
  \bibfield  {author} {\bibinfo {author} {\bibfnamefont {A.}~\bibnamefont {Taflove}}\ and\ \bibinfo {author} {\bibfnamefont {S.~C.}\ \bibnamefont {Hagness}},\ }\href@noop {} {\emph {\bibinfo {title} {Computational electrodynamics: the finite-difference time-domain method}}},\ \bibinfo {edition} {3rd}\ ed.\ (\bibinfo  {publisher} {Artech House},\ \bibinfo {address} {Norwood},\ \bibinfo {year} {2005})\BibitemShut {NoStop}%
\bibitem [{\citenamefont {Summers}(2018)}]{2018_Summers}%
  \BibitemOpen
  \bibfield  {author} {\bibinfo {author} {\bibfnamefont {J.~R.}\ \bibnamefont {Summers}},\ }\emph {\bibinfo {title} {Universal wavefront transmission through disordered media}},\ \href {https://scholarsmine.mst.edu/masters_theses/8065} {Master's thesis},\ \bibinfo  {school} {Missouri University of Science and Technology} (\bibinfo {year} {2018})\BibitemShut {NoStop}%
\bibitem [{\citenamefont {Groth}\ \emph {et~al.}(2014)\citenamefont {Groth}, \citenamefont {Wimmer}, \citenamefont {Akhmerov},\ and\ \citenamefont {Waintal}}]{Groth2014}%
  \BibitemOpen
  \bibfield  {author} {\bibinfo {author} {\bibfnamefont {C.~W.}\ \bibnamefont {Groth}}, \bibinfo {author} {\bibfnamefont {M.}~\bibnamefont {Wimmer}}, \bibinfo {author} {\bibfnamefont {A.~R.}\ \bibnamefont {Akhmerov}},\ and\ \bibinfo {author} {\bibfnamefont {X.}~\bibnamefont {Waintal}},\ }\bibfield  {title} {\bibinfo {title} {Kwant: a software package for quantum transport},\ }\href {https://doi.org/10.1088/1367-2630/16/6/063065} {\bibfield  {journal} {\bibinfo  {journal} {New Journal of Physics}\ }\textbf {\bibinfo {volume} {16}},\ \bibinfo {pages} {063065} (\bibinfo {year} {2014})}\BibitemShut {NoStop}%
\bibitem [{\citenamefont {Lin}\ \emph {et~al.}(2022)\citenamefont {Lin}, \citenamefont {Wang},\ and\ \citenamefont {Hsu}}]{2022_HoChun_MESTI}%
  \BibitemOpen
  \bibfield  {author} {\bibinfo {author} {\bibfnamefont {H.}~\bibnamefont {Lin}}, \bibinfo {author} {\bibfnamefont {Z.}~\bibnamefont {Wang}},\ and\ \bibinfo {author} {\bibfnamefont {C.~W.}\ \bibnamefont {Hsu}},\ }\bibfield  {title} {\bibinfo {title} {Fast multi-source nanophotonic simulations using augmented partial factorization},\ }\href {https://doi.org/10.1038/s43588-022-00370-6} {\bibfield  {journal} {\bibinfo  {journal} {Nature Computational Science}\ }\textbf {\bibinfo {volume} {2}},\ \bibinfo {pages} {815} (\bibinfo {year} {2022})}\BibitemShut {NoStop}%
\bibitem [{\citenamefont {Jara}\ \emph {et~al.}(2022)\citenamefont {Jara}, \citenamefont {Lin}, \citenamefont {Hsu}, \citenamefont {Cao},\ and\ \citenamefont {Yamilov}}]{2022_Jara_Simulation_Coherent_Remission}%
  \BibitemOpen
  \bibfield  {author} {\bibinfo {author} {\bibfnamefont {P.}~\bibnamefont {Jara}}, \bibinfo {author} {\bibfnamefont {H.~C.}\ \bibnamefont {Lin}}, \bibinfo {author} {\bibfnamefont {C.~W.}\ \bibnamefont {Hsu}}, \bibinfo {author} {\bibfnamefont {H.}~\bibnamefont {Cao}},\ and\ \bibinfo {author} {\bibfnamefont {A.}~\bibnamefont {Yamilov}},\ }\bibfield  {title} {\bibinfo {title} {Simulation of coherent remission in planar disordered medium},\ }\href {https://doi.org/10.23919/ACES57841.2023.10114726} {\bibfield  {journal} {\bibinfo  {journal} {2023 International Applied Computational Electromagnetics Society Symposium (ACES)}\ ,\ \bibinfo {pages} {1}} (\bibinfo {year} {2022})}\BibitemShut {NoStop}%
\bibitem [{\citenamefont {Goetschy}\ and\ \citenamefont {Stone}(2013)}]{2013_Goetschy_FRM}%
  \BibitemOpen
  \bibfield  {author} {\bibinfo {author} {\bibfnamefont {A.}~\bibnamefont {Goetschy}}\ and\ \bibinfo {author} {\bibfnamefont {A.}~\bibnamefont {Stone}},\ }\bibfield  {title} {\bibinfo {title} {Filtering random matrices: the effect of incomplete channel control in multiple scattering},\ }\href {https://doi.org/10.1103/PhysRevLett.111.063901} {\bibfield  {journal} {\bibinfo  {journal} {Physical Review Letters}\ }\textbf {\bibinfo {volume} {111}},\ \bibinfo {pages} {063901} (\bibinfo {year} {2013})}\BibitemShut {NoStop}%
\bibitem [{\citenamefont {Hsu}\ \emph {et~al.}(2017)\citenamefont {Hsu}, \citenamefont {Liew}, \citenamefont {Goetschy}, \citenamefont {Cao},\ and\ \citenamefont {Stone}}]{Hsu2017}%
  \BibitemOpen
  \bibfield  {author} {\bibinfo {author} {\bibfnamefont {C.~W.}\ \bibnamefont {Hsu}}, \bibinfo {author} {\bibfnamefont {S.~F.}\ \bibnamefont {Liew}}, \bibinfo {author} {\bibfnamefont {A.}~\bibnamefont {Goetschy}}, \bibinfo {author} {\bibfnamefont {H.}~\bibnamefont {Cao}},\ and\ \bibinfo {author} {\bibfnamefont {A.~D.}\ \bibnamefont {Stone}},\ }\bibfield  {title} {\bibinfo {title} {Correlation-enhanced control of wave focusing in disordered media},\ }\href {https://doi.org/10.1038/nphys4036} {\bibfield  {journal} {\bibinfo  {journal} {Nature Physics}\ }\textbf {\bibinfo {volume} {13}},\ \bibinfo {pages} {497–502} (\bibinfo {year} {2017})}\BibitemShut {NoStop}%
\bibitem [{\citenamefont {McIntosh}\ \emph {et~al.}(2024)\citenamefont {McIntosh}, \citenamefont {Goetschy}, \citenamefont {Bender}, \citenamefont {Yamilov}, \citenamefont {Hsu}, \citenamefont {Yılmaz},\ and\ \citenamefont {Cao}}]{McIntosh2024}%
  \BibitemOpen
  \bibfield  {author} {\bibinfo {author} {\bibfnamefont {R.}~\bibnamefont {McIntosh}}, \bibinfo {author} {\bibfnamefont {A.}~\bibnamefont {Goetschy}}, \bibinfo {author} {\bibfnamefont {N.}~\bibnamefont {Bender}}, \bibinfo {author} {\bibfnamefont {A.}~\bibnamefont {Yamilov}}, \bibinfo {author} {\bibfnamefont {C.~W.}\ \bibnamefont {Hsu}}, \bibinfo {author} {\bibfnamefont {H.}~\bibnamefont {Yılmaz}},\ and\ \bibinfo {author} {\bibfnamefont {H.}~\bibnamefont {Cao}},\ }\bibfield  {title} {\bibinfo {title} {Delivering broadband light deep inside diffusive media},\ }\href {https://doi.org/10.1038/s41566-024-01446-7} {\bibfield  {journal} {\bibinfo  {journal} {Nature Photonics}\ }\textbf {\bibinfo {volume} {18}},\ \bibinfo {pages} {744–750} (\bibinfo {year} {2024})}\BibitemShut {NoStop}%
\bibitem [{\citenamefont {Yamilov}\ and\ \citenamefont {Cao}(2006)}]{2006_Yamilov}%
  \BibitemOpen
  \bibfield  {author} {\bibinfo {author} {\bibfnamefont {A.}~\bibnamefont {Yamilov}}\ and\ \bibinfo {author} {\bibfnamefont {H.}~\bibnamefont {Cao}},\ }\bibfield  {title} {\bibinfo {title} {Effect of amplification on conductance distribution of a disordered waveguide},\ }\href {https://doi.org/10.1103/PhysRevE.74.056609} {\bibfield  {journal} {\bibinfo  {journal} {Physical Review E}\ }\textbf {\bibinfo {volume} {74}},\ \bibinfo {pages} {056609} (\bibinfo {year} {2006})}\BibitemShut {NoStop}%
\bibitem [{\citenamefont {Liew}\ \emph {et~al.}(2014)\citenamefont {Liew}, \citenamefont {Popoff}, \citenamefont {Mosk}, \citenamefont {Vos},\ and\ \citenamefont {Cao}}]{2014_Liew}%
  \BibitemOpen
  \bibfield  {author} {\bibinfo {author} {\bibfnamefont {S.~F.}\ \bibnamefont {Liew}}, \bibinfo {author} {\bibfnamefont {S.~M.}\ \bibnamefont {Popoff}}, \bibinfo {author} {\bibfnamefont {A.~P.}\ \bibnamefont {Mosk}}, \bibinfo {author} {\bibfnamefont {W.~L.}\ \bibnamefont {Vos}},\ and\ \bibinfo {author} {\bibfnamefont {H.}~\bibnamefont {Cao}},\ }\bibfield  {title} {\bibinfo {title} {Transmission channels for light in absorbing random media: From diffusive to ballistic-like transport},\ }\href {https://doi.org/10.1103/PhysRevB.89.224202} {\bibfield  {journal} {\bibinfo  {journal} {Physical Review B}\ }\textbf {\bibinfo {volume} {89}},\ \bibinfo {pages} {224202} (\bibinfo {year} {2014})}\BibitemShut {NoStop}%
\bibitem [{\citenamefont {Yamilov}\ \emph {et~al.}(2022)\citenamefont {Yamilov}, \citenamefont {Bender},\ and\ \citenamefont {Cao}}]{2022_Yamilov_Sum_rules}%
  \BibitemOpen
  \bibfield  {author} {\bibinfo {author} {\bibfnamefont {A.}~\bibnamefont {Yamilov}}, \bibinfo {author} {\bibfnamefont {N.}~\bibnamefont {Bender}},\ and\ \bibinfo {author} {\bibfnamefont {H.}~\bibnamefont {Cao}},\ }\bibfield  {title} {\bibinfo {title} {Sum rules for energy deposition eigenchannels in scattering systems},\ }\href {https://doi.org/10.1364/OL.468697} {\bibfield  {journal} {\bibinfo  {journal} {Optics Letters}\ }\textbf {\bibinfo {volume} {47}},\ \bibinfo {pages} {4889} (\bibinfo {year} {2022})}\BibitemShut {NoStop}%
\bibitem [{\citenamefont {Guan}\ \emph {et~al.}(2012)\citenamefont {Guan}, \citenamefont {Katz}, \citenamefont {Small}, \citenamefont {Zhou},\ and\ \citenamefont {Silberberg}}]{2012_Guan}%
  \BibitemOpen
  \bibfield  {author} {\bibinfo {author} {\bibfnamefont {Y.}~\bibnamefont {Guan}}, \bibinfo {author} {\bibfnamefont {O.}~\bibnamefont {Katz}}, \bibinfo {author} {\bibfnamefont {E.}~\bibnamefont {Small}}, \bibinfo {author} {\bibfnamefont {J.}~\bibnamefont {Zhou}},\ and\ \bibinfo {author} {\bibfnamefont {Y.}~\bibnamefont {Silberberg}},\ }\bibfield  {title} {\bibinfo {title} {Polarization control of multiply scattered light through random media by wavefront shaping},\ }\href {https://doi.org/10.1364/OL.37.004663} {\bibfield  {journal} {\bibinfo  {journal} {Optics Letters}\ }\textbf {\bibinfo {volume} {37}},\ \bibinfo {pages} {4663} (\bibinfo {year} {2012})}\BibitemShut {NoStop}%
\bibitem [{\citenamefont {de~Aguiar}\ \emph {et~al.}(2017)\citenamefont {de~Aguiar}, \citenamefont {Gigan},\ and\ \citenamefont {Brasselet}}]{2017_deAguiar}%
  \BibitemOpen
  \bibfield  {author} {\bibinfo {author} {\bibfnamefont {H.~B.}\ \bibnamefont {de~Aguiar}}, \bibinfo {author} {\bibfnamefont {S.}~\bibnamefont {Gigan}},\ and\ \bibinfo {author} {\bibfnamefont {S.}~\bibnamefont {Brasselet}},\ }\bibfield  {title} {\bibinfo {title} {Polarization recovery through scattering media},\ }\href {https://doi.org/10.1126/sciadv.1600743} {\bibfield  {journal} {\bibinfo  {journal} {Science Advances}\ }\textbf {\bibinfo {volume} {3}},\ \bibinfo {pages} {e1600743} (\bibinfo {year} {2017})}\BibitemShut {NoStop}%
\bibitem [{\citenamefont {Lin}\ and\ \citenamefont {Hsu}(2024)}]{2024_Wade_Transmission_eigenvalues}%
  \BibitemOpen
  \bibfield  {author} {\bibinfo {author} {\bibfnamefont {H.-C.}\ \bibnamefont {Lin}}\ and\ \bibinfo {author} {\bibfnamefont {C.~W.}\ \bibnamefont {Hsu}},\ }\bibfield  {title} {\bibinfo {title} {Full transmission of vectorial waves through 3d multiple-scattering media},\ }\href {https://doi.org/10.1364/OL.532642} {\bibfield  {journal} {\bibinfo  {journal} {Opt. Lett.}\ }\textbf {\bibinfo {volume} {49}},\ \bibinfo {pages} {5035} (\bibinfo {year} {2024})}\BibitemShut {NoStop}%
\bibitem [{202()}]{2024_Cluster}%
  \BibitemOpen
  \href {https://doi.org/10.71674/PH64-N397} {}\bibinfo {note} {The {M}ill {HPC} {C}luster, {M}issouri {U}niversity of {S}cience and {T}echnology (2024), https://scholarsmine.mst.edu/the-mill/1/}\BibitemShut {NoStop}%
\bibitem [{\citenamefont {Fang}\ \emph {et~al.}(2018)\citenamefont {Fang}, \citenamefont {Zhao},\ and\ \citenamefont {Tian}}]{2018_Fang}%
  \BibitemOpen
  \bibfield  {author} {\bibinfo {author} {\bibfnamefont {P.}~\bibnamefont {Fang}}, \bibinfo {author} {\bibfnamefont {L.}~\bibnamefont {Zhao}},\ and\ \bibinfo {author} {\bibfnamefont {C.}~\bibnamefont {Tian}},\ }\bibfield  {title} {\bibinfo {title} {Concentration-of-measure theory for structures and fluctuations of waves},\ }\href {https://doi.org/10.1103/PhysRevLett.121.140603} {\bibfield  {journal} {\bibinfo  {journal} {Physical Review Letters}\ }\textbf {\bibinfo {volume} {121}},\ \bibinfo {pages} {140603} (\bibinfo {year} {2018})}\BibitemShut {NoStop}%
\end{thebibliography}
%

\end{document}